\documentclass[12inch,a4paper]{article}
\usepackage[top=3cm, bottom=2.5cm, left=2.4cm, right=2.7cm]{geometry}
\usepackage{graphicx}
\usepackage{epstopdf}
\usepackage{booktabs}
\usepackage{blkarray}
\usepackage[british]{babel}

 \usepackage[sort,comma,authoryear]{natbib} 

    \setcitestyle{citesep={;}}
\setcitestyle{square}

\usepackage{amssymb}
\usepackage{setspace}
\usepackage{amsfonts}
\usepackage{amsmath}
\usepackage{float}
\usepackage{color}
\usepackage[dvipsnames]{xcolor}
 \usepackage{colortbl}
\usepackage{url}

\definecolor{Crimson}{rgb}{0.6471, 0.1098, 0.1882}

\usepackage{epstopdf}

\newtheorem{remark}{\normalfont \textit{Remark}}

\usepackage{tcolorbox}
\usepackage{epstopdf}
\usepackage{subcaption}
\providecommand{\keywords}[1]{\textit{Key words:} #1}

\usepackage[labelfont=bf]{caption}

\usepackage{titlesec}
\titleformat*{\subsection}{\normalfont \large}
\usepackage{abstract}
\usepackage{slashbox}
\begin{document}



\title{The effect of the spatial domain in FANOVA models with ARH(1) error term}

\author{Javier \'Alvarez-Li\'ebana$^1$ and M. Dolores Ruiz--Medina$^1$}
\maketitle
\begin{flushleft}
$^1$ Department of Statistics and O. R., University of Granada, Spain.

\textit{E-mail: javialvaliebana@ugr.es}
\end{flushleft}

\doublespacing



\renewcommand{\absnamepos}{flushleft}
\setlength{\absleftindent}{0pt}
\setlength{\absrightindent}{0pt}
\renewcommand{\abstractname}{Summary}
\begin{abstract}
Functional Analysis of Variance (FANOVA) from Hilbert-valued correlated data with spatial rectangular or circular supports is analyzed, when Dirichlet conditions are assumed on the boundary. Specifically, a Hilbert-valued fixed effect model with error term defined from an Autoregressive Hilbertian process of order one (ARH(1) process) is considered, extending the formulation given in Ruiz-Medina (2016). A new statistical test is also derived to contrast the significance of the functional fixed effect parameters. The Dirichlet conditions established at the boundary affect the dependence range of the correlated error term. While the rate of convergence to zero of the eigenvalues of the covariance kernels, characterizing the Gaussian functional error components, directly affects the stability of the generalized least-squares parameter estimation problem. A simulation study and a real-data application related to fMRI analysis are undertaken to illustrate the performance of the parameter estimator and statistical test derived.

\vspace{0.5cm}
Published in \textbf{Statistics and Its Interface, 10, pp. 607-628.}
 DOI: \url{doi.org/10.4310/SII.2017.v10.n4.a7}
\end{abstract}


\textcolor{Crimson}{\section{Introduction}
\label{A4:sec:1}}

In the last few decades, functional data analysis
techniques  have grown significantly given the new technologies
available, in particular, in the field of medicine (see, for
instance,  \cite{Sorensenetal13}). High--dimensional data, which are
functional in nature, are generated, for example, from measurements
in time, over spatial grids or images with many pixels (e.g., data
on electrical activity of the heart, data
on electrical activity along the scalp, data reconstructed from medical imaging,
expression profiles in genetics and genomics, monitoring of
continuity activity through accelerometers, etc). Effective
experimental design and modern functional statistics have led to
recent advances in medical imaging, improving, in particular, the
study of human brain function (see, for example, \cite{Delzelletal12}).  Magnetic Resonance Imaging (MRI) data have been analysed
with different aims.  For example, we refer to the studies related
with  cortical thickness (see \cite{LerchEvans05}), where magnetic
resonance imaging data are analysed to detect the spatial locations
of the surface  of the brain, where the cortical thickness
 is correlated with an
independent variable, such as age or gender (see also \cite{Shawetal06}).
Cortical thickness is usually previously smoothed along the surface
of the brain (see \cite{Chungetal05}). Thus, it can be considered as a
functional random variable with spatial circular support.  In
general, the following linear model is considered, for cortical thickness $Y_{i}(s)$ on subjects $i=1,\dots,n,$:
\begin{equation}
Y_{i}(\mathbf{s})=\boldsymbol{x}_{i}\boldsymbol{\beta} (\mathbf{s}) +
Z_{i}(\mathbf{s})\sigma_{i}(\mathbf{s}),\quad \mathbf{s}\in
S,\label{A4:ex1}
\end{equation}
\noindent where $\boldsymbol{x}_{i}$ is a vector of known $p$ regressors, and for each $\mathbf{s}\in S,$ with $S$ denoting the surface of the
brain, parameter  $\boldsymbol{\beta} (\mathbf{s})$ is an unknown $p$--vector of
regression coefficients. The errors $\left\lbrace Z_{1},\dots, Z_{n} \right\rbrace$ are
independent zero-mean Gaussian random fields. In \cite{TaylorWorsley07}, this
model is also considered to detect how the regressors are related to
the data at spatial location $\mathbf{s},$ by testing contrasts in
$\boldsymbol{\beta}(\mathbf{s}),$ for $\mathbf{s}\in S.$  The approach presented in
this paper allows the formulation of model (\ref{A4:ex1}) in a
functional (Hilbert--valued) framework, incorporating possible
correlations between subjects, due to genetic characteristics,
breed, geographic location, etc.

The statistical analysis of functional magnetic resonance image (fMRI) data has also generated  an important activity in research about brain
activity, where the functional statistical approach implemented in
this paper could lead to important spatio--temporal analysis
improvements. It is well--known that fMRI techniques have been
developed to address the unobserved effect of scanner noise in
studies of the auditory cortex.
A penalized likelihood approach to magnetic resonance image
reconstruction is presented in \cite{BulaevskayaOehlert07}. A new approach
which incorporates the spatial   information from neighbouring
voxels, as well as temporal correlation within each voxel, which
makes use of regional  kriging is derived in \cite{ChristensenYetkin05}.
 Conditional autoregressive and Markov random field modelling involves some restrictions in the characterization  of spatially contiguous
effect regions, and, in general, in the representation of the spatial dependence between spatially connected voxels (see, for example, \cite{Banerjeeetal04,Besag86}).
 Multiscale adaptive regression models assume spatial independence to construct a weighted likelihood parameter estimate.
   At each scale, the weights determine the amount of information that observations in a neighborhood voxel provides on the parameter vector to be estimated  at a given  voxel, under the assumption of independence between the conditional distributions of the responses  at the neighborhood voxels, for each scale. The weights are sequentially computed through different scales, for  adaptively update of the  parameter estimates and test statistics (see, for example, \cite{Lietal11}).

 In \cite{Zhuetal12}, a multivariate varying coefficient model is considered for neuroimaging data, under a mixed effect approach, to reflect dependence within--curve and between--curve, in the case where coefficients are one--parameter functions, although extension to higher dimension is straightforward. The approach presented in this paper adopts a functional framework to analyse multivariate varying coefficient models  in higher dimensions (two--dimensional design points), under the framework of multivariate fixed effect models in Hilbert spaces. Namely, the response  is a multivariate functional random variable reflecting dependence within-surface (between voxels), and between-surface (between different times), with Hilbert--valued multivariate Gaussian distribution. Hence, the varying coefficients are estimated from the application of an extended version of generalized least--squares estimation methodology, in
the multivariate Hilbert--valued context (see \cite{Ruiz16}), while, in \cite{Zhuetal12}, local linear regression is applied to estimate the coefficient functions. The dependence structure of the functional response is estimated here from the moment--based parameter estimation of the ARH(1) error term (see  \cite{Bosq00}). In \cite{Zhuetal12},  local linear regression technique is employed to estimate
the  random effects, reflecting dependence structure in the varying coefficient mixed effect model. An extended  formulation of the varying coefficient model considered in  \cite{Zhuetal12} is given in \cite{Zhuetal14}, combining a univariate measurement mixed effect model, a jumping surface model, and a functional component analysis model. In the approach presented in this paper, we have combined a nonparametric surface model with a multivariate  functional principal component approach in the ARH(1) framework. Thus,  a continuous spatial variation of the fMRI response is assumed,   incorporating
temporal and spatial
correlations (across voxels), with an important dimension reduction  in the estimation of the varying coefficient functions.

The above--referred  advances in medicine are supported by
the  extensive literature on linear models in function spaces developed in parallel in the last few decades. We particularly refer
to the functional linear regression context (see, for example,
\cite{CaiHall06,Cardotetal03,CardotSarda11,Chiouetal04,Crambesetal09,Cuevasetal02,Ferratyetal13,Kokoszkaetal08}, among others). See also \cite{Bosq00,Bosq07,Ruiz11,Ruiz12}, in the
functional time series context, and \cite{FerratyVieu06,FerratyVieu11} in the functional nonparametric regression
   framework. Functional
Analysis of Variance (FANOVA) techniques for  high--dimensional data
with a functional background have played  a crucial role, within the
functional linear model literature as well. Related work has been steadily
growing (see, for example, \cite{Angelinietal03,DetteDerbort01,Gu02,Huang98,KaufmanSain10,Kaziska11,Lin00,RamsaySilverman05,Spitzneretal03,Stoneetal97,Wahbaetal95}). The
paper  \cite{Ruiz16} extends the results in \cite{Zoglat08} from the
$L^{2}([0,1])$--valued context to the separable Hilbert--valued space
framework, and from the case of independent homocedastic error
components to the correlated heteroscedastic case. In the context of
hypothesis testing from functional data, tests of significance based
on wavelet thresholding are formulated in
 \cite{Fan96},  exploiting the
sparsity of the signal representation in the  wavelet domain,  for
dimension reduction.  A maximum likelihood ratio based test is
suggested for functional variance components in mixed--effect FANOVA
models in \cite{Guo02}. From classical ANOVA tests, an asymptotic
approach is derived in \cite{Cuevasetal04}, for studying the equality of
the functional means from $k$ independent samples of functional data.
The testing problem for mixed--effect functional analysis of variance
models is addressed in \cite{AbramovichAngelini06,Abramovichetal04},
developing asymptotically optimal (minimax) testing procedures for
the significance of functional global trend, and the functional
fixed effects. The wavelet transform of the data is again used in
the implementation of this approach (see also \cite{AntoniadisSapatinas07}).
 Recently, in the context of functional data defined by curves, considering the
$L^{2}$--norm,  an up--to--date overview of hypothesis testing methods
for functional data analysis is provided in \cite{Zhang13}, including
 functional ANOVA,
functional linear models with functional responses, heteroscedastic
ANOVA for functional data, and hypothesis tests for the equality of
covariance functions, among other related topics.

  In this paper, the model formulated in \cite{Ruiz16} is extended to the case where the error term is an  ARH(1) process. Furthermore,  an
  alternative  test to contrast the significance of the functional fixed
  effect parameters is formulated, based on a sharp form of
  the Cram\'er--Wold's Theorem derived in \cite{Cuestaetal07}, for
  Gaussian measures on a separable Hilbert space. The simulation
  study undertaken illustrates the effect of the boundary conditions and the geometry of the domain on the spatial dependence range of
  the functional vector error term. Specifically, in that simulations, we consider the  case where the Gaussian error components satisfy a stochastic partial differential equation, given in terms
  of a fractional power of the Dirichlet negative Laplacian operator. The autocovariance and cross--covariance operators  of the functional  error components are then defined in terms of the eigenvectors of the Dirichlet negative Laplacian operator. The eigenvectors of the Dirichlet negative Laplacian operator vanish continuously at the boundary, in the case of  the regular domains studied (the rectangle, disk and circular sector), with decay velocity determined by the boundary conditions and the geometry of the domain.  Thus, the boundary conditions and the geometry of the domain directly affect the dependence range of the error components,  determined by the rate of convergence to zero of the   Dirichlet negative Laplacian eigenvectors at the boundary. The influence of the truncation order is studied as well, since the rate of convergence to zero of the eigenvalues of the spatial covariance kernels, that define the matrix covariance operator of the error term, could affect the stability of the generalized least--squares estimation problem addressed here.
  Furthermore, in the fMRI  data problem considered, the presented  functional fixed effect model, with ARH(1) error term, is fitted. In that case,
  the temporal
  dependence range of the error term is controlled by the ARH(1) dynamics, while the
  spatial dependence range is controlled by the boundary conditions. Thus, the performance of the functional least--squares estimator and the functional significance test introduced in this paper is illustrated in both cases, the simulation study and the real--data example considered. A comparative study with the classical approach presented in
  \cite{Worsleyetal02} is also achieved for the fMRI data set analysed (freely available at  \url{http://www.math.mcgill.ca/keith/fmristat/}).

The outline of this paper is as follows. The  functional fixed
effect model with ARH(1) error term is formulated in \textcolor{Crimson}{Appendix} \ref{A4:sec:3}. The main results obtained  on generalized least--squares
estimation of the Hilbert--valued vector  of fixed effect parameters,
 and the functional analysis of variance
are also collected in this appendix. Linear hypothesis testing is
derived in \textcolor{Crimson}{Appendix} \ref{A4:scwth}.
 The results obtained from the
simulation study undertaken are displayed in \textcolor{Crimson}{Appendix} \ref{A4:sec:6}.
Functional statistical analysis of fMRI data is given in \textcolor{Crimson}{Appendix}
\ref{A4:sec: fmri}. Conclusions and open research lines are provided in \textcolor{Crimson}{Appendix} \ref{A4:sec:64}. Finally, the \textcolor{Crimson}{Supplementary Material} in \textcolor{Crimson}{Appendix} \ref{A4:Supp} introduces the required preliminary elements  on eigenvectors and eigenvalues of the Dirichlet negative Laplacian operator on the
 rectangle, disk and circular sector.

%
%

\textcolor{Crimson}{\section{Multivariate Hilbert--valued fixed effect model with ARH(1) error term}
\label{A4:sec:3}}

  This section provides the extended formulation of the
 multivariate Hilbert--valued  fixed effect model studied in \cite{Ruiz16}, to the case where
 the correlated functional components of the error term  satisfy an ARH(1) state equation.  In that formulation,  compactly supported non--separable   autocovariance
 and cross--covariance kernels are considered for the functional error components, extending the separable case studied in \cite{Ruiz16}.

Denote by $H$ a real separable Hilbert space with the inner product
$\left\langle \cdot,\cdot\right\rangle_{H},$ and
 the associated norm $\|\cdot\|_{H}.$
Let us first introduce  the multivariate Hilbert--valued fixed effect
model with ARH(1) error term

\begin{equation}
\mathbf{Y} \left(\cdot\right) = \mathbf{X}\boldsymbol{\beta
}\left(\cdot\right) +
\boldsymbol{\varepsilon}\left(\cdot\right), \label{A4:26mhem}
\end{equation}
\noindent where $\mathbf{X}$ is  a real-valued $n\times p$ matrix,
the fixed effect design matrix, $$\boldsymbol{\beta
}(\cdot)=[\beta_{1}(\cdot),\dots,\beta_{p}(\cdot)]^{T}\in H^{p}$$
represents the vector of fixed effect parameters,
$$\mathbf{Y}(\cdot)=[Y_{1}(\cdot),\dots,Y_{n}(\cdot)]^{T}$$ is the
 $H^{n}$-valued Gaussian  response, with
${\rm E} \left\lbrace \mathbf{Y} \right\rbrace=\mathbf{X}\boldsymbol{\beta }$. The $H^{n}$-valued
error term
$$\boldsymbol{\varepsilon}(\cdot)=[\varepsilon_{1}(\cdot),\dots,\varepsilon_{n}(\cdot)]^{T}$$
is assumed to be an ARH(1) process on the basic probability space
$(\Omega, \mathcal{A},\mathcal{P})$;  i.e.,  a stationary in time Hilbert--valued Gaussian process satisfying  (see \cite{Bosq00})
\begin{equation}
\varepsilon_{m} \left( \cdot \right) = \rho \left(\varepsilon_{m-1} \right) \left( \cdot \right) + \nu_{m} \left( \cdot \right),\quad m \in
\mathbb{Z}, \label{A4:24arh1} 
\end{equation}
\noindent  where  ${\rm E} \left\lbrace \varepsilon_{m} \right\rbrace=0,$ for each $m\in \mathbb{Z},$ and $\rho $ denotes the autocorrelation operator of the error
process $\varepsilon,$ which belongs to the space of bounded linear operators on $H.$ Here, $\nu= \left\lbrace \nu_{m }, \ m\in
\mathbb{Z} \right\rbrace$ is assumed to be a   Gaussian strong white noise; i.e.,
$\nu $ is a Hilbert--valued zero--mean stationary process,
with independent and identically  distributed components in time,
and with $\sigma^{2}={\rm E} \left\lbrace \|\nu_{m}\|_{H}^{2} \right\rbrace <\infty,$ for all
$m\in \mathbb{Z}.$ Thus, in (\ref{A4:26mhem}), the components of the vector error term $[\varepsilon_{1}(\cdot),\dots,\varepsilon_{n}(\cdot)]^{T}$ corresponding to observations at times $t_{1},\dots,t_{n},$ obey the functional state equation  (\ref{A4:24arh1}), under suitable conditions on the  point spectrum of the autocorrelation operator $\rho$. Hence, the non--null functional
entries of the matrix covariance   operator $\mathbf{R}_{\boldsymbol{\varepsilon}\boldsymbol{\varepsilon}}$ of $$\boldsymbol{\varepsilon}(\cdot)=[\varepsilon_{1}(\cdot),\dots,\varepsilon_{n}(\cdot)]^{T}$$\noindent are then constituted by the elements located at the three main diagonals. Specifically,  $${\rm E} \left\lbrace \varepsilon_{i}\otimes
\varepsilon_{j} \right\rbrace =R_{1}, \quad \text{if } j-i=1, \quad {\rm E} \left\lbrace \varepsilon_{i}\otimes
\varepsilon_{j} \right\rbrace=R_{1}^{*},\quad \text{if }  i-j=1,$$ and  $${\rm E} \left\lbrace \varepsilon_{i}\otimes
\varepsilon_{i} \right\rbrace=R_{0},\quad \text{if }  i=j,$$ where $R_{1}$ and $R_{1}^{*}$ denote, respectively, the cross--covariance operator and its adjoint for the ARH(1)
process $\varepsilon = \left\lbrace \varepsilon_{i},\ i\in \mathbb{Z} \right\rbrace,$ and $R_{0}$ represents its autocovariance operator. Note that, in this appendix, it is assumed that $\rho$ is sufficiently regular. In particular, $\rho$ is  such that $ \|\rho^{2}\|_{\mathcal{L}(H)} \simeq 0$.

Equivalently, the matrix covariance   operator $\mathbf{R}_{\boldsymbol{\varepsilon}\boldsymbol{\varepsilon}}$ is given by
\begin{eqnarray}\mathbf{R}_{\boldsymbol{\varepsilon}\boldsymbol{\varepsilon}}&=&{\rm E} \left\lbrace \left[ \varepsilon_{1}(\cdot),\dots,\varepsilon_{n}(\cdot)\right]^{T}
\left[\varepsilon_{1}(\cdot),\dots,\varepsilon_{n}(\cdot)\right] \right\rbrace
\nonumber\\
&= &\left(\begin{array}{ccc}{\rm E} \left\lbrace \varepsilon_{1}\otimes
\varepsilon_{1} \right\rbrace & \ldots &
{\rm E} \left\lbrace\varepsilon_{1}\otimes \varepsilon_{n}\right\rbrace\\
\vdots & \ddots & \vdots \\
{\rm E} \left\lbrace\varepsilon_{n}\otimes \varepsilon_{1} \right\rbrace&\ldots&
{\rm E} \left\lbrace\varepsilon_{n}\otimes \varepsilon_{n} \right\rbrace
\end{array}\right)\nonumber\\
&\simeq &\left(\begin{array}{cccccccc}R_{0}& R_{1}& 0_H & 0_H & \ldots &
0_H & 0_H & 0_H\\
R_{1}^{*} & R_{0} & R_{1} & 0_H & \ldots &
0_H & 0_H & 0_H \\
\vdots & \vdots & \vdots & \vdots &\ddots &\vdots & \vdots & \vdots\\
0_H& 0_H& 0_H & 0_H& \ldots & R_{1}^{*}& R_{0} & R_{1} \\
0_H& 0_H& 0_H & 0_H& \ldots & 0_H& R_{1}^{*}&
R_{0} \\
\end{array}\right), \nonumber 
\end{eqnarray}
\noindent where $0_H$ denotes the approximation by zero in the corresponding operator norm, given the conditions imposed on $\rho$.

 In the space $\mathcal{H}=H^n,$ we  consider the  inner product
$$\langle\mathbf{f}, \mathbf{g}\rangle_{H^n} = \displaystyle \sum_{i=1}^{n}
\langle f_i, g_i \rangle_H,~\quad \mathbf{f},\mathbf{g} \in H^n.$$

It is well--known that the autocovariance operator $R_{0}$ of an ARH(1) process is in the trace class  (see \cite[pp. 27--36]{Bosq00}).
Therefore, it admits a
diagonal spectral decomposition
$$R_{0}=\displaystyle \sum_{k=1}^{\infty} \lambda_{k}\phi_{k}\otimes \phi_{k},$$
in terms of a  complete orthogonal
eigenvector system $\left\lbrace \phi_{k}, \ k\geq 1 \right\rbrace,$  defining in  $H$ a
resolution of the identity $\displaystyle \sum_{k=1}^{\infty }\phi_{k}\otimes
\phi_{k}$. Here, for each $k\geq 1,$ $\lambda_{k}=\lambda_{k}(R_{0})$ is the $k$--th eigenvalue of $R_{0},$ with \linebreak $R_{0} \left(\phi_{k} \right)=\lambda_{k}(R_{0})\phi_{k}.$
 The following series expansion then holds, in the mean--square sense:
\begin{equation}
\varepsilon_i = \displaystyle \sum_{k=1}^{\infty} \left\langle
\varepsilon_i, \phi_k \right\rangle_H \phi_k  = \displaystyle
\sum_{k=1}^{\infty} \sqrt{\lambda_{k}}\eta_{k}(i)
\phi_k, \quad i=1,\dots,n, \nonumber 
\end{equation}
\noindent where $\eta_{k}(i) = \frac{1}{\sqrt{\lambda_{k}}}\langle
\varepsilon_i, \phi_k \rangle_{H},$ for $k\geq 1$ and $i\in \mathbb{N}.$

\bigskip

The following assumption is made:

\bigskip

\noindent  \textcolor{Aquamarine}{\textbf{Assumption A0.}} The
standard Gaussian random variable sequences $\{\eta_{k}(i),\ k\geq 1, ~i \in \mathbb{N} \},$ with, for each $k\geq 1,$
 $$\sqrt{\lambda_{k}}\eta_{k}(i)=\left\langle \varepsilon_{i},
\phi_{k}\right\rangle_{H},$$ for every  $i\in \mathbb{N},$
satisfy the following orthogonality condition, for every $i,j\in \mathbb{N},$
\begin{equation}
{\rm E} \left\lbrace \eta_{k}(i)\eta_{p}(j) \right\rbrace =\delta_{k,p},\quad k,p\in \mathbb{N}, \nonumber 
\end{equation} 
\noindent where $\delta $ denotes the
Kronecker delta function, and 
\begin{eqnarray}
R_{1}&=&\sum_{k=1}^{\infty}\lambda_{k}(R_{1})\phi_{k}\otimes \phi_{k}, \quad  R_{1}^{*} = \sum_{k=1}^{\infty}\lambda_{k}(R_{1}^{*})\phi_{k}\otimes \phi_{k}.\nonumber  
\end{eqnarray}

\bigskip

Under  \textcolor{Aquamarine}{\textbf{Assumption A0}}, the computation of the generalized
least--squares estimator of
$[\beta_{1}(\cdot),\dots,\beta_{p}(\cdot)]^{T}$ is achieved by
projection into the orthogonal basis of eigenvectors
$\left\lbrace \phi_k, \ k \geq 1 \right\rbrace$ of the autocovariance operator
$R_{0}$ of the ARH(1) process $\varepsilon = \left\lbrace \varepsilon_{i},\ i\in \mathbb{Z} \right\rbrace.$
 Denote by $\boldsymbol{\Phi}^{*}$ the projection operator  into the eigenvector  system $\left\lbrace \phi_k, \ k \geq 1 \right\rbrace,$
 acting on a vector function $\mathbf{f}\in \mathcal{H}=H^{n}$  as follows:  
 \begin{eqnarray}
 \boldsymbol{\Phi}^* \left( \mathbf{f}\right) &=& \left\lbrace \boldsymbol{\Phi}_{k}^{*}
 \left( \mathbf{f} \right), \ k \geq 1 \right\rbrace =  \left\lbrace\left( \langle f_1,\phi_k\rangle,\dots,\langle f_n,\phi_k \rangle \right)^{T}, \ k \geq 1 \right\rbrace
 \nonumber\\
&=&  \left\lbrace \left(f_{k1},\dots,f_{kn} \right)^{T}, \ k \geq 1 \right\rbrace = \left\lbrace \mathbf{f}_{k}^{T}, \ k \geq 1 \right\rbrace, \label{A4:eqinnprod}
\end{eqnarray}
 \noindent where  $\boldsymbol{\Phi} \boldsymbol{\Phi}^{*} = \mathbf{Id}_{\mathcal{H}=H^n},$ with

$$\boldsymbol{\Phi} \left( \left\lbrace  \mathbf{f}_{k}^{T}, \ k \geq 1 \right\rbrace \right)  = \left(\displaystyle \sum _{k=1}^{\infty} f_{k1}\phi_k,\ldots, \displaystyle \sum _{k=1}^{\infty} f_{kn}
\phi_k \right)^{T}.$$

\bigskip

 For  $\mathbf{A} = \left\lbrace A_{i,j} \right\rbrace_{i=1,\ldots,n}^{j=1,\ldots,n}$ be a matrix operator  such that, for each $i,j=1,\dots,n,$ its functional entries are given by $$A_{i,j} = \displaystyle \sum _{k=1}^{\infty} \gamma_{kij} \phi_k \otimes \phi_k$$ 
  with $\displaystyle \sum _{k=1}^{\infty} \gamma_{kij}^{2} < \infty.$ The following identities are straightforward:

\begin{eqnarray}
\boldsymbol{\Phi}^{*}\mathbf{A}\boldsymbol{\Phi} &=& \left\lbrace
\boldsymbol{\Gamma}_k, \ k \geq 1 \right\rbrace, \quad \boldsymbol{\Phi}\left(\left\lbrace
\boldsymbol{\Gamma}_k, \ k \geq 1 \right\rbrace \right)
\boldsymbol{\Phi}^{*} = \mathbf{A}, \label{A4:29}
\end{eqnarray}
\noindent where, for each $k\geq 1,$ the entries of $\boldsymbol{\Gamma}_k$ are $\Gamma_{kij} = \gamma_{kij},$ for $i,j=1,\dots,n.$

\bigskip

 Applying  (\ref{A4:eqinnprod})--(\ref{A4:29}),  we directly obtain

\begin{eqnarray}
\boldsymbol{\Phi}^{*}\mathbf{R}_{\boldsymbol{\varepsilon}
\boldsymbol{\varepsilon}}\boldsymbol{\Phi} &=&
\left\lbrace\boldsymbol{\Lambda}_k, \ k \geq 1 \right\rbrace, \quad
\boldsymbol{\Phi}^{*}\mathbf{R}_{\boldsymbol{\varepsilon}
\boldsymbol{\varepsilon}}^{-1}\boldsymbol{\Phi} =
\left\lbrace\boldsymbol{\Lambda}_{k}^{-1}, \ k \geq 1 \right\rbrace, \nonumber \\
\mathbf{R}_{\boldsymbol{\varepsilon}
\boldsymbol{\varepsilon}}^{-1}\left(\mathbf{f},\mathbf{g} \right)
&=& \boldsymbol{\Phi}^{*}\mathbf{R}_{\boldsymbol{\varepsilon}
\boldsymbol{\varepsilon}}^{-1} \boldsymbol{\Phi}
\left(\boldsymbol{\Phi}^{*}\mathbf{f},\boldsymbol{\Phi}^{*}\mathbf{g}\right)
=
\langle \mathbf{f}, \mathbf{g} \rangle_{\mathbf{R}_{\boldsymbol{\varepsilon} \boldsymbol{\varepsilon}}^{-1}} = 
\displaystyle \sum_{k=1}^{\infty}\mathbf{f}_{k}^{T}
\boldsymbol{\Lambda}_{k}^{-1} \mathbf{g}_k, \quad
\mathbf{f},~\mathbf{g} \in \mathbf{R}_{\boldsymbol{\varepsilon}
\boldsymbol{\varepsilon}}^{1/2}\left(H^n \right), \nonumber \\ 
\Vert \mathbf{f}\Vert_{\mathbf{R}_{\boldsymbol{\varepsilon}
\boldsymbol{\varepsilon}}^{-1}}^{2} &=& \displaystyle
\sum_{k=1}^{\infty} \mathbf{f}_{k}^{T}
\boldsymbol{\Lambda}_{k}^{-1}\mathbf{f}_{k}, \quad  \mathbf{f}\in
\mathbf{R}_{\boldsymbol{\varepsilon}
\boldsymbol{\varepsilon}}^{1/2}\left(H^n\right), \label{A4:33}
\end{eqnarray}
\noindent where, for each
$k\geq 1$, $\boldsymbol{\Lambda}_k=\boldsymbol{\Phi}^{*}_{k}\mathbf{R}_{\boldsymbol{\varepsilon}
\boldsymbol{\varepsilon}}\boldsymbol{\Phi}_{k}$ is given by
\begin{equation}\boldsymbol{\Lambda}_k=\left[\begin{array}{cccccc}\lambda_{k}(R_{0})& \lambda_{k}(R_{1})& 0 &\dots & 0 & 0 \\
\lambda_{k}(R_{1}^{*}) & \lambda_{k}(R_{0}) & \lambda_{k}(R_{1}) &\dots & 0 & 0 \\
\vdots & \vdots & \vdots &\ddots &\vdots &\vdots\\
0& 0 & 0 & \dots & \lambda_{k}(R_{1}^{*})&
\lambda_{k}(R_{0})\\
\end{array}\right],
\label{A4:covest2bb}
\end{equation}
\noindent with
$\boldsymbol{\Lambda}_k^{-1}$ denoting its inverse matrix.

\bigskip

\begin{remark}
\label{A4:stdre}
\textit{In \textcolor{Crimson}{Appendix} \ref{A4:sec:6}, we restrict our attention to the functional error model studied in \cite{Ruiz16}, considering the  Hilbert--valued stochastic partial differential equation system framework. In that framework, matrices  $\left\lbrace \boldsymbol{\Lambda}_{k}, \ k\geq 1 \right\rbrace,$ are known, since they are defined  from the eigenvalues of the differential operators involved in the equation system. Particularly, in that section,
 for each $k\geq 1,$ matrix $\boldsymbol{\Lambda}_{k}$ is considered to have entries $\Lambda_{kij}$ given by
\begin{eqnarray}\Lambda_{kij} &=& \exp\left( -\frac{\vert i - j
\vert}{\lambda_{ki} +
\lambda_{kj}}\right), \quad \text{if } i\neq j,\nonumber\\
\Lambda_{kii}
&=&\lambda_{ki}=\lambda_{k}\left([f_{i}(-\boldsymbol{\Delta}_{D_{l}})]^{2}\right)=\lambda_{k}\left((-\boldsymbol{\Delta}_{D_{1}})^{-2(d-\gamma_{i})}\right) = \left[\lambda_{k}\left((-\boldsymbol{\Delta}_{D_{1}})\right)\right]^{-2(d-\gamma_{i})}, \nonumber \\
\label{A4:covopf}
\end{eqnarray}
\noindent with $$\gamma_{i} \in (0,d/2), \quad i=1,\dots,n,$$ and
$(-\boldsymbol{\Delta}_{D_{l}})$ representing the Dirichlet negative
Laplacian operator on domain $D_{l},$ for $l=1$ (the rectangle),
$l=2$ (the disk) and $l=3$ (the circular sector). However, in practice, as shown in \textcolor{Crimson}{Appendix} \ref{A4:sec: fmri}  in the analysis of  fMRI data, matrices   $\left\lbrace \boldsymbol{\Lambda}_{k}, \ k\geq 1 \right\rbrace,$ are not known, and  should be estimated from the data. Indeed, in that real--data example, we approximate the entries of $\left\lbrace \boldsymbol{\Lambda}_{k}, \ k\geq 1 \right\rbrace,$ from the coefficients (eigenvalues and singular values), that define the diagonal spectral expansion   of the empirical autocovariance $\widehat{R_{0}}$ and cross covariance $\widehat{R_{1}}$ operators, given by   (see  \cite{Bosq00})
\begin{eqnarray}
\widehat{R_{0}} &=& \frac{1}{n}\sum_{i=1}^{n}\varepsilon_{i}\otimes \varepsilon_{i}, \quad \widehat{R_{1}} = \frac{1}{n-1}\sum_{i=1}^{n-1}\varepsilon_{i}\otimes \varepsilon_{i+1}, \quad 
\widehat{R_{1}^{\ast}} = \frac{1}{n-1}\sum_{i=2}^{n}\varepsilon_{i}\otimes \varepsilon_{i-1}.
\label{A4:eqempiricalcov}
\end{eqnarray}}
\end{remark}

\bigskip

We also consider here the following semi--orthogonal condition for the non-square design matrix $\mathbf{X}:$

\bigskip

\textcolor{Aquamarine}{\textbf{Assumption A1.}} The fixed effect design matrix $\mathbf{X}$
is a semi--orthogonal non--square matrix. That is,
\begin{equation}
\mathbf{X}^{T} \mathbf{X} = \mathbf{Id}_p, \quad \mathbf{Id}_p \in \mathbb{R}^{p \times p}. \nonumber 
\end{equation}

\bigskip

\begin{remark}
\textit{\textcolor{Aquamarine}{\textbf{Assumption A1}} implies (see \cite{Ruiz16})
\begin{equation}
\displaystyle \sum_{k=1}^{\infty} {\rm Tr} \left( \mathbf{X}^T
\boldsymbol{\Lambda}_{k}^{-1} \mathbf{X}\right)^{-1} < \infty. \nonumber  
\end{equation}}
\end{remark}

\bigskip

  The generalized least--squares estimation of
$[\beta_{1}(\cdot),\ldots,\beta_{p}(\cdot)]^{T}$ is achieved by minimizing the loss \linebreak quadratic function in the norm
of the Reproducing Kernel
 Hilbert Space (RKHS norm). Note that, for an $\mathcal{H}$--valued  zero--mean Gaussian random variable with autocovariance operator $R_Z,$ the RKHS of $Z$  is defined by
$$ \mathcal{H} \left(Z \right) = R_Z^{1/2} \left(\mathcal{H}\right)$$
(see, for example,  \cite{PratoZabczyk02}).

From equation (\ref{A4:33}) we get

\begin{equation}
{\rm E} \left\lbrace \Vert \mathbf{Y} - \mathbf{X}\boldsymbol{\beta }
\Vert_{\boldsymbol{R}_{\boldsymbol{\varepsilon}
\boldsymbol{\varepsilon}}^{-1}}^{2} \right\rbrace =  \boldsymbol{R}_{\boldsymbol{\varepsilon}
\boldsymbol{\varepsilon}}^{-1} \left(\boldsymbol{\varepsilon}
\right)\left(\boldsymbol{\varepsilon} \right) = \displaystyle
\sum_{k=1}^{\infty} {\rm E} \left \lbrace \Vert \boldsymbol{\varepsilon}_k
\left(\boldsymbol{\beta}_k \right)
\Vert_{\boldsymbol{\Lambda}_{k}^{-1}}^{2} \right\rbrace \simeq  \displaystyle
\sum_{k=1}^{\infty} {\rm E} \left\lbrace \Vert \boldsymbol{\varepsilon}_k
\left(\boldsymbol{\beta}_k \right)
\Vert_{\widehat{\boldsymbol{\Lambda}}_{k}^{-1}}^{2} \right\rbrace, \label{A4:45}
\end{equation}
\noindent where, in the last identity,  for each $k\geq 1,$ matrix $\widehat{\boldsymbol{\Lambda}}_{k}$  represents the empirical counterpart of $\boldsymbol{\Lambda}_{k},$ constructed from the eigenelements of  $\widehat{R_{0}},$  $\widehat{R_{1}}$ and $\widehat{R_{1}}^{\ast},$ considered when   $R_{0}$ and $R_{1}$ are unknown. Here,  $$\boldsymbol{\varepsilon} = \mathbf{Y} - \mathbf{X}
\boldsymbol{\beta }, \quad \boldsymbol{\varepsilon}_k
\left(\boldsymbol{\beta }_k \right) = \boldsymbol{\Phi}_{k}^{*}
\left( \mathbf{Y}- \mathbf{X}\boldsymbol{\beta } \right), \quad k
\geq 1.$$ The minimum of equation (\ref{A4:45}) is attached if, for each $k\geq
1,$  the expectation $${\rm E} \left\lbrace \Vert \boldsymbol{\varepsilon}_k
\left(\boldsymbol{\beta }_k \right)
\Vert_{\boldsymbol{\Lambda}_{k}^{-1}}^{2} \right\rbrace$$ is minimized, with, as before, $\boldsymbol{\Lambda}_{k}^{-1}$ defining the inverse of matrix
 $\boldsymbol{\Lambda}_{k}$ given in (\ref{A4:covest2bb}) (and approximated by $\widehat{\boldsymbol{\Lambda}}_{k},$ when $R_{0}$ and $R_{1}$ are unknown). That is,
 \begin{eqnarray}
 \widehat{\boldsymbol{\beta}}_{k} &=&
\left(\widehat{\beta_{k1}},\dots,\widehat{\beta_{kp}} \right)^T =
\left(\mathbf{X}^T \boldsymbol{\Lambda}_{k}^{-1} \mathbf{X}
\right)^{-1} \mathbf{X}^T\boldsymbol{\Lambda}_{k}^{-1}
\mathbf{Y}_k, \nonumber 
\end{eqnarray}
\noindent and given by
\begin{equation}
\left(\widetilde{\beta_{k1}},\dots,\widetilde{\beta_{kp}} \right)^T =\left(\mathbf{X}^T \widehat{\boldsymbol{\Lambda}}_{k}^{-1} \mathbf{X}
\right)^{-1} \mathbf{X}^T\widehat{\boldsymbol{\Lambda}}_{k}^{-1}
\mathbf{Y}_k, \label{A4:ppe}
\end{equation} 
\noindent in the case where  $R_{0}$ and $R_{1}$ are unknown. Here, $\mathbf{Y}_k
= \boldsymbol{\Phi}^{*}_{k}\left( \mathbf{Y} \right)$ is the
vector of projections into  $\phi_k$ of the components of
$\mathbf{Y},$ for each $k\geq 1.$

 In the remaining of this  section, we restrict our attention to the case where  $R_{0}$ and $R_{1}$ are known. In that case,
\begin{equation}
\widehat{\boldsymbol{\beta }} = \boldsymbol{\Phi} \left( \left\lbrace
\widehat{\boldsymbol{\beta }_k}, \ k \geq 1 \right\rbrace \right) =
\left(\displaystyle \sum_{k=1}^{\infty} \widehat{\beta_{k1}}\phi_k,
\dots,\displaystyle \sum_{k=1}^{\infty} \widehat{\beta_{kp}}\phi_k
\right)^{T}. \nonumber 
\end{equation}

 The estimated response is
then given by $\widehat{\mathbf{Y} } = \mathbf{X}
\widehat{\boldsymbol{\beta }}$. Under \textcolor{Aquamarine}{\textbf{Assumption A1}},

\begin{eqnarray}
{\rm E} \left\lbrace \displaystyle \sum_{k=1}^{\infty}\sum_{i=1}^{p}\widehat{\beta}_{ki}^{2}\right\rbrace &=& \displaystyle \sum_{k=1}^{\infty} {\rm Tr} (\mathbf{X}^{T}\boldsymbol{\Lambda
}_{k}^{-1} \mathbf{X})^{-1}+\| \boldsymbol{\beta }\|^{2}_{H^{p}}
<\infty,
 \label{A4:proofpr11}
\end{eqnarray}
\noindent  i.e.,  $\widehat{\boldsymbol{\beta}} \in H^p$ almost
surely (see  \cite{Ruiz16} for more details).

\bigskip

\begin{remark}
\label{A4:B00} 
\textit{In the case where $R_{0}$ and $R_{1}$ are unknown, under
the conditions assumed in \cite[Corollary 4.2, pp. 101--102]{Bosq00}, strong consistency of the empirical autocovariance
operator $\widehat{R_{0}}$ holds. Moreover, under  the conditions
assumed in \cite[Theorem 4.8, pp. 116--117]{Bosq00}, the
empirical cross--covariance operator $\widehat{R_{0}}$ is strongly--consistent. Therefore, the plug--in functional parameter estimator
 (\ref{A4:ppe}) satisfies (\ref{A4:proofpr11}), for $n$ sufficiently large.}
 \end{remark}

\bigskip

The Functional Analysis of Variance  of model in (\ref{A4:26mhem})--(\ref{A4:24arh1}) can be achieved as described in \cite{Ruiz16}. Specifically,
a linear transformation of the functional data should be considered, for the almost surely finiteness of the  functional components of variance, in the following way:
\begin{equation}
\mathbf{W}\mathbf{Y}=\mathbf{W}\mathbf{X}\boldsymbol{\beta
}+\mathbf{W}\boldsymbol{\varepsilon},\label{A4:tfdm} 
\end{equation}
\noindent where $\mathbf{W}$ is such that

\[\mathbf{W} =\left( \begin{array}{ccc}
\displaystyle \sum_{k=1}^{\infty} w_{k11} \phi_k \otimes \phi_k  & \dots  & \displaystyle \sum_{k=1}^{\infty} w_{k1n} \phi_k \otimes \phi_k \\
\vdots & \ddots  & \vdots  \\
\displaystyle \sum_{k=1}^{\infty} w_{kn1} \phi_k \otimes \phi_k  &
\dots & \displaystyle \sum_{k=1}^{\infty} w_{knn} \phi_k
\otimes \phi_k
\end{array} \right),\]

\noindent and satisfies

\begin{equation}
\sum_{k=1}^{\infty} {\rm Tr} \left( \boldsymbol{\Lambda}_{k}^{-1}
\mathbf{W}_{k} \right) < \infty.\label{A4:64}
\end{equation}

 Here, for each $k\geq 1,$  $\boldsymbol{\Lambda}_{k}$ is defined  in (\ref{A4:covest2bb}).
The  functional components of variance associated with the transformed model (\ref{A4:tfdm}) are then given by
\begin{eqnarray}
\widetilde{SST} &=& \langle \mathbf{W} \mathbf{Y},\mathbf{W}
\mathbf{Y} \rangle_{R_{\boldsymbol{\varepsilon} \boldsymbol{\varepsilon}}^{-1}} = \displaystyle \sum_{k=1}^{\infty}
\mathbf{Y}_{k}^{T} \mathbf{W}_{k}^{T} \boldsymbol{\Lambda}_{k}^{-1}
\mathbf{W}_k \mathbf{Y}_k,\nonumber \\ 
\widetilde{SSE} &=& \langle \mathbf{W} \left(\mathbf{Y} -
\widehat{\mathbf{Y}} \right),\mathbf{W} \left(\mathbf{Y}-
\widehat{\mathbf{Y}} \right) \rangle_{R_{\boldsymbol{\varepsilon} \boldsymbol{\varepsilon}}^{-1}} = 
\displaystyle \sum_{k=1}^{\infty}\left(
\mathbf{M}_k
\mathbf{W}_k \mathbf{Y}_k \right)^T \boldsymbol{\Lambda}_{k}^{-1}\mathbf{M}_k \mathbf{W}_k \mathbf{Y}_k,\nonumber\\
 \widetilde{SSR} &=& \widetilde{SST} - \widetilde{SSE}. \nonumber 
\end{eqnarray}

\noindent where $\mathbf{M}_k = \mathbf{Id}_{n \times n} -
\mathbf{X} \left(\mathbf{X}^T \boldsymbol{\Lambda}_{k}^{-1}
\mathbf{X}\right)^{-1} \mathbf{X}^T \boldsymbol{\Lambda}_{k}^{-1},$ for
each $k \geq 1.$

The statistics
\begin{equation}
F = \frac{\widetilde{SSR}}{\widetilde{SSE}}, \label{A4:800}
\end{equation}
\noindent provides information on  the relative magnitude between the empirical variability explained by the functional transformed model and the residual variability (see \textcolor{Crimson}{Appendix} \ref{A4:sec:6}).

%
%
%

\textcolor{Crimson}{\section{Significance test from the Cram\'er--Wold's Theorem}
\label{A4:scwth}}

 In \cite{Ruiz16}, a linear functional statistical test
is formulated, with explicit definition of the probability
distribution of the derived functional statistics under the null
hypothesis:
\begin{equation}
H_0:~\mathbf{K} \boldsymbol{\beta }= \mathbf{C}, \nonumber 
\end{equation}
\noindent against $$H_{1}:~\mathbf{K}\boldsymbol{\beta }\neq
\mathbf{C},$$  where $\mathbf{C} \in H^m$ and $$\mathbf{K}: H^p
\longrightarrow H^m$$ is a matrix operator  such that its functional
entries $\mathbf{K}  = \left\lbrace K_{ij} \right\rbrace_{i=1,\dots,m}^{j=1,\dots,p},$ are given, for each $f,g \in H,$ by
\begin{equation}
K_{ij}\left(f\right)\left(g\right) = \displaystyle
\sum_{k=1}^{\infty} \lambda_k \left(K_{ij}\right) \langle \phi_k,g
\rangle_H \langle \phi_k,f \rangle_H. \nonumber 
\end{equation}

In particular, $$\left\lbrace \left( \boldsymbol{\Phi}_{k}^{*}
\mathbf{K} \boldsymbol{\Phi}_{k} \right), \ k \geq 1 \right\rbrace =
\left\lbrace \mathbf{K}_k, \ k \geq 1 \right\rbrace$$ with

\[\mathbf{K}_k = \left( \begin{array}{ccc}
\lambda_k \left(K_{11}\right) & \dots & \lambda_k \left(K_{1p}\right) \\
\vdots & \ddots & \vdots\\
\lambda_k \left(K_{m1}\right) & \dots & \lambda_k
\left(K_{mp}\right) \end{array} \right) \in \mathbb{R}^{m \times p}.\]

 At level
$\alpha ,$ there exists a test $\psi $ given by:
$$\psi=\left\{\begin{array}{l}
1\quad \mbox{if}\quad S_{H_{0}}(\mathbf{Y})>C(H_{0},\alpha ),\\
0\quad \mbox{otherwise,}\\
\end{array}\right.$$
\noindent where $$S_{H_{0}}(\mathbf{Y})=\left\langle
\mathbf{K}\widehat{\boldsymbol{\beta
}}-\mathbf{C},\mathbf{K}\widehat{\boldsymbol{\beta
}}-\mathbf{C}\right\rangle_{\mathcal{H}=H^{n}}.$$ The constant $C(H_{0},\alpha )$ is
such that 
\begin{eqnarray} \mathcal{P}\left\{
S_{H_{0}}(\mathbf{Y})>C(H_{0},\alpha ),\quad \mathbf{K}\boldsymbol{\beta
}=\mathbf{C}\right\} = 1-\mathcal{P}\left\{ S_{H_{0}}(\mathbf{Y})\leq
C(H_{0},\alpha ),~\mathbf{K}\boldsymbol{\beta
}=\mathbf{C}\right\} =  1-\mathbf{F}_{\alpha }=\alpha ,\nonumber
\end{eqnarray}
 \noindent where the
probability distribution  $\mathbf{F}$ on $\mathcal{H}=H^{n}$ has
characteristic functional given in \cite[Proposition 4, Eq. (66)]{Ruiz16}.

Alternatively, as an application of \cite[Theorem 4.1]{Cuestaetal07},  a multivariate version of the significance test formulated  in
\cite{CuestaFebrero10} is considered here, for the fixed effect parameters (see, in particular, \cite[Theorem 2.1]{CuestaFebrero10}.    Specifically, we consider

\begin{equation}
H_0^{\mathbf{h}}:~\mathbf{K}\boldsymbol{\beta }(\mathbf{h})=
\mathbf{C},\label{A4:hoproy}
\end{equation} \noindent for $\mathbf{h}=(h,\dots,h)^{T}_{p\times 1}$
defining  a random vector in $H^{p},$ with $h$ generated
from a zero--mean Gaussian measure  $\mu $ in $H,$ with trace
covariance operator $R_{\mu}$ (see, for example, \cite{PratoZabczyk02}). Here, $$\boldsymbol{\beta
}(\mathbf{h})=\left(\left\langle \beta_{1},h\right\rangle_{H},\dots,
\left\langle \beta_{p},h \right\rangle_{H}\right)^{T}_{p\times 1},$$ $\mathbf{K}$ is given by

\begin{equation}\mathbf{K} =   \left(
\begin{array}{ccccc}
1 & -1 & 0 & \dots & 0 \\
1 & 0 & -1 & \dots & 0 \\
\vdots & \vdots & \vdots & \ddots & \vdots \\
1 & 0 & 0 & \dots & -1 \end{array} \right) \in \mathbb{R}^{\left(p-1 \right) \times
p},\label{A4:79b} \end{equation} \noindent and $\mathbf{C}$ is a null
$\left(p-1\right)\times 1$ functional vector; i.e.,
\begin{equation}
\mathbf{C}=  \left(0, 0, \dots, 0\right)^{T} \in \mathbb{R}^{\left(p-1\right)\times 1}. \label{A4:80}
\end{equation}

From equations (\ref{A4:79b})--(\ref{A4:80}), for any $\left(p\times 1\right)$--dimensional
functional random  vector \linebreak $\mathbf{h}=(h,\dots,h)_{p\times 1}^{T}$
generated from a Gaussian measure $\mu$ on $H,$ $H_0^{\mathbf{h}}$
can then be equivalently expressed as
\begin{equation}H_0^{\mathbf{h}}:~\left\langle
\beta_{1},h\right\rangle_{H} =\left\langle
\beta_{2},h\right\rangle_{H}=\dots=\left\langle
\beta_{p},h\right\rangle_{H}.\label{A4:eq1rt}
\end{equation}

The test statistic  to contrast  (\ref{A4:eq1rt}) is defined as
\begin{equation}
T_{\boldsymbol{h}} = \left(\mathbf{K}\widehat{\boldsymbol{\beta}}(\mathbf{h}) -
\mathbf{C}  \right)^T \left(\mathbf{K} \mathbf{Q}_{\mathbf{h}}
\mathbf{K}^{T} \right)^{-1} \left( \mathbf{K}
\widehat{\boldsymbol{\beta}}(\mathbf{h})- \mathbf{C} \right),
\label{A4:78}
\end{equation}
\noindent where $\mathbf{K}$ and $\mathbf{C}$ are respectively given
in equations (\ref{A4:79b})--(\ref{A4:80}), and
\begin{eqnarray}
\mathbf{Q}_{\mathbf{h}}&=&(\mathbf{X}^{T}\boldsymbol{\Lambda }_{\mathbf{h}}
\mathbf{X})^{-1}, \quad 
\widehat{\boldsymbol{\beta}}(\mathbf{h}) =
\left(\mathbf{X}^T \boldsymbol{\Lambda}_{\mathbf{h}}^{-1} \mathbf{X}
\right)^{-1} \mathbf{X}^T\boldsymbol{\Lambda}_{\mathbf{h}}^{-1}
\mathbf{Y}(\mathbf{h}),
\label{A4:ts}
\end{eqnarray}
\noindent with $$\mathbf{Y}(\mathbf{h})=\left(\left\langle
Y_{1},h\right\rangle_{H},\dots,\left\langle
Y_{n},h\right\rangle_{H}\right).$$ Here,
$\boldsymbol{\Lambda}_{\mathbf{h}}$ is a $\left(n\times n\right)$--dimensional matrix with
entries $\left\lbrace \Lambda_{\mathbf{h}}(i,j) \right\rbrace_{i=1,\dots,n}^{j=1,\ldots,n},$
given by
$$\Lambda_{\mathbf{h}}(i,j)=\sum_{k=1}^{\infty}\left[\left\langle
h,\phi_{k}\right\rangle_{H}\right]^{2}\lambda_{k}(R_{ij}),\quad
i,j=1,\dots,n,$$   where, as before, $\lambda_{k}(R_{ij})$
denotes the $k$--th coefficient in the diagonal expansion of the
covariance operator $R_{ij}$ with respect to the basis
$\{\phi_{k}\otimes \phi_{k},\ k\geq 1\}$; i.e., in the diagonal expansion
$$R_{ij}=\sum_{k=1}^{\infty}\lambda_{k}(R_{ij})\phi_{k}\otimes
\phi_{k}, \quad i,j=1,\dots,n.$$ Note that in the ARH(1) error term case described in \textcolor{Crimson}{Appendix} \ref{A4:sec:3}, from
equation (\ref{A4:covest2bb}), $$\lambda_{k}(R_{ij})=0,\quad \text{for } \left| i-j \right|>1,~k\geq 1.$$

Assuming that the autocovariance and cross--covariance operator of
the ARH(1) error terms are known,  under the null hypothesis
$H_0^{\mathbf{h}},$ the conditional distribution of $T_h$ in
(\ref{A4:78}), given $Y=h,$  is a  chi--square distribution with $p-1$  degrees of freedom.
Here,  $Y$ is a zero-mean $H$--valued
random variable
 with  Gaussian probability  measure $\mu $ on $H,$ having trace covariance operator $R_{\mu}.$
 Note that the last assertion directly follows from
 the fact that, in equation (\ref{A4:ts}),  the conditional distribution of $\widehat{\boldsymbol{\beta}}(\mathbf{h})$ given $Y=h$ is $$\widehat{\boldsymbol{\beta}}(\mathbf{h})\sim \mathcal{N}(\boldsymbol{\beta}(\mathbf{h}),\mathbf{Q}_{\mathbf{h}}),$$
 with $\mathbf{Q}_{\mathbf{h}}$ being introduced in equation (\ref{A4:ts}); i.e., the conditional distribution of $\widehat{\boldsymbol{\beta}}(\mathbf{h})$, given $Y=h$,  is a multivariate
 Gaussian distribution with mean vector $\boldsymbol{\beta}(\mathbf{h})$ and covariance matrix $\mathbf{Q}_{\mathbf{h}}.$

From \cite[Theorem 4.1]{Cuestaetal07} and \cite[Theorem 2.1]{CuestaFebrero10},  if $$H_0:~\beta_{1}(\cdot )=
\beta_{2}(\cdot )=\dots=\beta_{p}(\cdot )$$ fails, then, for
$\mu$-almost every function $h\in H,$ $H_0^{\mathbf{h}}$ in
(\ref{A4:hoproy}), or equivalently in (\ref{A4:eq1rt}), also fails. Thus,
a statistical test at level $\alpha $ to test $H_0^{\mathbf{h}}$ is
a statistical test at the same level $\alpha $ to test $H_0.$

%
%
%

\textcolor{Crimson}{\section{Simulation study}
\label{A4:sec:6}}

In this section, we consider the real separable Hilbert space $$H =
L_{0}^{2}\left(D_{l}\right) = \overline{\mathcal{C}_{0}^{\infty
}\left(D_{l}\right)}^{L^{2}\left(\mathbb{R}^2\right)},$$ the closure,
in the norm of the square integrable functions in $\mathbb{R}^{2},$ of
the space of infinitely differentiable functions with compact
support contained in $D_{l},$ for each $l=1,2,3.$ We restrict our
attention to the family of error covariance operators given in
(\ref{A4:covopf}).  Thus, for each $i,j=1,\dots,n,$

\begin{equation}
R_{\varepsilon_i \varepsilon_j} = R_{ij} = {\rm E} \left\lbrace  \varepsilon_i
\otimes \varepsilon_j  \right\rbrace = \displaystyle \sum_{k=1}^{\infty} \left(\delta_{i,j}^{*}
\exp\left(-\frac{\vert i - j \vert}{\lambda_{ki} + \lambda_{kj}}\right) +
\delta_{i,j}\sqrt{\lambda_{ki}\lambda_{kj}}\right) \phi_{k} \otimes
\phi_{k},   \label{A4:41}
\end{equation}
\noindent where $\delta_{i,j}^{*} = 1 - \delta_{i,j},$ and $\delta_{i,j}$ is the Kronecker
delta function. As before, for each $i,j=1,\dots,n$ and $k\geq 1,$
$$\lambda_{ki}=\lambda_{k}(R_{ii}), \quad \lambda_{k}(R_{ij})=\exp\left(-\frac{\vert i - j \vert}{\lambda_{ki} + \lambda_{kj}}\right).$$

Note that the above error covariance operator models correspond to define, for $i=1,\dots,n,$ the functional Gaussian error component $\varepsilon_{i}$ as the solution, in the mean--square sense,  of the stochastic
partial differential equation
$$(-\boldsymbol{\Delta}_{D_{l}})^{(d-\gamma_{i})}\varepsilon_{i}=\xi_{i},\quad \gamma_{i}\in (0,d/2),$$
 with $\xi_{i}$ being spatial Gaussian white noise on $L^{2}(D_{l}),$ for $l=1,2,3.$

 To approximate
\begin{eqnarray}
{\rm FMSE}_{\boldsymbol{\beta}} &=&  {\rm E} \left\lbrace \Vert \boldsymbol{\beta}
\left(\cdot\right) - \widehat{\boldsymbol{\beta}}\left(\cdot\right)
\Vert_{H^{p}}^{2} \right\rbrace, \nonumber 
\end{eqnarray}
\noindent  $\nu$ samples are generated for the computation of
\begin{equation}
{\rm EFMSE}_{\boldsymbol{\beta}} =\displaystyle \sum_{v=1}^{\nu}
\frac{\displaystyle \sum_{s=1}^{p} \Vert \boldsymbol{\beta}_{s}^{v}
\left(\cdot \right) - \widehat{\boldsymbol{\beta}}_{s}^{v}
\left(\cdot \right) \Vert^{2}_{H}}{\nu}, \label{A4:777}
\end{equation}

\noindent the empirical functional mean--square error
${\rm EFMSE}_{\boldsymbol{\beta}}$ associated with the functional
estimates $$\left\lbrace \widehat{\boldsymbol{\beta}}_{s}^{v}
\left(\cdot \right) = \left(\widehat{\beta}_{s}^{v}
\left(x_1,y_1 \right),\dots,\widehat{\beta}_{s}^{v}
\left(x_L,y_L \right) \right), \ s=1,
\dots,p,~v=1,\dots,\nu \right\rbrace$$ of $\boldsymbol{\beta},$  where $L$ is the
number of nodes considered in the regular grid constructed over the
domains \linebreak $\left\lbrace D_{l}, \ l=1,2,3 \right\rbrace.$

 Also, we
will compute the following statistics:
\begin{equation}
L_{\boldsymbol{\beta}}^{\infty} \left(\cdot \right) = \displaystyle
\sum_{v=1}^{\nu}  \frac{\left(\Vert
\boldsymbol{\varepsilon}^{2}_{\boldsymbol{\beta},v} \left(x_1,y_1
\right) \Vert_\infty,\dots,\Vert
\boldsymbol{\varepsilon}^{2}_{\boldsymbol{\beta},v} \left(x_L,y_L
\right) \Vert_\infty \right)}{\nu}, \nonumber 
\end{equation}
\noindent where
$$\boldsymbol{\varepsilon}^{2}_{\boldsymbol{\beta},v} \left(x_j,y_j
\right) = \left(\varepsilon^{2}_{\beta_{1}^{v}}
\left(x_j, y_j
\right),\dots,\varepsilon^{2}_{\beta_{p}^{v}}
\left(x_j, y_j \right) \right), \quad j=1,\dots,L,$$ and
$$\varepsilon_{\boldsymbol{\beta}_{s}^{v}} \left(x_j, y_j \right) =
\beta_{s}^{v} \left(x_j, y_j \right) -
\widehat{\beta}_{s}^{v} \left(x_j, y_j \right), \quad s=1,\dots,p,\quad j=1,\dots,L, \quad v=1,\dots,\nu,$$ with $\Vert \cdot \Vert_\infty$ denoting the
$L^\infty$--norm.

\bigskip

Let $$\left\lbrace \mathbf{Y}_{i}^{v} \left(\cdot \right) =
\left(Y_{i}^{v} \left(x_1,y_1
\right),\dots,Y_{i}^{v} \left(x_L,y_L \right)
 \right), \ i=1,\dots,n,~v=1,\dots,\nu \right\rbrace$$ be the generated functional samples.
The empirical approximation of $${\rm FMSE}_{\mathbf{Y}} = {\rm E} \left\lbrace \Vert
\mathbf{Y}\left(\cdot\right) -
\widehat{\mathbf{Y}}\left(\cdot\right) \Vert_{H^n}^{2} \right\rbrace,$$  with  ${\rm FMSE}_{\mathbf{Y}}$ being the ${\rm FMSE}$ of
$\mathbf{Y},$  can be computed as follows:

\begin{equation}
{\rm EFMSE}_{\mathbf{Y}} =\displaystyle \sum_{v=1}^{\nu}
\frac{\displaystyle \sum_{i=1}^{n} \Vert \mathbf{Y}_{i}^{v}
\left(\cdot \right) - \widehat{\mathbf{Y}}_{i}^{v} \left(\cdot
\right) \Vert^{2}_{H}}{\nu}. \label{A4:788}
\end{equation}

Also, we will consider the statistics
\begin{equation}
L_{\mathbf{Y}}^{\infty} \left(\cdot \right) = \displaystyle
\sum_{v=1}^{\nu}  \frac{\left(\Vert
\boldsymbol{\varepsilon}^{2}_{\mathbf{Y},v} \left(x_1,y_1 \right)
\Vert_\infty,\dots,\Vert \boldsymbol{\varepsilon}^{2}_{\mathbf{Y},v}
\left(x_L,y_L \right) \Vert_\infty \right)}{\nu}, \nonumber 
\end{equation}
\noindent where
$$\boldsymbol{\varepsilon}^{2}_{\mathbf{Y},v}\left(x_j,y_j \right) =
\left(\varepsilon^{2}_{\mathbf{Y}_{1}^{v}} \left(x_j, y_j
\right),\dots,\varepsilon^{2}_{\mathbf{Y}_{n}^{v}} \left(x_j, y_j
\right) \right), \quad \boldsymbol{\varepsilon}_{\mathbf{Y}_{i}^{v}} \left(x_j, y_j \right) =
\mathbf{Y}_{i}^{v} \left(x_j, y_j \right) -
\widehat{\mathbf{Y}}_{i}^{v} \left(x_j, y_j \right),$$ for
$i=1,\dots,n,$   $j=1,\dots,L,$ and $v=1,\dots,\nu.$

In the following numerical examples, the functional analysis of
variance is implemented from a transformed functional data model,
considering the matrix operator  $\mathbf{W}$ such that, for each
$k\geq 1,$  $\boldsymbol{\Phi}^{*}_{k}\mathbf{W}=\mathbf{W}_k$
compensates the divergence of the eigenvalues of
$\boldsymbol{\Lambda}_k^{-1}.$ Thus,  condition
(\ref{A4:64}) is satisfied. Hence, for all  $k \geq 1, \mathbf{W}_k$ can
be defined as

\begin{equation}
\mathbf{W}_k = \boldsymbol{\Psi}_k \boldsymbol{\Omega }\left(
\mathbf{W}_k \right) \boldsymbol{\Psi}_{k}^{T}, \label{A4:66}
\end{equation}
\noindent where $\boldsymbol{\Omega }\left(\mathbf{W}_k \right) =
diag\left(\omega_{k11},\dots,\omega_{knn}\right)$ denoting a diagonal matrix, which elements
are defined by $$w_{kii} = \omega_i \left( \boldsymbol{\Lambda}_k
\right) + \frac{1}{a_k},$$ under $$\displaystyle \sum_{k=1}^{\infty}
\frac{1}{a_k} < \infty.$$ We have chosen $a_k = k^2.$ Here, for each
$k\geq 1,$ $\boldsymbol{\Psi}_k$ denotes the projection operator
into the system \linebreak $\left\lbrace \psi_{lk}, l=1,\ldots, n \right\rbrace$ of eigenvectors of matrix
$\boldsymbol{\Lambda}_{k},$  and $\left\lbrace \omega_i \left(
\boldsymbol{\Lambda}_k \right), \ i=1,\ldots,n \right\rbrace$ are the associated eigenvalues (see \cite{Ruiz16}).

In practice, the infinite series defining the generalized
least--squares estimator, and the functional components of variance
is truncated at $TR.$ Specifically, in the rectangle,  we work with
a two--dimensional truncation parameter  $TR = TR_1 \times TR_2,$
and, for circular domains, we fix a one--dimensional parameter (the
order $k$ of Bessel functions), thus,
 $TR_{1}= 1,$
and move the second truncation parameter associated with the radius
$R$ (see \textcolor{Crimson}{Appendices} \ref{A4:secdisk}--\ref{A4:seccsec}). We then have

\begin{eqnarray}
\widehat{\boldsymbol{\beta}} &\simeq& \boldsymbol{\Phi} \left(
\left\lbrace \widehat{\boldsymbol{\beta}_k}, \ k=1,\dots,TR \right\rbrace \right),
\\\label{A4:67} 
\widetilde{SSE}  &\simeq& \displaystyle
\sum_{k=1}^{TR}\left( \mathbf{M}_k \mathbf{W}_k \mathbf{Y}_k
\right)^T \boldsymbol{\Lambda}_{k}^{-1}\mathbf{M}_k \mathbf{W}_k \mathbf{Y}_k,\\
\label{A4:68} \widetilde{SST} &\simeq& \displaystyle \sum_{k=1}^{TR}
\mathbf{Y}_{k}^{T} \mathbf{W}_{k}^{T} \boldsymbol{\Lambda}_{k}^{-1} \mathbf{W}_k \mathbf{Y}_k, \\
\label{A4:69}
\widetilde{SSR}  &=& \widetilde{SST} - \widetilde{SSE}, \\
\label{A4:70} \boldsymbol{\Lambda}_k &=&\boldsymbol{\Psi}_k
\boldsymbol{\Omega}\left(\boldsymbol{\Lambda}_k \right) \boldsymbol{\Psi}_{k}^{T}, \quad k=1,\dots,TR, \\
\label{A4:71} \mathbf{W}_k &=& \boldsymbol{\Psi}_k \boldsymbol{\Omega}
\left( \mathbf{W}_k \right) \boldsymbol{\Psi}_{k}^{T}, \quad k=1,\dots,TR.
\label{A4:72}
\end{eqnarray}

From the transformed  model (\ref{A4:tfdm}),
the finite--dimensional approximations (\ref{A4:67})--(\ref{A4:72}) of $\widetilde{SSE},$ $\widetilde{SST},$ and
$\widetilde{SSR},$
 respectively, are computed to obtain the values of  the statistics (\ref{A4:800}),
reflecting  the relative magnitude between the empirical functional
variability explained by the model and the residual variability.

In the computation of the test statistics $T_{\boldsymbol{h}},$ a truncation order is also considered in the
calculation of the elements defining matrix $\boldsymbol{\Lambda}_{\boldsymbol{h}}.$

\bigskip

In all the subsequent sections,  the truncation order $TR$ has been
selected according to the following criteria:

\bigskip

\begin{itemize}
\item[(i)] The percentage of explained functional variance. In all the subsequent numerical examples, the $TR$ values considered always ensure a percentage
of explained functional variance larger or equal than $95\%.$
 \item[(ii)] The rate of convergence to zero of the eigenvalues of the covariance operators, defining the functional entries of the matrix covariance  operator of the  $H^{n}$--valued error term. Specifically, in the simulation study undertaken, according to the  asymptotic order (rate of convergence to zero) of such eigenvalues, we have selected the optimal $TR$ to remove divergence of the spectra of the corresponding inverse covariance operators.
\item[(iii)] The functional form of the eigenvectors, depending on
the  geometry of the domain and the Dirichlet conditions on the boundary. Small truncation orders or values of  $TR$ are considered, when  fast decay velocity to zero is displayed at the boundary, by the common eigenvectors of the autocovariance operators of the error components, since, in that case,   the error dependence range is shorter.
\end{itemize}

 Summarizing,  lower truncation orders are required when  a fast decay velocity to zero
 is displayed by the covariance kernel eigenvalues, since a sufficient percentage of explained variability is achieved with a few terms. Note that
larger truncation orders can lead to a ill--posed nature of the
functional parameter estimation problem, and associated response plug-in prediction.
In the subsequent sections, applying criteria
 (i)--(iii),   a smaller number of terms is required in   circular domains than in  rectangular domains.

\textcolor{Crimson}{\subsection{Rectangular domain}
\label{A4:sec:61}}

The $H^{n}$--valued   zero--mean Gaussian error
term  is generated  from the matrix covariance
 operator
$\boldsymbol{R}_{\boldsymbol{\varepsilon}\boldsymbol{\varepsilon}},$ whose functional entries $\left\lbrace \boldsymbol{R}_{\varepsilon_{i}\varepsilon_{j}} \right\rbrace_{i=1,\dots,n}^{j=1,\dots,n},$  are
defined in equation (\ref{A4:41}), with for $i=1,\dots,n,$ \linebreak $\lambda_{ki}=\lambda_{k}(R_{ii})$ being  given in equations (\ref{A4:covopf}) and (\ref{A4:19}). Specifically,    $\{\phi_{k},\ k\geq 1\}$ are the eigenvectors  of the Dirichlet negative Laplacian operator
on the rectangle,  associated with the eigenvalues of such an operator (see equation (\ref{A4:19}) in the \textcolor{Crimson}{Supplementary Material} in \textcolor{Crimson}{Appendix} \ref{A4:Supp}), arranged in decreasing order of their modulus magnitude.

Let us now define the scenarios studied for the rectangular domain $$D_{1} =
\displaystyle \prod_{i=1}^{2} \left[ a_i , b_i \right],$$ where $\nu
= 20$ functional samples of size  $n=200$ have been considered, for
a given  semi--orthogonal design matrix $$\mathbf{X} \in
\mathbb{R}^{n\times p}, \quad \mathbf{X}^T \mathbf{X} =
\mathbf{Id}_p.$$
 These scenarios are determined from the possible values  of the vector variable  $(P_i, u, C_i),$
where $P_i$  refers to the number of components of
$\boldsymbol{\beta },$ specifically, for
$i=1,$ $p=4$ components,  and for $i=2,$ $p=9$ components. Here, $u$ takes the values $a,b,c,d$
respectively corresponding to the truncation orders \linebreak $TR=16$ ($u=a$),
$TR=36$ ($u=b$), $TR=64$ ($u=c$) and $TR=144$ ($u=d$).
 In addition, $\left\lbrace C_i, \ i=1,2\right\rbrace$ indicate the shape of $\boldsymbol{\beta}.$
 Specifically, we have considered

\bigskip

\begin{itemize}
\item $\beta_s\left(x,y\right) = \sin\left(\frac{\pi s x_{b_1}}{l_1}\right)\sin\left(\frac{\pi s y_{b_2}}{l_2}\right)$\quad  (\textcolor{Crimson}{C1})

\item $\beta_s\left(x,y\right) = \cos\left(\frac{x_{b_1} + x_{a_1}}{l_1}\right)\cos\left(\frac{y_{b_2}+ y_{a_2}}{l_2}\right)$ \quad (\textcolor{Crimson}{C2}),
\end{itemize}
\noindent where $$x_{b_1} = \frac{\pi}{2}\left(2s+1\right)\left(b_1 - x\right),~x_{a_1} = \left(x - a_1 \right), \quad
y_{b_2} = \frac{\pi}{2}\left(2s+1\right)\left(b_2 - y\right), \quad
y_{a_2} = \left(y - a_2 \right)$$ and $s=1,\dots,p$.

\bigskip

A summary of the generated and analysed scenarios are displayed in
Table \ref{A4:table:1} below.

\bigskip

\begin{table}[H]
\caption[\hspace{0.7cm} Scenarios in the FANOVA simulation study for rectangular domain.]{\small{Scenarios for rectangular domain.}}
\centering
\begin{small}
\begin{tabular}{|c||c|c|c|c|c|}
  \hline
  \textbf{Cases} & $a_1 = a_2$ & $b_1 = b_2$ & $h_x = h_y$ & $p$ & $TR$ \\
  \hline  \hline
  (P$_{1}$,a,C$_{1}$) & $-2$ & $3$ & $0.05$ & $4$ & $4 \times 4$  \\
  \hline
  (P$_{1}$,b,C$_{2}$) & $-2$ & $3$ & $0.05$ & $4$ & $6 \times 6$  \\
  \hline
 (P$_{1}$,c,C$_{2}$) & $-2$ & $3$ & $0.05$ & $4$ & $8 \times 8$  \\
 \hline
(P$_{1}$,d,C$_{1}$) & $-2$ & $3$ & $0.05$ & $4$ & $12 \times 12$  \\
  \hline
 (P$_{2}$,a,C$_{2}$) & $-2$ & $3$ & $0.05$ & $9$ & $4 \times 4$ \\
  \hline
(P$_{2}$,b,C$_{1}$) & $-2$ & $3$ & $0.05$ & $9$ & $6 \times 6$  \\
  \hline
(P$_{2}$,c,C$_{1}$) & $-2$ & $3$ & $0.05$ & $9$ & $8 \times 8$ \\
 \hline
 (P$_{2}$,d,C$_{2}$) & $-2$ & $3$ & $0.05$ & $9$ & $12 \times 12$ \\
  \hline
\end{tabular}
\end{small}
 \label{A4:table:1}
\end{table}

\bigskip

\begin{figure}[H]
  \centering
    \includegraphics[width=0.8\textwidth]{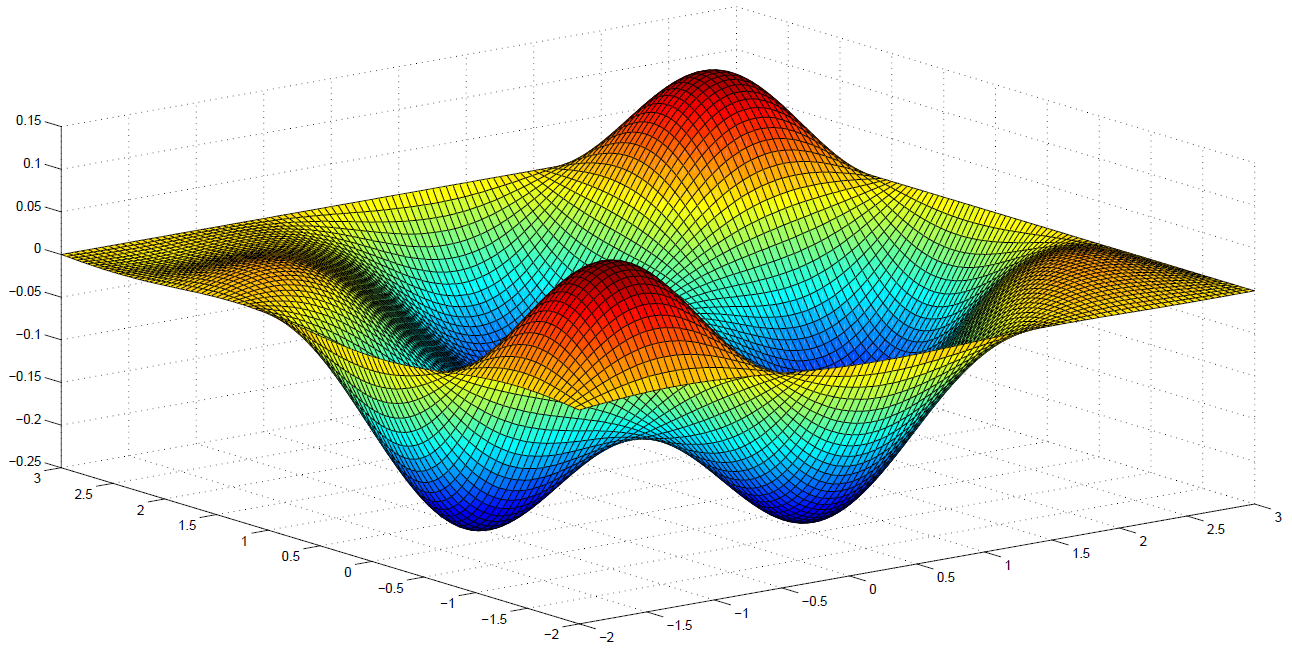}
    
     \vspace{-0.1cm}
    \caption[\hspace{0.7cm} Simulated response for rectangular domains (case C1).]{\small{Case (P$_{1}$,a,C$_1$). Simulated response with $p=4$, $TR=16$ and $\boldsymbol{\beta}$ of type C$_{1}$.}}
    \label{A4:fig:1}
\end{figure}

\begin{figure}[H]
  \centering
    \includegraphics[width=0.8\textwidth]{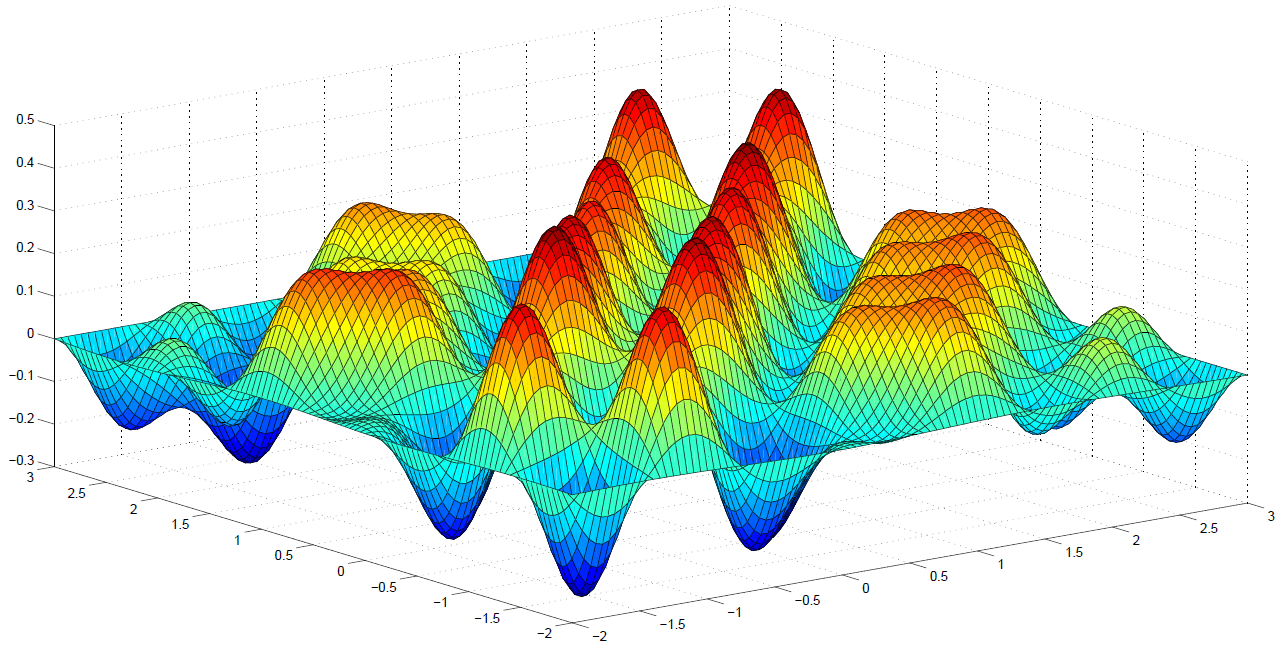}
    
     \vspace{-0.1cm}
    \caption[\hspace{0.7cm} Simulated response for rectangular domains (case C2).]{\small{Case (P$_{2}$,a,C$_2$). Simulated response with $p=9$, $TR=16$ and $\boldsymbol{\beta}$ of type C$_{2}$.}}
  \label{A4:fig:2}
\end{figure}

In Table \ref{A4:table:1}, $h_x$ and $h_y$ refer to the discretization step size at each
dimension.
 In the  cases (P$_{1}$,a,C$_{1}$)
and (P$_{2}$,a,C$_2$), a generation of a
functional value (surface) of the response is respectively represented in Figures \ref{A4:fig:1}--\ref{A4:fig:2}.

\bigskip

Figures \ref{A4:fig:3}--\ref{A4:fig:4} below show the respective functional
estimates $\widehat{\mathbf{Y}}=
\mathbf{X}\widehat{\boldsymbol{\beta }}$ of the responses
displayed in Figures \ref{A4:fig:1}--\ref{A4:fig:2} above.

\bigskip

\begin{figure}[H]
  \centering
    \includegraphics[width=0.8\textwidth]{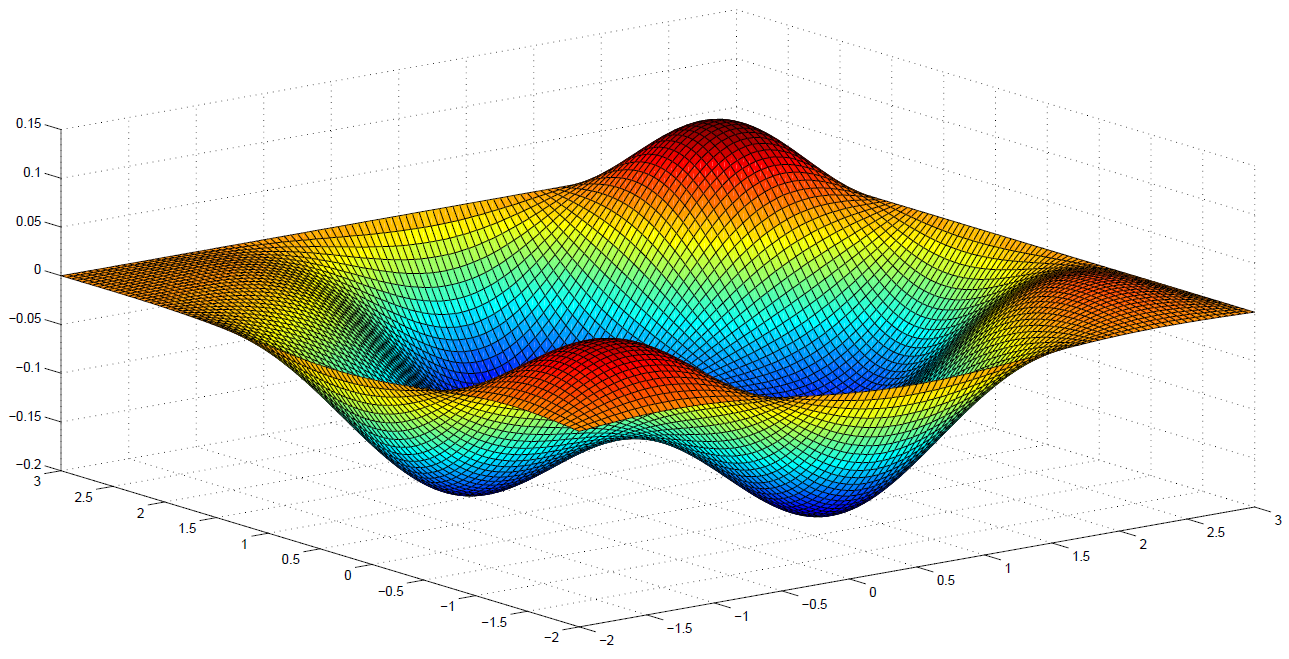}
    
      \vspace{-0.1cm}
     \caption[\hspace{0.7cm} Estimated response for rectangular domains (case C1).]{\small{Case (P$_{1}$,a,C$_{1}$). Estimated response with $p=4$, $TR=16$ and $\boldsymbol{\beta }$ of type C$_{1}$.}}
    \label{A4:fig:3}
\end{figure}

\begin{figure}[H]
  \centering
    \includegraphics[width=0.8\textwidth]{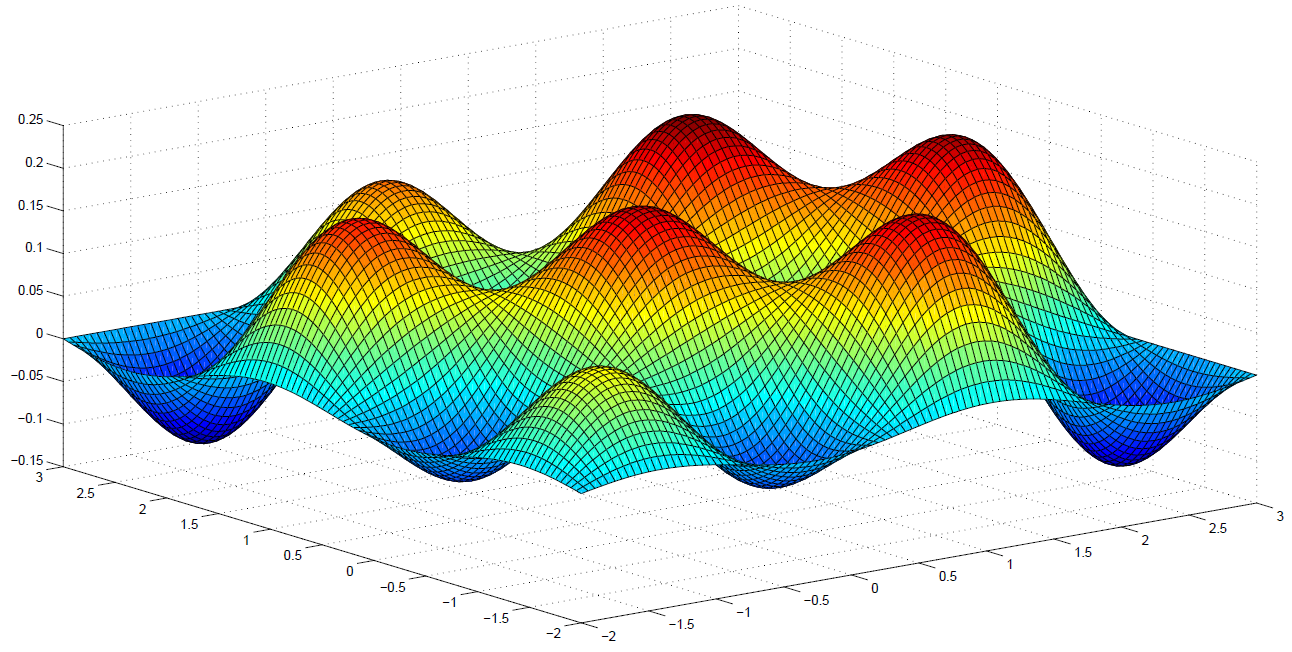}
    
     \vspace{-0.1cm}
     \caption[\hspace{0.7cm} Estimated response for rectangular domains (case C2).]{\small{Case (P$_{2}$,a,C$_2$). Estimated response with $p=9$, $TR=16$ and $\boldsymbol{\beta }$ of type C$_{2}$.}}
  \label{A4:fig:4}
\end{figure}

 \bigskip
 
 The statistics (\ref{A4:777})--(\ref{A4:788})
are evaluated in all the cases displayed in Table \ref{A4:table:1} (see
 Tables   \ref{A4:table:3}--\ref{A4:table:2} for the statistics  $L_{\boldsymbol{\beta}}^{\infty}$ and $L_{\mathbf{Y}}^{\infty},$ respectively).

\bigskip

\begin{table}[H]
\caption[\hspace{0.7cm} Empirical functional mean square errors on the estimation of the fixed effect parameters for rectangular domain.]{\small{$EFMSE_{\boldsymbol{\beta}}$ for rectangular domain.}}
\centering
\begin{small}
$EFMSE_{\boldsymbol{\beta}}$ \\
\begin{tabular}{|c|c|c|c|}
  \hline
   (P$_{1}$,a,C$_{1}$) & (P$_{1}$,b,C$_{2}$) & (P$_{1}$,c,C$_{2}$) & (P$_{1}$,d,C$_{2}$)  \\
  \hline 
  $1.070 \left(10 \right)^{-3}$  & $1.060 \left(10 \right)^{-3}$ &  $ 1.040 \left(10 \right)^{-3}$ & $1.040 \left(10 \right)^{-3}$ \\
  \hline
  \hline
  (P$_{2}$,a,C$_{2}$) & (P$_{2}$,b,C$_{1}$) & (P$_{2}$,c,C$_{1}$) & (P$_{2}$,d,C$_{2}$) \\ 
  \hline
$  9.400\left(10 \right)^{-4}$   & $9.300  \left(10 \right)^{-4}$ &  $ 9.300  \left(10 \right)^{-4}$ & $9.100  \left(10 \right)^{-4}$ \\
  \hline
\end{tabular}
\end{small}
\label{A4:table:3}
\end{table}

\bigskip

\begin{table}[H]
\caption[\hspace{0.7cm} Empirical functional mean square errors on the estimation of the response for rectangular domain.]{\small{$EFMSE_{\mathbf{Y}}$ for rectangular domain.}}
\centering
\begin{small}
$EFMSE_{\mathbf{Y}}$ \\
\begin{tabular}{|c|c|c|c|c|c|c|c|}
  \hline
  (P$_{1}$,a,C$_{1}$) & (P$_{1}$,b,C$_{2}$) & (P$_{1}$,c,C$_{2}$) & (P$_{1}$,d,C$_{2}$) &  (P$_{2}$,a,C$_{2}$) & (P$_{2}$,b,C$_{1}$) & (P$_{2}$,c,C$_{1}$) & (P$_{2}$,d,C$_{2}$) \\
  \hline  \hline
  $0.014$  &  $0.013$ &  $0.010$ & $0.009$  & $0.011$  &  $0.011$ &  $0.009$ & $0.007$ \\
  \hline
\end{tabular}
\end{small}
\label{A4:table:2}
\end{table}

\bigskip

As expected, the results
displayed in Table \ref{A4:table:3}, corresponding to the empirical
functional mean \linebreak quadratic errors associated with the estimation of
$\boldsymbol{\beta },$ are less than the ones obtained in Table
\ref{A4:table:2} for the response, with
 order of magnitude
$10^{-3}$  in all the scenarios generated.
In Table \ref{A4:table:2}, we can appreciate a better performance of
the generalized least--squares estimator for the higher truncation
orders. However, we have to note that, even for the smallest
truncation order considered; i.e., for  $TR=4\times 4=16,$ a good
performance is observed with associated empirical functional mean
quadratic errors having order of  magnitude $10^{-2}$ in all the
cases displayed in Table \ref{A4:table:1} (see the above truncation order criteria (i)--(iii)).  It can also be observed that the number of components of
parameter $\boldsymbol{\beta },$ and their  functional  shapes  do
not affect the accuracy of the least--squares generalized estimations
of the functional values of the response. It can also be observed that the number of components of
parameter $\boldsymbol{\beta },$ and their  functional  shapes  do
not affect the accuracy of the least--squares generalized estimations
of the functional values of the response.

 \bigskip

The statistics (\ref{A4:800}) is now computed, as an empirical
approximation of the relative magnitude between the explained
functional variability  and the residual variability, after fitting
the transformed Hilbert--valued fixed effect  model (\ref{A4:tfdm}). The
results obtained are given in Table \ref{A4:table:4}. It can be
observed that, in all the cases studied,  the explained functional
variability exceeds the residual functional variability. The
truncation order, the number of components of $\boldsymbol{\beta },$
and the functional shape of such components do not substantially
affect the goodness of fit of the  transformed Hilbert--valued fixed
effect  model in (\ref{A4:tfdm}).

\bigskip

\begin{table}[H]
\caption[\hspace{0.7cm} $F$ statistics for rectangular domain.]{\small{$F$ statistics (\ref{A4:800}) for rectangular domain.}}

\centering
\begin{small}
\begin{tabular}{|c|c|c|c|c|c|c|c|c|}
  \hline
  Cases & (P$_{1}$,a,C$_{1}$) & (P$_{1}$,b,C$_{2}$) & (P$_{1}$,c,C$_{2}$) & (P$_{1}$,d,C$_{1}$)  & (P$_{2}$,a,$C_{2}$) & ($P_{2}$,b,$C_{1}$) & (P$_{2}$,c,C$_{1}$) & (P$_{2}$,d,C$_{2}$) \\
  \hline  \hline
  $F$ & $1.926$ & $1.717$ & $1.673$ & $1.626$ & $1.898$ & $1.845$ & $1.761$ & $1.606$ \\
  \hline
\end{tabular}
\end{small}
\label{A4:table:4}
\end{table}

\bigskip

 Let us now compute the statistics $T_{\boldsymbol{h}}$ in (\ref{A4:78}) to contrast the significance of parameter vector $\boldsymbol{\beta}$
 in Case C$_{1},$  when $p=4.$ To
 apply \cite[Theorem 4.1]{Cuestaetal07} and \cite[Theorem 2.1]{CuestaFebrero10}, we have generated
eight realizations of a Gaussian  random function $\boldsymbol{h},$ from the
trajectories of the Gaussian random field $\xi,$ solution, in the
mean--square sense, of the following boundary value problem:

\begin{eqnarray} 
(-\Delta )\xi (\mathbf{x})&=&\varsigma
(\mathbf{x}), \quad \mathbf{x}=(x_{1},x_{2})\in [-2,3]\times [-2,3], \nonumber \\
  \xi (-2,x_{2}) &=& \xi (3,x_{2})=\xi
(x_{1},-2)=\xi(x_{1},3)=0, \quad
x_{1},x_{2}\in [-2,3]\times [-2,3], \nonumber \\\label{A4:eqdlrect}
\end{eqnarray}
\noindent where $\varsigma $ denotes a zero--mean Gaussian white noise
on $L^{2}([-2,3]\times [-2,3])$; i.e., a zero--mean generalized
Gaussian process satisfying 

\begin{eqnarray} 
\displaystyle \int_{[-2,3]\times
[-2,3]}f(\mathbf{x}) {\rm E} \left\lbrace \varsigma (\mathbf{y})\varsigma
(\mathbf{x}) \right\rbrace d\mathbf{x}=f(\mathbf{y}), \quad \mathbf{y}\in [-2,3]\times [-2,3], \quad \forall f\in L^{2}([-2,3]\times
[-2,3]).\nonumber
\end{eqnarray}

Table \ref{A4:tab:Tabla1} below reflects the percentage of successes, for
$\alpha =0.05,$ and the averaged $p$--values over the $150$ samples
of the response generated with parameter $\boldsymbol{\beta}$
 of C$_{1}$ type having  $p=4$ components, and    with size  $n=150,$ for
 $TR= 4\times 4.$

\bigskip

\begin{table}[H]
\caption[\hspace{0.7cm} Significance of the fixed effect parameters for rectangle domains.]{\small{\emph{\textcolor{Crimson}{Rectangle}}. Percentage  of successes for $\alpha
=0.05,$ at the left--hand side, and averaged $p$--values at the
right--hand side, for each one of the eight realizations considered of the Gaussian
function $h\in L^{2}([-2,3]\times [-2,3]).$}}
\centering
\begin{small}
\begin{tabular}{|c||c|c|}
  \hline
 $D$ & $\%$ Success & $p$\\
  \hline \hline
   1 & $100 \%$  & $0$ \\
  \hline
    2 & $100 \%$  & $0$ \\
  \hline
     3 & $99.75 \%$ & $1.998(10)^{-8}$\\
  \hline
     4 & $100 \%$  &  $0$ \\
  \hline
     5 & $99.8 \%$  &  $7.541(10)^{-7}$\\
  \hline
      6 & $100 \%$  &  $0$\\
  \hline
     7 & $100 \%$ &  $0$\\
  \hline
    8 & $100 \%$ & $6.441(10)^{-10}$\\
  \hline
\end{tabular}
\end{small}
  \label{A4:tab:Tabla1}
\end{table}

\bigskip

    A high percentage of
successes and very small  $p$--values are observed in Table \ref{A4:tab:Tabla1}; i.e.,  a good performance of the test statistics is observed.

 \textcolor{Crimson}{\subsection{Disk domain}
\label{A4:sec:62}}

In the disk  domain $$D_{2} = \left\lbrace \mathbf{x} \in \mathbb{R}^2:~0 <
\Vert \mathbf{x} \Vert < R \right\rbrace,$$  the zero--mean Gaussian $H^{n}$--valued  error
term is generated  from the matrix covariance
 operator
$\boldsymbol{R}_{\boldsymbol{\varepsilon}\boldsymbol{\varepsilon}},$ whose functional entries  are
defined in equation (\ref{A4:41}), considering the eigenvectors $\{\phi_{k},\ k\geq 1\}$  of  the Dirichlet negative Laplacian operator
on the disk   (see equation (\ref{A4:21}) in the \textcolor{Crimson}{Supplementary Material} in \textcolor{Crimson}{Appendix} \ref{A4:Supp}), arranged in decreasing order of the modulus magnitude of their associated eigenvalues. Specifically,  for $i=1,\dots,n,$ $\lambda_{ki}=\lambda_{k}(R_{ii})$ in (\ref{A4:41}) is defined in equations (\ref{A4:covopf}) and (\ref{A4:21}).
Again, $\nu =20$ functional
 samples of size $n=200$ of the response have been generated.   The cases studied are
 summarized in terms of the vector  (P$_i$, u, C$_j$), $i=1,2,$ $j=1,2,3,$ with variable
 $u=a,b,c,d,e,f.$ Namely, it is considered $u=a$ for $TR=3,$ $u=b$ for $TR=5,$ $u=c$ for $TR=7,$ $u=d$ for $TR=15,$ $u=e$ for $TR=31,$ and $u=f$ for
 $TR=79.$ Furthermore,  P$_i$ indicates  the number of components of $\boldsymbol{\beta},$ with $p=4$ for $i=1,$ and $p=9$ for $i=2.$ Finally,
 the values of C$_j,$ $j=1,2,3,$ refer to the shape of the components of $\boldsymbol{\beta} ,$ defined from  their projections, in terms of  the following equations:

\begin{eqnarray}
\beta_{ks} &=& \frac{\left(-1\right)^s}{k^{3.5}}
e^{\left(\frac{k}{TR}\right)^{7.5 + 2s - 1}}P\left(s,k\right)^{2.5 +
2s - 1} \nonumber \\
& + &  e^{\left(\frac{k}{TR}\right)^{6.5 + 2s -
1}}P\left(s,k\right)^{3.5 + 2s - 1}, \quad  k=1,\dots,TR,\quad
s=1,\dots,p \quad
 \mbox{(\textcolor{Crimson}{C1})}\nonumber\\
\beta_{ks} & = & \frac{1}{R}e^{\frac{s + \frac{k}{R}}{n}}+ k \cos
\left( \left(-1 \right)^k 2 \pi \frac{R}{k} \right),\quad k=1,\dots,TR, \quad s=1,\dots,p\quad  \mbox{(\textcolor{Crimson}{C2})}\nonumber\\
\beta_{ks} &=& \frac{1}{k^{2.5 + 2s - 1}}P\left(s,k\right)^{1.5+ 2s
- 1}, \quad k=1,\dots,TR,\quad s=1,\dots,p\quad
\mbox{(\textcolor{Crimson}{C3})}\nonumber\\ 
 P\left(s,k\right) &=& 1 +
 \left(\frac{k}{TR}\right)^2
+ \left(\frac{TR - k +1}{TR}\right)^4, \quad k=1,\dots,TR,\quad s=1,\dots,p.\nonumber
\end{eqnarray}

\bigskip

Table \ref{A4:table:6} reflects a summary with all the cases analysed.

\bigskip

\begin{table}[H]
\caption[\hspace{0.7cm} Scenarios in the FANOVA simulation study for disk domain.]{\small{Scenarios for disk domain.}} 

\centering
\begin{small}
\begin{tabular}{|c||c|c|c|c|c|}
  \hline
  \textbf{Cases} & $R$ & $h_R$ & $h_\phi$ & $TR$ & $p$\\
  \hline \hline
(P$_{1}$,a,C$_{3}$) & $12$ & $\frac{R}{145}$ & $\frac{2 \pi}{135}$ & $3$ & $4$ \\
  \hline
(P$_{1}$,b,C$_{2}$) & $18$ & $\frac{R}{145}$ & $\frac{2 \pi}{135}$ & $5$ & $4$\\
  \hline
(P$_{1}$,c,C$_{1}$) & $25$ & $\frac{R}{145}$& $\frac{2 \pi}{135}$ & $7$ & $4$ \\
 \hline
(P$_{1}$,d,C$_{1}$) & $50$ & $\frac{R}{145}$ & $\frac{2 \pi}{135}$ & $15$ & $4$\\
  \hline
(P$_{1}$,e,C$_{2}$) & $100$ & $\frac{R}{145}$ & $\frac{2 \pi}{135}$& $31$ & $4$\\
  \hline
(P$_{1}$,f,C$_{3}$) & $250$ & $\frac{R}{145}$ & $\frac{2 \pi}{135}$ & $79$ & $4$\\
    \hline
      \hline
(P$_{2}$,a,C$_{1}$) & $12$ & $\frac{R}{145}$ & $\frac{2 \pi}{135}$ & $3$ & $9$ \\
  \hline
(P$_{2}$,b,C$_{2}$) & $18$ & $\frac{R}{145}$ & $\frac{2 \pi}{135}$ & $5$ & $9$ \\
  \hline
(P$_{2}$,c,C$_{3}$) & $25$ & $\frac{R}{145}$& $\frac{2 \pi}{135}$ & $7$ & $9$ \\
 \hline
(P$_{2}$,d,C$_{3}$) & $50$ & $\frac{R}{145}$ & $\frac{2 \pi}{135}$ &  $15$ & $9$ \\
  \hline
(P$_{2}$,e,C$_{2}$) & $100$ & $\frac{R}{145}$ & $\frac{2 \pi}{135}$& $31$  & $9$ \\
  \hline
(P$_{2}$,f,C$_{1}$) & $250$ & $\frac{R}{145}$ & $\frac{2 \pi}{135}$ & $79$ & $9$ \\
  \hline
\end{tabular}
\end{small}
\label{A4:table:6}
\end{table}

\bigskip

\begin{figure}[H]
  \centering
    \includegraphics[width=0.8\textwidth]{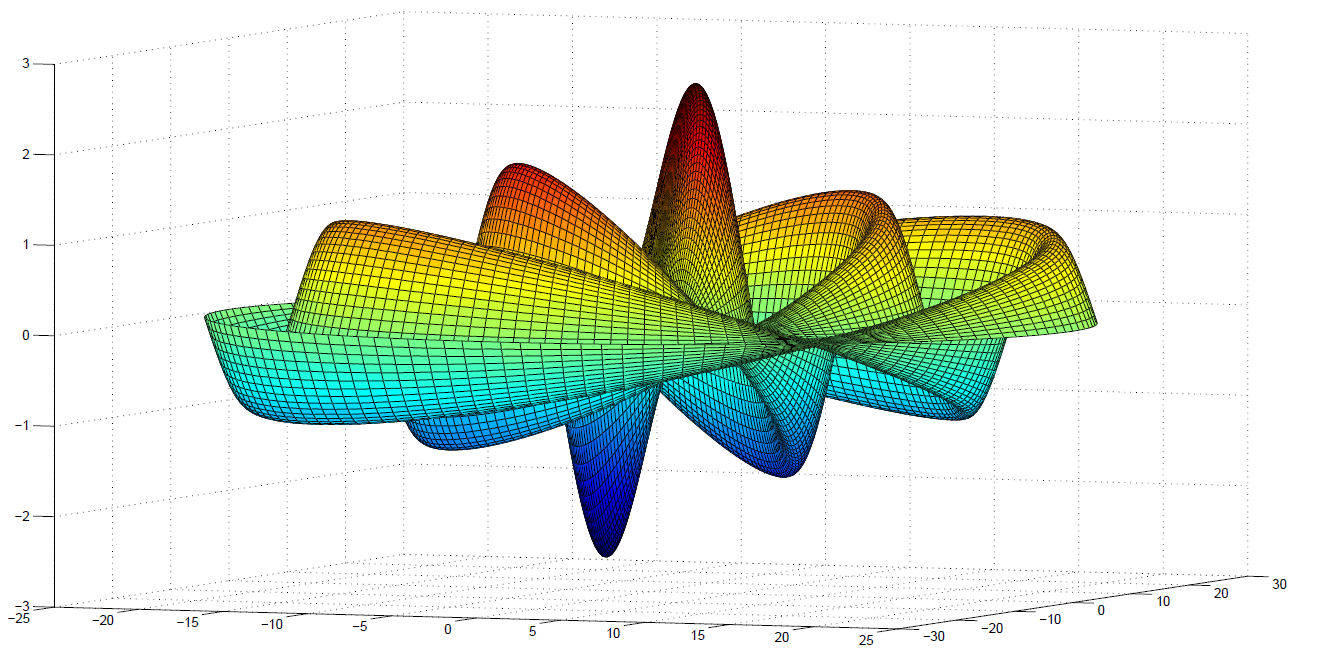}
    
    \vspace{-0.1cm}
    \caption[\hspace{0.7cm} Simulated response for disk domains (case C1).]{\small{Case (P$_{1}$,c,C$_{1}$). Simulated response with $p=4$, $R=25$ and $\boldsymbol{\beta}$ of type C$_{1}$.}}
  \label{A4:fig:5}
\end{figure}

\begin{figure}[H]
  \centering
    \includegraphics[width=0.8\textwidth]{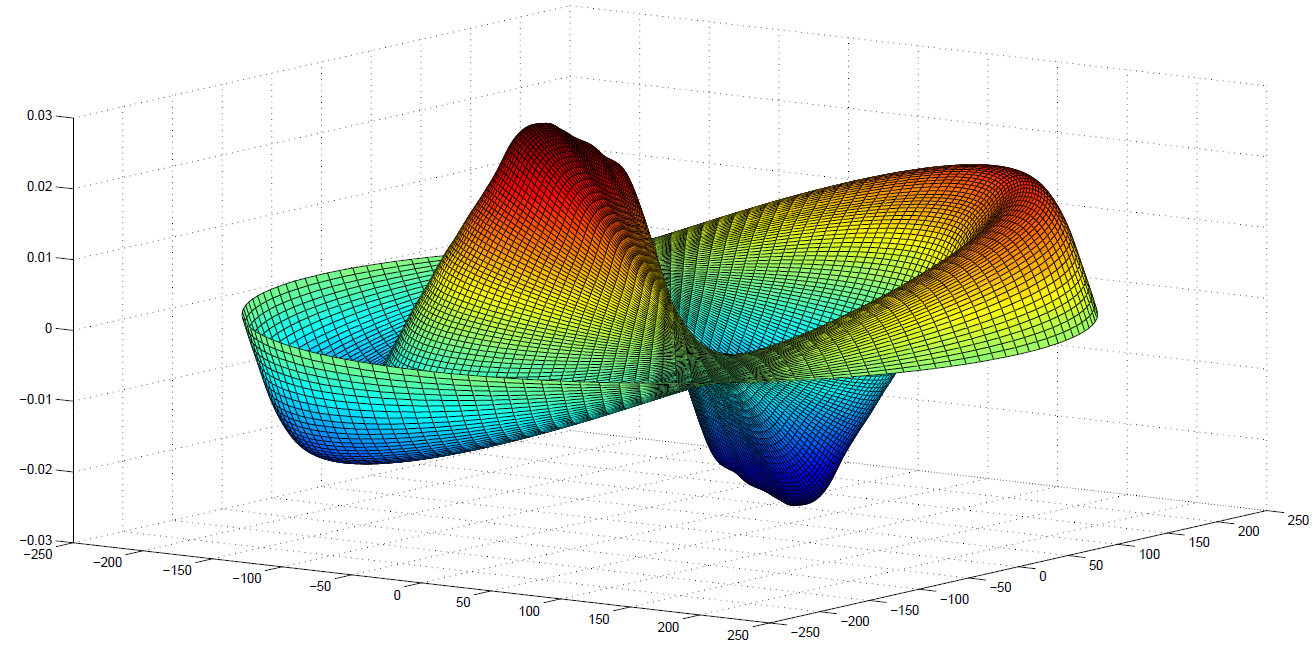}
        
    \vspace{-0.1cm}
\caption[\hspace{0.7cm} Simulated response for disk domains (case C3).]{\small{Case (P$_{1}$,f,C$_{3}$). Simulated response with $p=4$, $R=250$ and $\boldsymbol{\beta}$ of type C$_{3}$.}}
  \label{A4:fig:7cc}
\end{figure}

Figures \ref{A4:fig:5}--\ref{A4:fig:7cc}  respectively reflect the
generation of a functional value of the response in the cases
 (P$_{1}$,c,C$_{1}$) and  (P$_{1}$,f,C$_{3}$). 
 
 \bigskip

\bigskip 

\begin{figure}[H]
  \centering
    \includegraphics[width=0.8\textwidth]{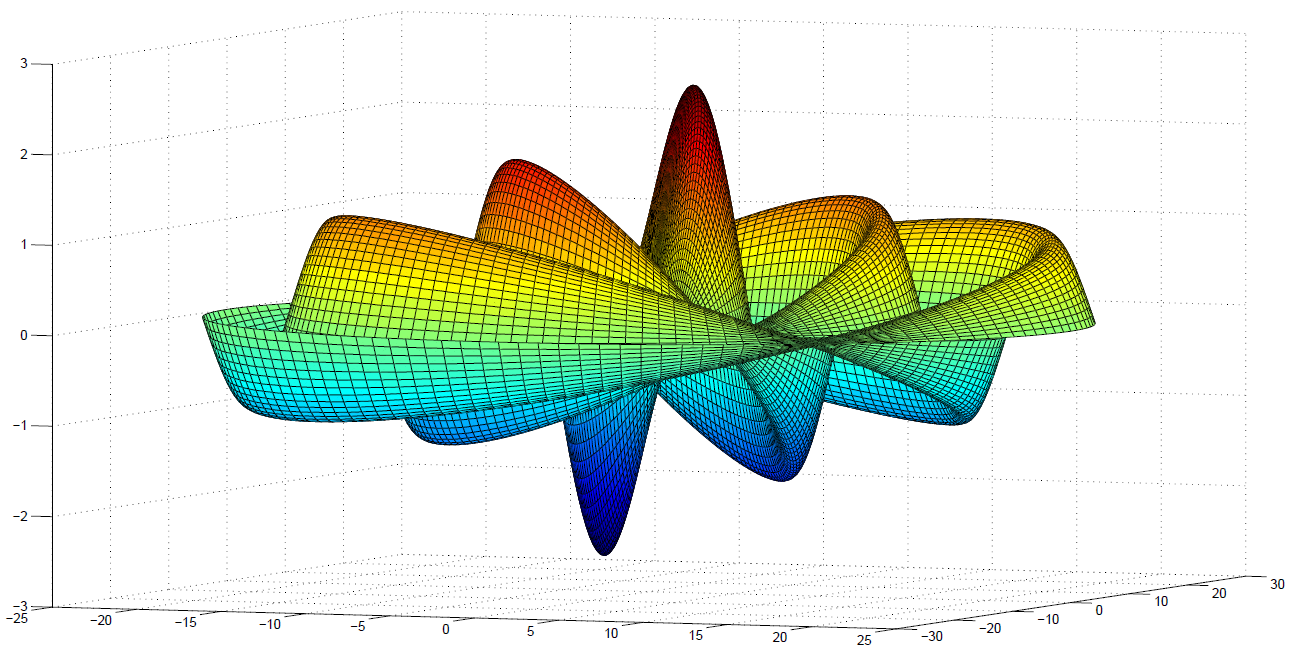}
    
        \vspace{-0.1cm}
\caption[\hspace{0.7cm} Estimated response for disk domains (case C1).]{\small{Case (P$_{1}$,c,C$_{1}$). Estimated response with $p=4$, $R=25$ and $\boldsymbol{\beta}$ of type $C_1$.}}
 \label{A4:fig:8hh}
\end{figure}

\begin{figure}[H]
  \centering
    \includegraphics[width=0.8\textwidth]{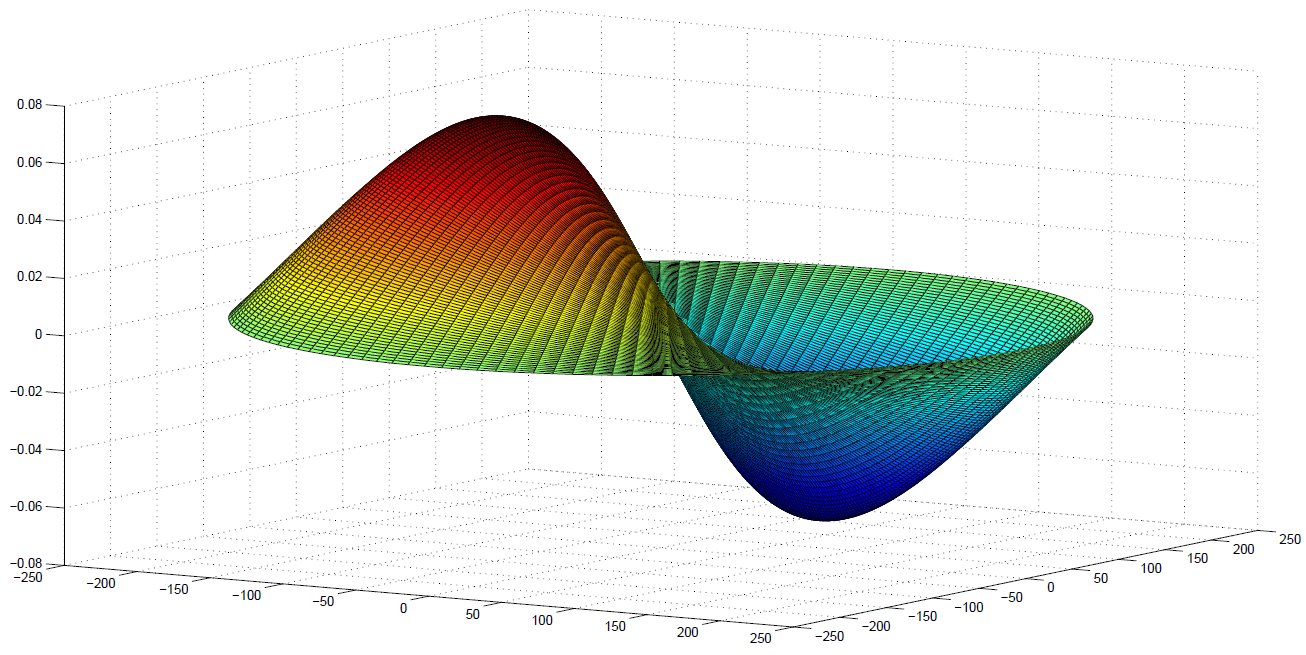}
    
        \vspace{-0.1cm}
\caption[\hspace{0.7cm} Estimated response for disk domains (case C3).]{\small{Case (P$_{1}$,f,C$_{3}$). Estimated response with $p=4$, $R=250$ and $\boldsymbol{\beta}$ of type $C_3$.}}
  \label{A4:fig:10cc}
\end{figure}

The respective generalized least--squares  functional estimates are
displayed  in Figures \ref{A4:fig:8hh}--\ref{A4:fig:10cc}.

\bigskip

\bigskip

 The empirical functional mean quadratic errors  (see equations (\ref{A4:777})--(\ref{A4:788})) are displayed  in   Table \ref{A4:table:8}, for the estimation of the  functional parameter vector $\boldsymbol{\beta },$ and
 in Table \ref{A4:table:7} for the estimation
 of the response $\mathbf{Y}$. 
It can be observed, as in the rectangular domain, that the order of
magnitude of the empirical  functional mean quadratic errors
associated with  the estimation of $\boldsymbol{\beta }$ is of order
$10^{-3},$ and for   the estimation of the response is $10^{-2}$.  However, the number of terms considered is less than in
the case of the rectangle; i.e.,  a
finite dimensional space with lower dimension than in the rectangle is required, according to criterion  (iii) reflected in \textcolor{Crimson}{Appendix} \ref{A4:sec:6}. It can also be appreciated that the number of components of $\boldsymbol{\beta }$   does not
substantially affect the accuracy of the estimates.

\bigskip

\begin{table}[H]
\caption[\hspace{0.7cm} Empirical functional mean square errors on the estimation of the fixed effect parameters for disk domain.]{\small{$EFMSE_{\boldsymbol{\beta}}$ for disk domain.}}
\centering
\begin{small}
$EFMSE_{\boldsymbol{\beta}}$ \\
\begin{tabular}{|c|c|c|}
  \hline
   (P$_{1}$,a,C$_{3}$) & (P$_{1}$,b,C$_{2}$) & (P$_{1}$,c,C$_{1}$) \\
  \hline
  $7.500 \left(10 \right)^{-4}$  &  $7.500 \left(10 \right)^{-4}$ &  $7.400 \left(10 \right)^{-4}$ \\
  \hline
   (P$_{1}$,d,C$_{1}$) & (P$_{1}$,e,C$_{2}$) & (P$_{1}$,f,C$_{3}$)  \\
  \hline
$7.500 \left(10 \right)^{-4}$&  $7.600 \left(10 \right)^{-4}$ &  $7.500 \left(10 \right)^{-4}$ \\
  \hline
  \hline
   (P$_{2}$,a,C$_{1}$) & (P$_{2}$,b,C$_{2}$) & (P$_{2}$,c,C$_{3}$)  \\
  \hline
  $7.000 \left(10 \right)^{-4}$  &  $7.100 \left(10 \right)^{-4}$&  $7.100 \left(10 \right)^{-4}$ \\
  \hline
  (P$_{2}$,d,C$_{3}$) & (P$_{2}$,e,C$_{2}$) & (P$_{2}$,f,C$_{1}$) \\
  \hline
  $7.900 \left(10 \right)^{-4}$  &  $8.000 \left(10 \right)^{-4}$ &  $8.000 \left(10 \right)^{-4}$ \\
  \hline
\end{tabular}
\end{small}
\label{A4:table:8}
\end{table}

\bigskip

\begin{table}[H]
\caption[\hspace{0.7cm} Empirical functional mean square errors on the estimation of the response for disk domain.]{\small{$EFMSE_{\mathbf{Y}}$ for disk domain.}}
\centering
\begin{small}
$EFMSE_{\mathbf{Y}}$ \\
\begin{tabular}{|c|c|c|c|c|c|}
  \hline
   (P$_{1}$,a,C$_{3}$) & (P$_{1}$,b,C$_{2}$) & (P$_{1}$,c,C$_{1}$) & (P$_{1}$,d,C$_{1}$) & (P$_{1}$,e,C$_{2}$) & (P$_{1}$,f,C$_{3}$) \\
  \hline
  $0.048$  &  $0.048$ &  $0.048$ &   $0.048$ &  $0.048$ &  $0.048$\\
  \hline
  \hline
  (P$_{2}$,a,C$_{1}$) & (P$_{2}$,b,C$_{2}$) & (P$_{2}$,c,C$_{3}$) & (P$_{2}$,d,C$_{3}$) & (P$_{2}$,e,C$_{2}$) & (P$_{2}$,f,C$_{1}$) \\
  \hline
  $0.050$  &  $0.050$ &  $0.050$ &  $0.049$  &  $0.050$ &  $0.050$ \\
  \hline
\end{tabular}
\end{small}
\label{A4:table:7}
\end{table}

\bigskip

The statistics (\ref{A4:800}) is now computed (see Table
\ref{A4:table:9}), as an empirical approximation of the relative
magnitude between the explained functional variability  and the
residual variability, after fitting the transformed Hilbert-valued
fixed effect  model (\ref{A4:tfdm}). It can be noticed that the values
of $\frac{\widetilde{SSR}}{\widetilde{SST}}$ are very close to one
in all the scenarios analysed. This fact induces large values of
(\ref{A4:800}) (see Table \ref{A4:table:9}), since
$$F=\frac{\widetilde{SSR}}{\widetilde{SSE}}=\frac{\widetilde{SSR}/\widetilde{SST}}{1-\widetilde{SSR}/\widetilde{SST}}.$$
It can be observed, one time more, from criterion  (iii), reflected in \textcolor{Crimson}{Appendix} \ref{A4:sec:6}, that the boundary conditions and the geometry of the domain
allows in this case a more substantial dimension reduction than in the rectangular domain case, since with lower truncation orders a better model fitting is obtained.

\bigskip

\begin{table}[H]
\caption[\hspace{0.7cm} $F$ statistics for the  disk domain.]{\small{$F$ statistics (\ref{A4:800}) over the  disk domain.}}
\centering
\begin{small}
\begin{tabular}{|c||c|c|c|}
  \hline
  Cases & (P$_{1}$,a,C$_{3}$) & (P$_{1}$,b,C$_{2}$) & (P$_{1}$,c,C$_{1}$) \\
  \hline 
  $F$ & $1.100(10^{2})$ & $4.100(10^{3})$ & $1.200(10^{5})$ \\
  \hline
  Cases & (P$_{1}$,d,C$_{1}$) & (P$_{1}$,e,C$_{2}$) & (P$_{1}$,f,C$_{3}$)  \\
  \hline
   $F$ & $3.900(10^{6})$ & $6.300(10^{6})$ & $4.200(10^{6})$ \\
  \hline
   \hline
  Cases & (P$_{2}$,a,C$_{1}$) & (P$_{2}$,b,C$_{2}$) & (P$_{2}$,c,C$_{3}$) \\
  \hline
  $F$ & $2.200(10^{3})$ & $8.200(10^{3})$ & $7.600(10^{7})$ \\
  \hline
  Cases & (P$_{2}$,d,C$_{3}$) & (P$_{2}$,e,C$_{2}$) & (P$_{2}$,f,C$_{1}$)  \\
  \hline
   $F$ & $2.500(10^{7})$ & $1.400(10^{7})$ & $8.500(10^{7})$ \\
  \hline
\end{tabular}
\end{small}
\label{A4:table:9}
\end{table}

\bigskip

The statistics $T_{\boldsymbol{h}}$ in (\ref{A4:78}) is computed  to contrast the
significance of the parameter vector $\boldsymbol{\beta}$ in case
  C$_{1},$  with $p=4$ components.  Again, eight realizations of Gaussian   random functions $\boldsymbol{h}$ are considered,  generated from   a Gaussian
random field $\xi,$ solution, in the mean--square sense, of the
following boundary value problem on the disk:

\begin{eqnarray} 
 (-\Delta )\xi (\mathbf{x})&=&\varsigma
(\mathbf{x}), \quad \mathbf{x}=(x_{1},x_{2})\in D_{25}=\{\mathbf{x}\in \mathbb{R}^{2}; \ 0<\|\mathbf{x}\|< 25\}, \nonumber \\
 \xi (\theta ,25)&=&0,\quad \forall \theta \in [0,2\pi]\nonumber
\end{eqnarray}
\noindent where $\varsigma$ denotes a  zero--mean Gaussian white
noise on $L^{2}(D_{25})$; i.e., a zero--mean generalized Gaussian
process satisfying 

\begin{eqnarray} 
\displaystyle \int_{[0,2\pi]\times
[0,25]}f(\varphi, v){\rm E} \left\lbrace \varsigma (\theta, r)\varsigma (\varphi,
v) \right\rbrace d\varphi dv=f(\theta, r), \quad (\theta, r)\in
[0,2\pi]\times [0,25], \quad f\in L^{2}(D_{25}).\nonumber
\end{eqnarray}

Table \ref{A4:tab:Tabla2} reflects the percentage of successes, for
$\alpha =0.05,$ and the averaged $p$--values over the $150$ samples, generated with size $n=150,$
of the functional response having parameter vector $\boldsymbol{\beta}$ of type
  C$_{1}$ with $p=4$ components, for  $TR= 7.$

\bigskip

\begin{table}[H] 

\caption[\hspace{0.7cm} Significance of the fixed effect parameters for disk domain.]{\small{\emph{\textcolor{Crimson}{Disk}}. Percentage  of successes for $\alpha =0.05,$
at the left--hand side, and averaged $p$--values at the right--hand
side, for each one of the eight realizations of the Gaussian function
$h\in L^{2}(D_{25}).$}}
\centering
\begin{small}
\begin{tabular}{|c||c|c|}
  \hline
 $D$& \% Success & $p$\\
  \hline \hline
   1 & $99.95\%$  &  $1.672(10)^{-8}$\\
  \hline
    2 & $99.5 \%$  & $9.746(10)^{-7}$\\
  \hline
     3 & $100  \%$  & $0$ \\
  \hline
     4 & $99.9 \%$  & $8.546(10)^{-8}$ \\
  \hline
     5 & $97.45 \%$  & $7.400(10)^{-7}$  \\
  \hline
      6 & $100 \%$  & $0$  \\
  \hline
     7 & $100 \%$ & $8.775(10)^{-9}$ \\
  \hline
      8 & $100 \%$ & $0$ \\
  \hline
\end{tabular}
\end{small}
  \label{A4:tab:Tabla2}
\end{table}

\bigskip

Table \ref{A4:tab:Tabla2} again illustrates a good performance of the
statistics  $T_{\boldsymbol{h}}$ in (\ref{A4:78}). Indeed, we can appreciate a high
percentage of successes, and very small $p$--values, very close to
zero, that support the significance of the functional parameter
vector, considered in the generation of the data set analysed.

\textcolor{Crimson}{\subsection{Circular sector domain}
\label{A4:sec:63}}

In the circular sector $$D_{3}= \left\lbrace
\left(r \cos\left(\varphi \right),~ r\sin\left(\varphi
\right)\right):~0 < \Vert r \Vert < R,~0 < \varphi < \pi \theta
\right\rbrace$$   of radius $R$ and angle $\pi \theta,$ the zero-mean Gaussian vector error
term  is generated  from the matrix covariance
 operator
$\boldsymbol{R}_{\boldsymbol{\varepsilon}\boldsymbol{\varepsilon}},$ whose functional entries  are
defined in equation (\ref{A4:41}). The eigenvectors  $\{\phi_{k},\ k\geq 1\}$  of  the Dirichlet negative Laplacian operator
on the circular sector  are considered (see equation (\ref{A4:eqcsa}) in the \textcolor{Crimson}{Supplementary Material} in \textcolor{Crimson}{Appendix} \ref{A4:Supp}), arranged in decreasing order of the modulus magnitude of their  associated eigenvalues. Specifically, here, $\boldsymbol{R}_{\boldsymbol{\varepsilon}\boldsymbol{\varepsilon}}$ is defined in equation (\ref{A4:41}), with for $i=1,\dots,n,$ $\lambda_{ki}=\lambda_{k}(R_{ii})$ being given in equations (\ref{A4:covopf}) and (\ref{A4:eqcsa}).

As in
the above examples, $\nu=20$ functional samples of size $n=200$ are
generated. The  cases studied are also summarized in terms of the
values of the vector  $(P_i, u, C_j)$, $i=1,2,~u=a,b,c,d,e,f,$
and $j=1,2,3,$ with the values of $u$ having the same meaning as in the
disk domain. Again, values of $P_i$ provide the number $p$  of components
 of $\boldsymbol{\beta};$ i.e., $p=4$ if $i=1$, and $p=9$ if
$i=2$. The values $C_1,$ $C_{2}$ and $C_3$ respectively correspond
to the following functions defining the components of
 $\boldsymbol{\beta },$ whose projections are given by:

\begin{eqnarray}
\beta_{sk} &=& 1+ (k-1)s, \quad k=1,\dots,TR, \quad s=1,\dots,p\quad
\mbox{(\textcolor{Crimson}{C1})}\nonumber\\
\beta_{sk} &=& \frac{1}{R}e^{\frac{s + \frac{k}{R}}{n}}+ k \cos
\left( \left(-1 \right)^k 2 \pi \frac{R}{k} \right),\quad
k=1,\dots,TR, \quad s=1,\dots,p\quad  \mbox{(\textcolor{Crimson}{C2})}\nonumber\\
\beta_{sk} &=& \cos \left(\pi \frac{TR - k}{k} \right)\cos
\left(\pi \frac{p - s}{s} \right), \quad k=1,\dots,TR, \quad
s=1,\dots,p\quad \mbox{(\textcolor{Crimson}{C3})}.\nonumber
\end{eqnarray}

\bigskip

A summary of the cases analysed is given in  Table \ref{A4:table:12}.

\bigskip

\begin{table}[H]
\caption[\hspace{0.7cm} Scenarios in the FANOVA simulation study for circular sector domain.]{\small{Scenarios for circular sector domain.}} 

\centering
\begin{small}
\begin{tabular}{|c||c|c|c|c|c|c|}
  \hline
  \textbf{Cases} & $R$ & $h_R$ & $h_\phi$ & $TR$ & $\theta$ & $p$\\
  \hline \hline
(P$_{1}$,a,C$_{3}$) & $12$ & $\frac{R}{145}$ & $\frac{2 \pi}{115}$ & $3$ & $\frac{2}{3}$ & $4$\\
  \hline
(P$_{1}$,b,C$_{2}$) & $18$ & $\frac{R}{145}$ & $\frac{2 \pi}{115}$ & $5$ & $\frac{2}{3}$ & $4$\\
  \hline
(P$_{1}$,c,C$_{1}$) & $25$ & $\frac{R}{145}$& $\frac{2 \pi}{115}$ & $7$ & $\frac{2}{3}$ & $4$\\
 \hline
(P$_{1}$,d,C$_{1}$) & $50$ & $\frac{R}{145}$ & $\frac{2 \pi}{115}$ & $15$ & $\frac{2}{3}$ & $4$\\
  \hline
(P$_{1}$,e,C$_{2}$) & $100$ & $\frac{R}{145}$ & $\frac{2 \pi}{115}$& $31$ &$\frac{2}{3}$ & $4$\\
  \hline
(P$_{1}$,f,C$_{3}$) & $250$ & $\frac{R}{145}$ & $\frac{2 \pi}{115}$ & $79$ & $\frac{2}{3}$ & $4$\\
    \hline
    \hline
(P$_{2}$,a,C$_{1}$) & $12$ & $\frac{R}{145}$ & $\frac{2 \pi}{115}$ & $3$ & $\frac{2}{3}$ & $9$ \\
  \hline
(P$_{2}$,b,C$_{2}$) & $18$ & $\frac{R}{145}$ & $\frac{2 \pi}{115}$ & $5$ & $\frac{2}{3}$ & $9$ \\
  \hline
(P$_{2}$,c,C$_{3}$) & $25$ & $\frac{R}{145}$& $\frac{2 \pi}{115}$ & $7$ & $\frac{2}{3}$ & $9$ \\
 \hline
(P$_{2}$,d,C$_{3}$) & $50$ & $\frac{R}{145}$ & $\frac{2 \pi}{115}$ &  $15$ & $\frac{2}{3}$ & $9$ \\
  \hline
(P$_{2}$,e,C$_{2}$) & $100$ & $\frac{R}{145}$ & $\frac{2 \pi}{115}$& $31$  & $\frac{2}{3}$ & $9$ \\
  \hline
(P$_{2}$,f,C$_{1}$) & $250$ & $\frac{R}{145}$ & $\frac{2 \pi}{115}$ & $79$ & $\frac{2}{3}$ & $9$ \\
  \hline
\end{tabular}
\end{small}
 \label{A4:table:12}
\end{table}

\bigskip

\begin{figure}[H]
  \centering
    \includegraphics[width=0.8\textwidth]{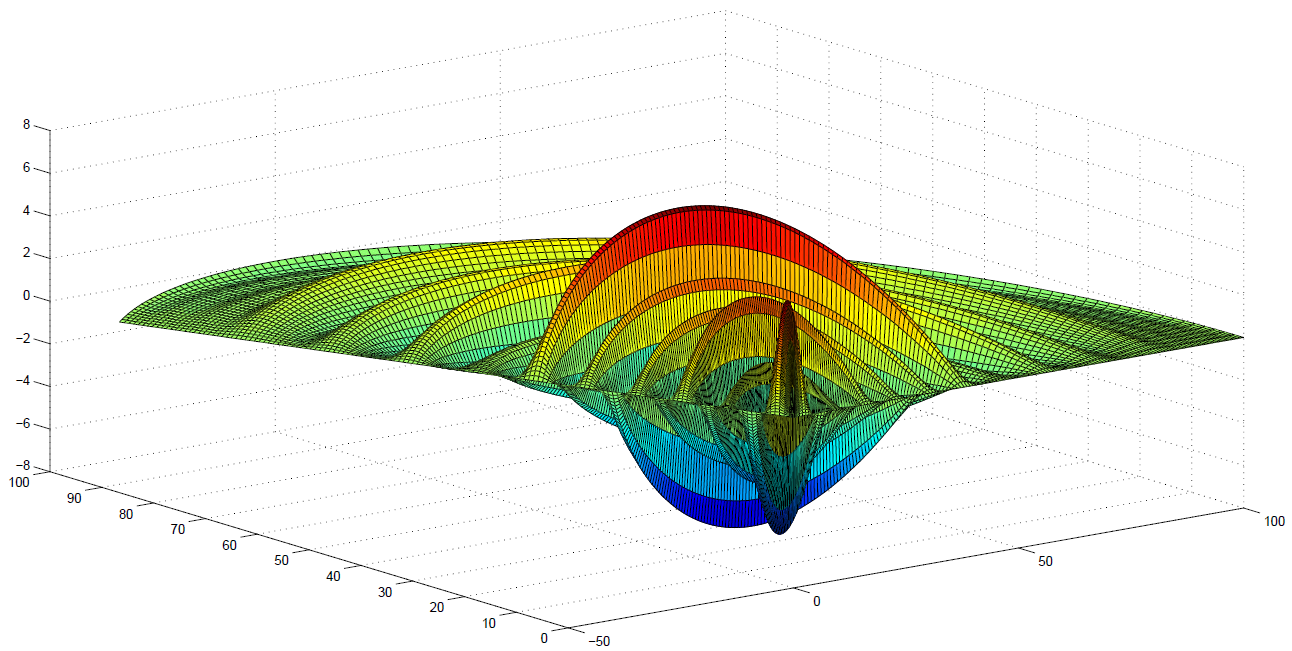}
    
       \vspace{-0.1cm}
\caption[\hspace{0.7cm} Simulated response for circular sector domains (case C2).]{\small{Case (P$_{2}$,e,C$_{2}$). Simulated response with $p=9$, $R=100$ and $\boldsymbol{\beta}$ of type C$_{2}$.}}
  \label{A4:fig:12hh}
\end{figure}

\begin{figure}[H]
  \centering
    \includegraphics[width=0.8\textwidth]{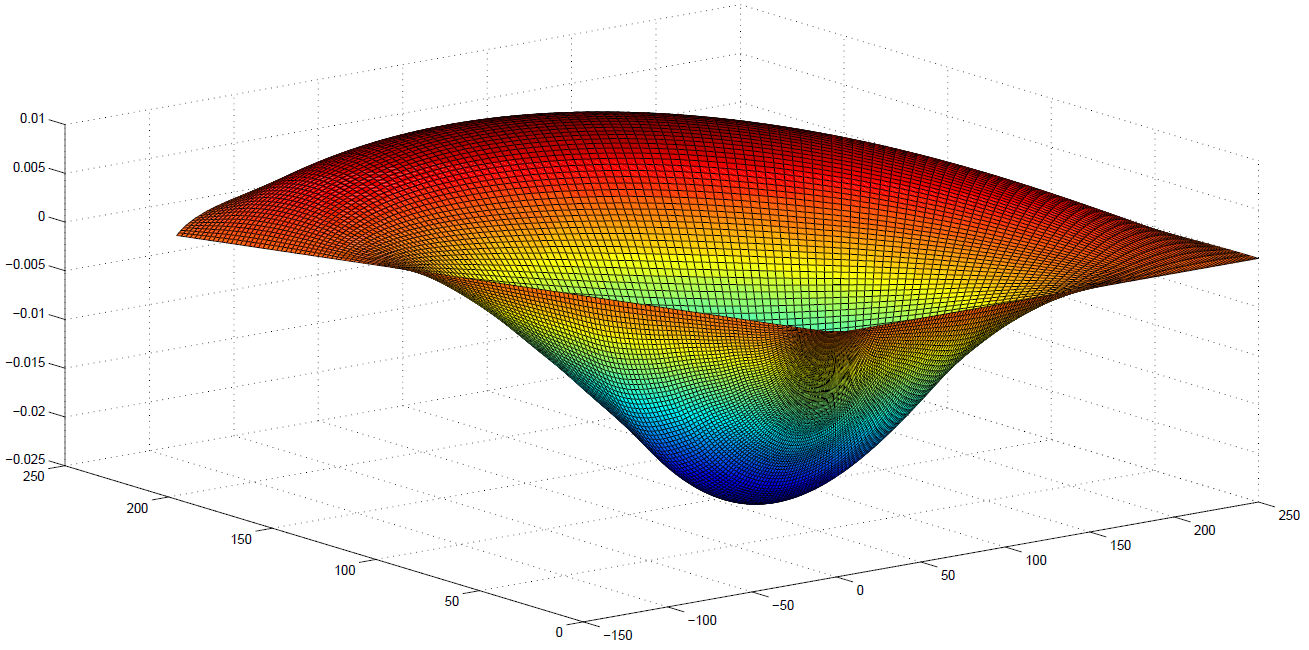}
    
       \vspace{-0.1cm}
\caption[\hspace{0.7cm} Simulated response for circular sector domains (case C3).]{\small{Case (P$_{1}$,f,C$_{3}$). Simulated response with $p=4$, $R=250$ and $\boldsymbol{\beta}$ of type C$_{3}$.}}
  \label{A4:fig:13}
\end{figure}

 Figures  \ref{A4:fig:12hh}--\ref{A4:fig:13} display the generation of a functional value of the response in the cases
(P$_{2}$,e,C$_{2}$) and (P$_{1}$,f,C$_{3}$),  respec\-tively.

\bigskip

\bigskip

\begin{figure}[H]
  \centering
    \includegraphics[width=0.8\textwidth]{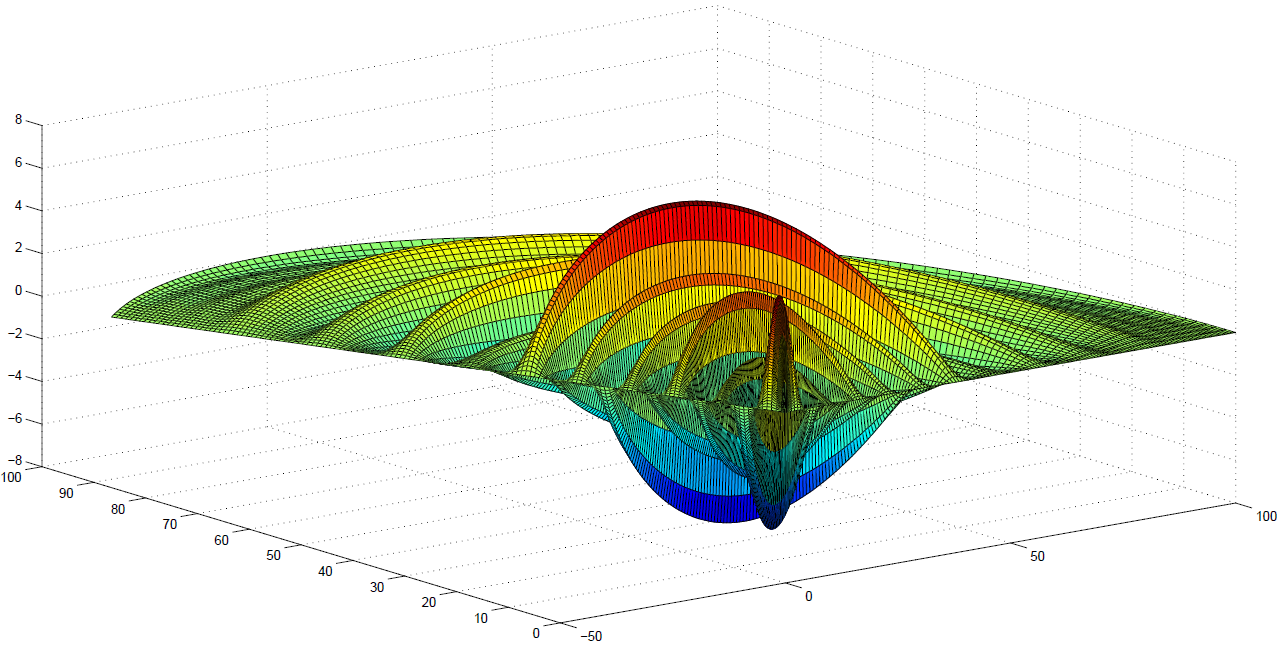}
    
           \vspace{-0.1cm}
\caption[\hspace{0.7cm} Estimated response for circular sector domains (case C2).]{\small{Case (P$_{2}$,e,C$_{2}$). Estimated response with $p=9$, $R=100$ and $\boldsymbol{\beta}$ of type C$_{2}$.}}
  \label{A4:fig:15}
\end{figure}

\begin{figure}[H]
  \centering
    \includegraphics[width=0.8\textwidth]{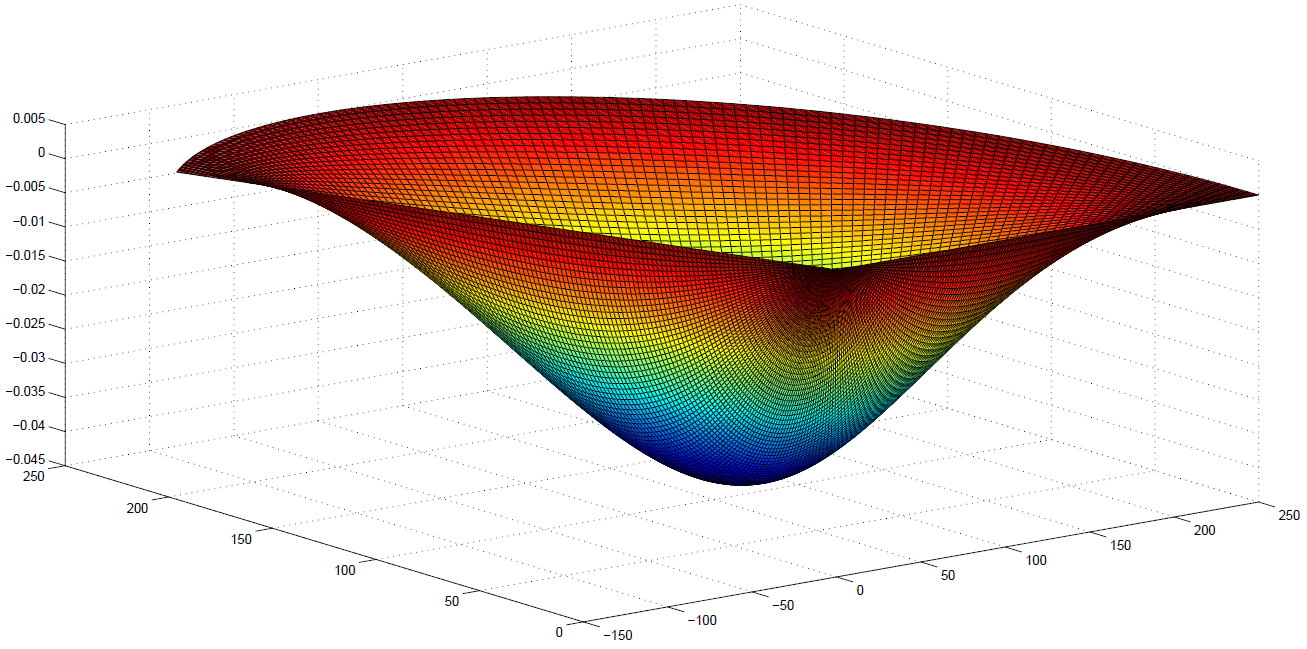}
    
           \vspace{-0.1cm}
\caption[\hspace{0.7cm} Estimated response for circular sector domains (case C3).]{\small{Case (P$_{1}$,f,C$_{3}$). Estimated response with $p=4$, $R=250$ and $\boldsymbol{\beta}$ of type C$_{3}$.}}
  \label{A4:fig:16}
\end{figure}

The functional estimates obtained from the finite--dimensional
approximation of the generalized least--squares estimator of
$\boldsymbol{\beta }$ are now given in Figures \ref{A4:fig:15}--\ref{A4:fig:16}, for the cases (P$_{2}$,e,C$_{2}$) and (P$_{1}$,f,C$_{3}$), respectively.

\bigskip

\bigskip

As in the previous sections, the empirical functional mean quadratic
errors, associated with the estimation of $\boldsymbol{\beta }$ and
$\mathbf{Y},$ are computed from equations (\ref{A4:777})--(\ref{A4:788}). They are  shown in Table \ref{A4:table:14}, for $\boldsymbol{\beta},$ and in Table \ref{A4:table:13}, for
$\mathbf{Y}.$

These empirical functional mean quadratic errors are very stable through the different cases considered, and their  order of magnitude is again
 $10^{-3}$ for the parameter $\boldsymbol{\beta},$ and
 $10^{-2}$ for the response.
   Here, the results displayed also correspond to the projection into  lower
 finite--dimensional spaces than in the case of the rectangle, according to the functional form of the eigenvectors (see truncation order criterion (iii) in \textcolor{Crimson}{Appendix} \ref{A4:sec:6}).

\bigskip

\begin{table}[H]
 \caption[\hspace{0.7cm} Empirical functional mean square errors on the estimation of the fixed effect parameters for circular sector domain.]{\small{$EFMSE_{\boldsymbol{\beta}}$ for the circular sector.}}
\centering
\begin{small}
$EFMSE_{\boldsymbol{\beta}}$ \\
\begin{tabular}{|c||c|c|}
  \hline
  (P$_{1}$,a,C$_{3}$) & (P$_{1}$,b,C$_{2}$) & (P$_{1}$,c,C$_{1}$) \\
  \hline
  $1.200 \left(10 \right)^{-4}$  & $1.100 \left(10 \right)^{-4}$ &  $1.200 \left(10 \right)^{-4}$ \\
  \hline
  (P$_{1}$,d,C$_{1}$) & (P$_{1}$,e,C$_{2}$) & (P$_{1}$,f,C$_{3}$)  \\
  \hline
 $1.200 \left(10 \right)^{-4}$ &  $1.200 \left(10 \right)^{-4}$ & $1.100 \left(10 \right)^{-4}$\\
  \hline
  \hline
  (P$_{2}$,a,C$_{1}$) & (P$_{2}$,b,C$_{2}$) & (P$_{2}$,c,C$_{3}$) \\
  \hline
$1.900 \left(10 \right)^{-4}$  &  $2.000 \left(10 \right)^{-4}$&  $2.000 \left(10 \right)^{-4}$ \\
  \hline
  (P$_{2}$,d,C$_{3}$) & (P$_{2}$,e,C$_{2}$) & (P$_{2}$,f,C$_{1}$)  \\
  \hline
$1.900 \left(10 \right)^{-4}$  &  $1.900 \left(10 \right)^{-4}$&  $2.000 \left(10 \right)^{-4}$ \\
  \hline
\end{tabular}
\end{small}
\label{A4:table:14}
\end{table}

\bigskip

\begin{table}[H]
\caption[\hspace{0.7cm} Empirical functional mean square errors on the estimation of the response for circular sector domain.]{{\small $EFMSE_{\mathbf{Y}}$ for the circular sector.}}
\centering
\begin{small}
$EFMSE_{\mathbf{Y}}$ \\
\begin{tabular}{|c||c|c|}
  \hline
  (P$_{1}$,a,C$_{3}$) & (P$_{1}$,b,C$_{2}$) & (P$_{1}$,c,C$_{1}$)  \\
  \hline
  $8.770 \left(10 \right)^{-3}$ & $8.810 \left(10 \right)^{-3}$  & $8.820 \left(10 \right)^{-3}$\\
  \hline
   (P$_{1}$,d,C$_{1}$) & (P$_{1}$,e,C$_{2}$) & (P$_{1}$,f,C$_{3}$)  \\
  \hline
  $8.820 \left(10 \right)^{-3}$ & $8.820 \left(10 \right)^{-3}$ &  $8.810 \left(10 \right)^{-3}$ \\
  \hline
  \hline
  (P$_{2}$,a,C$_{1}$) & (P$_{2}$,b,C$_{2}$) & (P$_{2}$,c,C$_{3}$)  \\
  \hline
 $9.630 \left(10 \right)^{-3}$ &  $9.670 \left(10 \right)^{-3}$ & $9.670 \left(10 \right)^{-3}$ \\
  \hline
  (P$_{2}$,d,C$_{3}$) & (P$_{2}$,e,C$_{2}$) & (P$_{2}$,f,C$_{1}$)   \\
  \hline
$9.670 \left(10 \right)^{-3}$ & $9.680 \left(10 \right)^{-3}$ &  $9.660 \left(10 \right)^{-3}$\\
  \hline
\end{tabular}
\end{small}
 \label{A4:table:13}
\end{table}

\bigskip

 Statistics (\ref{A4:800}) is now computed. Its values are displayed in   Table \ref{A4:table:15}.
 Again, as in the disk,  the proportion of explained functional variability is very close to
 one leading to large values of  statistics (\ref{A4:800}), as it can be observed in  Table \ref{A4:table:15} for all the cases analysed.

\bigskip

\begin{table}[H]
\caption[\hspace{0.7cm} $F$ statistics for the circular sector domain.]{{\small $F$ statistics (\ref{A4:800}) for the circular sector.}}

\centering
\begin{small}
\begin{tabular}{|c||c|c|c|c|c|c|c|}
  \hline
  Cases & (P$_{1}$,a,C$_{3}$) & (P$_{1}$,b,C$_{2}$) & (P$_{1}$,c,C$_{1}$)  & (P$_{1}$,d,C$_{1}$) & (P$_{1}$,e,C$_{2}$) & (P$_{1}$,f,C$_{3}$)  \\
  \hline
  $F$ & $9.2(10^{2})$ & $3.1(10^{3})$ & $4.2(10^{6})$  & $4.8(10^{8})$ & $5.8(10^{6}))$ & $7.3(10^{8})$ \\
  \hline
   \hline
  Cases & (P$_{2}$,a,C$_{1}$) & (P$_{2}$,b,C$_{2}$) & (P$_{2}$,c,C$_{3}$) &  (P$_{2}$,d,C$_{3}$) & (P$_{2}$,e,C$_{2}$) & (P$_{2}$,f,C$_{1}$)  \\
  \hline
  $F$ & $1.8(10^{3})$ & $4.1(10^{3})$ & $2.6(10^{7})$  & $3.1(10^{9})$ & $6.8(10^{6})$ & $1.8(10^{9})$ \\
  \hline
\end{tabular}
\end{small}
\label{A4:table:15}
\end{table}

\bigskip

The statistics $T_{\boldsymbol{h}}$ in (\ref{A4:78}) is computed  to contrast the
significance of the parameter vector $\boldsymbol{\beta}$ in case
  C$_{1}$ with  $p=4$ functional components. Eight realizations of a Gaussian random function $\boldsymbol{h}$ are considered from  a Gaussian random field
$\xi,$ solution, in the mean-square sense, of the following boundary
value problem on the circular sector
\begin{eqnarray} 
(-\Delta )\xi (\mathbf{x})&=&\varsigma
(\mathbf{x}), \quad \mathbf{x} =\left(r
\cos\left(\varphi \right),~ r\sin\left(\varphi
\right)\right),~0 < \Vert r \Vert < R,~0 < \varphi <
\pi \theta,
\nonumber\\
 \xi (\varphi  ,25)&=&0,\quad  \varphi  \in [0,\pi \theta
], \nonumber
\end{eqnarray}
\noindent where $\theta =2/3,$ $\varsigma$ denotes a  zero--mean
Gaussian white noise on the circular sector such that
\begin{eqnarray} 
\displaystyle \int_{[0,\pi \theta]\times [0,25]}f(\varphi,
v) {\rm E} \left\lbrace \varsigma(\gamma, r)\varsigma (\varphi, v) \right\rbrace d\varphi
dv=f(\gamma, r), \quad (\gamma, r)\in [0,\pi
\theta]\times [0,25],\quad f\in L^{2}(CS),\nonumber
\end{eqnarray}
\noindent with  $L^{2}(CS)$ denoting the space of square--integrable
functions on the circular sector. Table \ref{A4:tab:Tabla3} reflects
the percentage of successes, for $\alpha =0.05,$ and the averaged
$p$--values over the $150$ samples, with
size \linebreak $n=150,$ of the response, having  C$_{1}$--type  functional parameter vector $\boldsymbol{\beta}$ with  $p=4$ components, considering  $TR= 7.$

\bigskip

\begin{table}[H]
\caption[\hspace{0.7cm} Significance of the fixed effect parameters for circular sector domains.]{\small{\emph{\textcolor{Crimson}{Circular Sector.}} Percentage  of successes for
$\alpha =0.05,$ at the left--hand side, and averaged $p$--values at
the right--hand side, for each one of the eight realizations of the
Gaussian function $h\in L^{2}(CS).$}}
\centering
\begin{small}
\begin{tabular}{|c||c|c|}
  \hline
 $D$ & \% Success& $p$\\
  \hline \hline
   1 &  $97.5 \%$   &  $6.504(10)^{-6}$ \\
  \hline
    2 & $100 \%$  & $0$\\
  \hline
     3 & $100  \%$  & $3.600(10)^{-8}$ \\
  \hline
     4 & $100  \%$  & $0$ \\
  \hline
     5 &  $98 \%$  & $2.006(10)^{-6}$ \\
  \hline
      6 &  $99.5 \%$  & $9.807(10)^{-8}$ \\
  \hline
     7 & $100  \%$  & $0$ \\
  \hline
      8 & $99.5  \%$  & $4.111(10)^{-7}$ \\
  \hline
\end{tabular}
\end{small}
  \label{A4:tab:Tabla3}
\end{table}

\bigskip

Table \ref{A4:tab:Tabla3} again confirms the good performance of the
test statistics $T_{\boldsymbol{h}},$ showing a high percentage of successes, and
very small magnitudes for the averaged $p$--value (almost
zero values), according to the significance of the parameter vector $\boldsymbol{\beta}$ considered in the generation of the analysed
functional data set.

%
%

\textcolor{Crimson}{\section{Functional statistical analysis of fMRI data}
\label{A4:sec: fmri}}

 In this section, we compare the  results obtained from the application of the MatLab function \textit{fmrilm.m} (see \cite{Liaoetal02} and \cite{Worsleyetal02}) from   \textit{fmristat.m} function set (available   at \url{http://www.math.mcgill.ca/keith/fmristat}), with those ones provided by the implementation of our proposed functional statistical methodology, based on the Hilbert-valued fixed effect models with ARH(1) error term above introduced. The fMRI data set analysed is also freely available in AFNI format at \url{http://www.math.mcgill.ca/keith/fmristat/}.
(AFNI Matlab toolbox can be applied to read such a data set). In the next section, structural information about such fMRI data is provided (see \textit{BrikInfo.m} Matlab function).

The first step in the statistical analysis of fMRI data
is to modeling the data response to an
external stimulus. Specifically, at each voxel, denote by  $x(t)$ the (noise-free) fMRI response
 at time $t,$ and by $s(t)$ the external
 stimulus at that time. It is well--known that the corresponding
fMRI response is not instantaneous, suffering a blurring and a delay
of the peak response by about $6s$ (see, for example,
\cite{Liaoetal02}). This fact is usually modelled by assuming that  the
fMRI response depends on the external stimulus by convolution with a
hemodynamic response function $h(t)$ (which is usually assumed to be
independent of the voxel), as follows:
\begin{equation}
x(t)=\int_{0}^{\infty}h(u)s(t-u)du.
\label{A4:eqsthr}
\end{equation}

Several models have been proposed in the literature for the hemodynamic response
function (hrf). For example, the gamma function (see \cite{LangeZeger97}), or the
difference of two gamma functions, to model the slight
intensity dip after the response has fallen back to zero
(see \cite{Fristonetal98}).

The  effects  $\left(x_{i,1},\ldots, x_{i,p}\right)$ of $p$ different  types of stimuli
on data, in scan $i,$ is combined  in terms of an additive model with
different multiplicative coefficients $\left( \beta_{1},\ldots, \beta_{p} \right)$
that vary from voxel to voxel. The combined fMRI response is then
modeled as the linear model (see \cite{Fristonetal95})
$$x_{i,1}\beta_{1}(v)+\dots +x_{i,p}\beta_{p}(v),$$   for each voxel $v.$

An important drift over time can be observed in fMRI time series data in some voxels. Such a drift is usually
 linear, or a
more general slow variation function. In the first case, i.e., for a
linear function $$x_{i,k+1}\beta_{k+1}(v)+\dots
+x_{i,m}(v)\beta_{m}(v),$$ when the drift is not
removed, it can be confounded with the fMRI response. Otherwise, it
can be added to  the estimate of the random noise $\varepsilon ,$
which, in the  simplest case is assumed to be an AR(1) process at
each voxel. In that case,  the linear model
fitted to the observed fMRI data is usually given by
\begin{equation}
Y_{i}(v)=x_{i,1}\beta_{1}(v)+\dots +x_{i,p}\beta_{p}(v) + x_{i,k+1}\beta_{k+1}(v)+\dots
+x_{i,m}\beta_{m}(v)+\varepsilon_{i}(v),\quad i=1,\dots,n,\label{A4:Wm}
\end{equation}
\noindent for each one of the voxels $v,$ in the real--valued
approach presented in  \cite{Worsleyetal02}. In (\ref{A4:Wm}), $$ \varepsilon_{i}(v)=\rho (v)\varepsilon_{i-1}(v)+\xi_{i}(v),\quad |\rho(v)|<1,$$
where  $\{\xi_{i}(v),\ i=1,\dots,n\}$ are $n$ random components of   Gaussian white
noise in time, for each voxel $v.$ This  temporal correlation
structure for the noise has sense, under the assumption that the
scans are equally spaced in time, and that the error from the
previous scan is combined with fresh noise to produce the error for
the current scan. In the presented Hilbert--valued   approach, a
similar reasoning can be applied to arrive to the  fixed effect
model with ARH(1) error term, introduced in \textcolor{Crimson}{Appendix} \ref{A4:sec:3}. This model allows the representation
of  fMRI data in a functional spatially continuous form. Specifically,   for the
scan $i,$  a  continuous spatial variation  is assumed underlying to
the values of the noise across the voxels, reflected in   the
functional value of the ARH(1) process, representing the error term.  In
the same way, the $H$--valued  components of the parameter vector
$\boldsymbol{\beta }(\cdot)$ provide a continuous model to represent
spatial variation over the voxels of the multiplicative coefficients
$\beta_{1}(\cdot),\dots, \beta_{p}(\cdot),$ independently of time.
Since the fMRI response is subsampled at the $n$ scan acquisition
times $t_{1},\dots,t_{n},$ the fixed effect design matrix $\boldsymbol{X},$
constituted by the values of the fMRI response (\ref{A4:eqsthr}) at
such times, under the $p$ different types of stimuli  considered, has
dimension $n\times p.$ Note that in  (\ref{A4:eqsthr}) $x$ is assumed
to be independent of the voxel, according to the definition of the
hrf.

\textcolor{Crimson}{\subsection{Description of the data set and the fixed effect design matrix}
\label{A4:sec:fmri2}}

  Brain scan measurements are represented on a set
of $64 \times 64 \times 16$ voxels. Each one of such voxels
represents a cube of $3.75 \times 3.75 \times 7~mm.$
At each one of the $16$ depth levels or slices $\left\lbrace S_i, \ i=1,\dots,16 \right\rbrace$,
the brain is scanned in $68$ frames, $\left\lbrace Fr_h, \ h=1,\dots,68 \right\rbrace$.
Equivalently, for $i=1,\dots,16,$ on the slice $S_i,$  a $64 \times 64$
rectangular grid is considered, where measurements at  each one of the $68$ frames are collected.

 We restrict our attention to
the case $p=2,$ where two type of events are considered,
respectively representing  scans hot stimulus (with a height $h_h$)
and scans warm stimulus (with a height $h_w$).
 The default parameters, chosen by \cite{Glover99}, to
generate the hrf as the difference of two gamma densities is the row
vector $r = \left[5.4, 5.2, 10.8, 7.35, 0.35 \right]$, where the
first and third parameters represent the time to peak of the first
and second gamma densities ($\Gamma_1$ and $\Gamma_2$),
respectively; the second and fourth parameters represent the
approximate full width at half maximum (FWHM) of the first and
second gamma densities, respectively; and the fifth parameter
(called also $DIP$) denotes the coefficient of the second gamma
density, for more details, see \cite{Glover99}, about modelling the
hrf as the difference of two gamma density functions, in the
following way:
\begin{equation}
hrf = \frac{\Gamma_1}{max(\Gamma_1)} - DIP\left(\frac{\Gamma_2}{max(\Gamma_2)}\right). \nonumber
\end{equation}

\begin{figure}[H]
  \centering
    \includegraphics[width=0.7\textwidth]{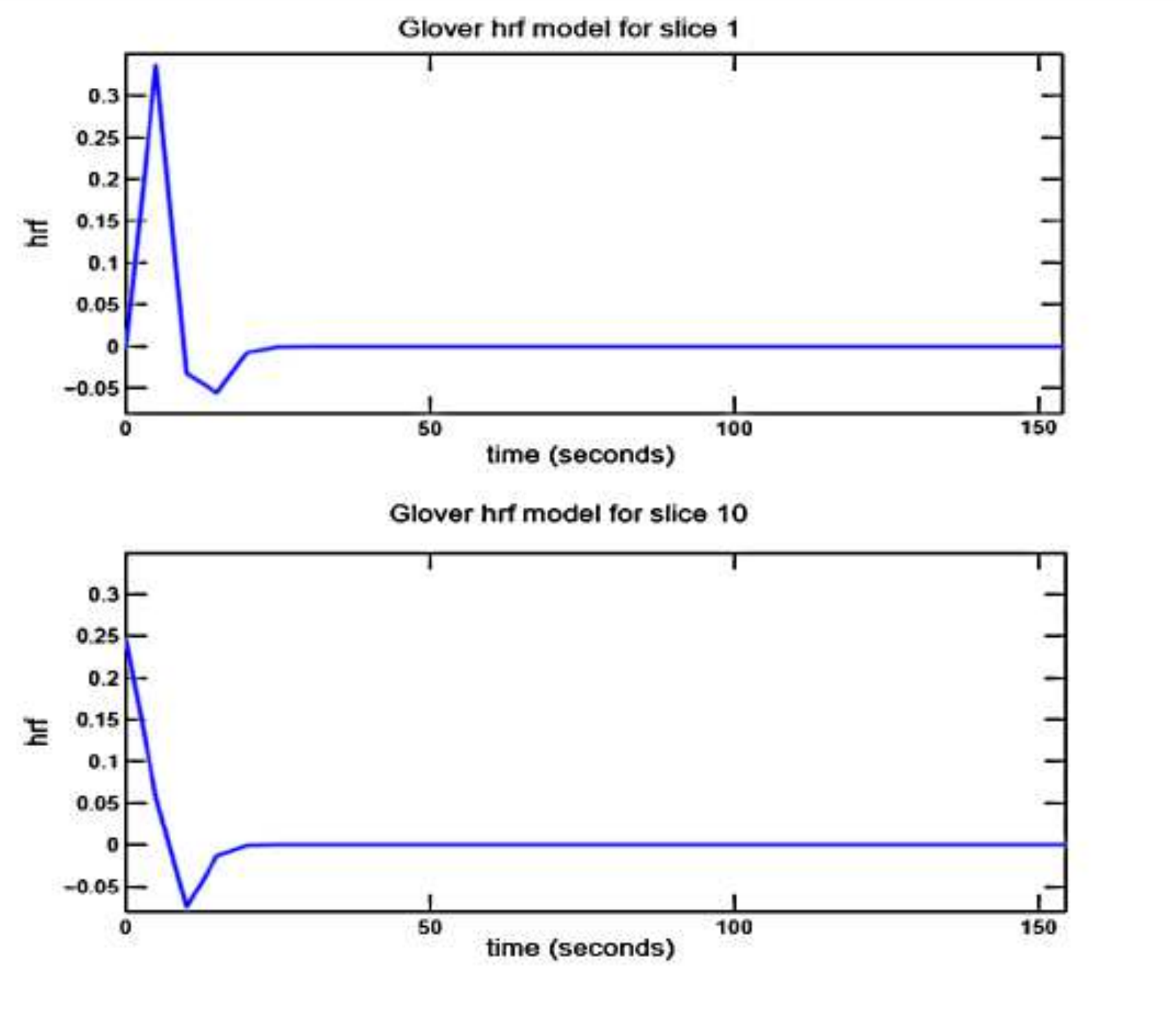}
    
    \vspace{-0.25cm}
    \caption[\hspace{0.7cm} Glover's hrf model (without convoluting) obtained by \textit{fmridesign.m}.]{\small{hrf model in \cite{Glover99} (without convoluting) obtained by \textit{fmridesign.m} Matlab function, for slices $S_i$, with $i=1$ (top) and $i=10$ (bottom), until frame time $Fr_{times} = 150$ (i.e., the Glover's hrf continues to be zero).}}
  \label{A4:fig:23}
\end{figure}

 Considering $TR_t = 5$ seconds  as the temporal step between each frame $Fr_h,~h=1,\dots,68$,
 in which all slices are scanned, frame times will be $Fr_{times} = \left(0, 5, 10 ,\dots, 330, 335 \right)$ (see Figure \ref{A4:fig:23}). Remark that, for any of the $68$ scans, separated by $TR_t = 5$ seconds, keeping in mind that the first $4$ frames are removed, $16$ slices $\left\lbrace S_i, \ i=1,\dots,16 \right\rbrace$,
 are interleaved every $0.3125$ seconds, approximately.

\bigskip

\bigskip

The events matrix $\boldsymbol{E}$, which will be convoluted with the hrf, is a matrix whose rows are the events, and whose columns are the identifier of the event type, the   starting event time, the duration of the event, and the  height of the response for the event, respectively. In our example, we have considered a block design of $4$ scans rest, $4$ scans hot stimulus, $4$ scans rest, $4$ scans warm stimulus, repeating $4$ times this block with $4$ last scans rest ($68$ scans total). As noted before, we remove the first 4 frames. The hot event is identified by $1$ and the warm event by $2,$ such that $h_h = 0.5$ and $h_w = 1$. Event times, for hot and warm stimulus, will be $\left[20, 60,  \dots, 260, 300 \right]$, since there are $8$ frames between the beginning of events ($4$ frames for the previous event and $4$ frames rest). Then, our events matrix $\boldsymbol{E}$ considered is

\begin{small}
\begin{equation}\boldsymbol{E} = \left( \begin{array}{cccc}
1 & 20 & 5 & 0.5 \\
2 & 60 & 5 & 1 \\
1 & 100 & 5 & 0.5  \\
2 & 140 & 5 & 1 \\
1 & 180 & 5 & 0.5 \\
2 & 220 & 5 & 1 \\
1 & 260 & 5 & 0.5 \\
2 & 300 & 5 & 1
\end{array} \right).\label{A4:Eventm}
\end{equation}
\end{small}

\begin{figure}[H]
  \centering
    \includegraphics[width=0.7\textwidth]{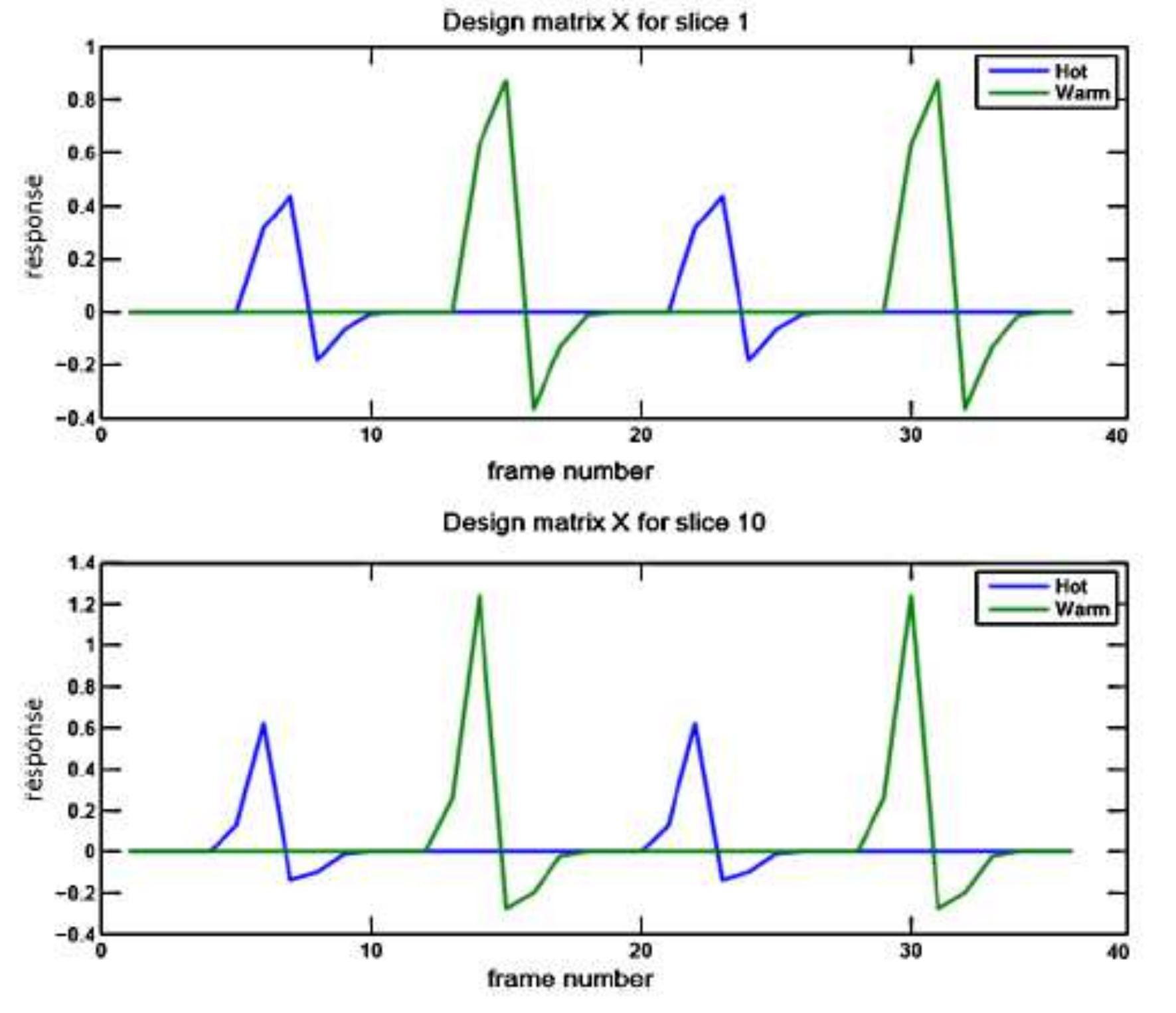}
    
    \vspace{-0.25cm}
    \caption[\hspace{0.7cm} Design matrix  for the first 40 frames, and slices $S_i$, with $i=1$ (top)  and $i=10$ (bottom).]{\small{Design matrix $\boldsymbol{X}_i$ for the first 40 frames, and slices $S_i$, with $i=1$ (top)  and $i=10$ (bottom), obtained by \textit{fmridesign.m} Matlab function through the convolution of our events matrix with the hrf model in \cite{Glover99}.}}
  \label{A4:fig:24}
\end{figure}

Convolution of matrix $\boldsymbol{E}$, in (\ref{A4:Eventm}), with the hrf leads to the  set of real--valued $64 \times 2$ design matrices $$\left\lbrace \boldsymbol{X}_i, \ i=1,\dots,16 \right\rbrace, \quad \boldsymbol{X}_i \in \mathbb{R}^{64 \times 2},$$
implemented by \textit{fmridesgin.m} Matlab function (see Figure \ref{A4:fig:24}).

\bigskip

\textcolor{Crimson}{\subsection{Hilbert--valued fixed effect model fitting to FMRI data. A comparative study}
\label{A4:sec:fmri4}}

The estimation results obtained with the implementation of the
classical and Hilbert--valued linear model methodology are now compared.
Specifically, in the classical case, from the linear model approach presented in  \cite{Worsleyetal02}, we consider  a fixed--effect model fitting, in the case where  the error term  is an AR(1)  process,
 ignoring spatial correlation across the voxels. In particular,  the MatLab function \textit{fmrilm.m}  is implemented to fit model (\ref{A4:Wm}) to a single run of fMRI data, allowing for spatially
varying temporal correlated errors.   The parameters of the  spatial varying  AR(1)
models (from voxel to voxel) are estimated from the sample autocorrelation of the
residuals, obtained after estimation of the fixed effect parameter
by ordinary least--squares, ignoring temporal correlation of the
errors, at each voxel. This procedure could be iterated. That is, the estimated
autocorrelation coefficient can be used to   pre--whitening the data at each voxel.
Hence, the fixed effect parameter is estimated by ordinary least-squares,
from such data. This iterative estimation procedure can be repeated  several times.  However, as  pointed out in \cite{Worsleyetal02},  such iterations do not lead to a substantial  improvement in practice.
A  variance  reduction technique is then applied in \cite{Worsleyetal02} to the estimated autocorrelation coefficient
(reduced bias sample autocorrelation), consisting of
spatial smoothing of the sample autocorrelations. This technique reduces variability,
although slightly increases the bias.

 In this subsection, we also implement the approach
introduced in \textcolor{Crimson}{Appendix} \ref{A4:sec:3}, from the fMRI data set described
in \textcolor{Crimson}{Appendix} \ref{A4:sec:fmri2}.  As commented before, our approach
presents the advantage of providing a continuous spatial  description
of the variation of the fixed effect parameters, as well as of the
parameters characterizing the temporal correlated error term, with
autoregressive dynamics. Furthermore, the spatial correlations are also
incorporated to our functional statistical analysis, computed from
the spatial autocovariance and cross-covariance kernels,
respectively defining the operators $R_{0}$ and $R_{1},$
characterizing the functional dependence structure  of the ARH(1)
error term.

Functional fixed effect model fitting is independently  performed at
each slice $S_i,$ for $i=1,\dots,16.$ Specifically, for  $i=1,\dots,
16,$ as commented before,   a real-valued $n\times p,$ with $p=2,$ fixed effect
design matrix $\mathbf{X}_i$ is considered (see \textcolor{Crimson}{Appendix} \ref{A4:sec:fmri2}). The effects of the two different events studied are combined by the vector of functional
fixed effect parameters $$\boldsymbol{\beta}_i
(\cdot)=[\beta_{1,i}(\cdot),\beta_{2,i}(\cdot)]^{T}\in H^{2}.$$
Here, $H^{2}$ is the Hilbert space of $2$--dimensional vector
functions, whose components are  square--integrable over the spatial
rectangular grid considered at each slice. Furthermore,  for
$i=1,\dots,16,$ $$\mathbf{Y}_i
(\cdot)=[Y_{1,i}(\cdot),\dots,Y_{n,i}(\cdot)]^{T}$$ is the
$H^{n}$--valued Gaussian  fMRI data response,  with $n$ representing
the number of frames ($n=64,$ since the first $4$ frames are
removed because they do not represent steady--state images).
 In the computation of the generalized least--squares estimate of $\boldsymbol{\beta },$ the empirical   matrices $\left\lbrace
\widehat{\boldsymbol{\Lambda}}_k, \ k=1,\dots,TR \right\rbrace$ are
computed from the empirical covariance operators
(\ref{A4:eqempiricalcov}), where $TR$ is selected according to the
required conditions specified, in relation to  the  sample size $n,$ in \cite{Bosq00} (see, in particular,  \cite[pp. 101--102 and pp. 116--117]{Bosq00}, and    \textcolor{Crimson}{Remark} \ref{A4:B00}).

In the subsequent developments, in the results obtained by applying
the Hilbert--valued multivariate fixed effect approach, we will
distinguish between cases A and B, respectively corresponding to the
projection into two and five empirical eigenvectors. For each one of
the $16$ slices, the temporal and spatial averaged empirical
quadratic errors, associated with the estimates of the response,
computed with the \textit{fmrilm.m} MatLab function, and with the proposed
multivariate Hilbert--valued mixed effect approach,
 respectively denoted as $EFMSE_{\boldsymbol{Y}_{i}^{fMRI}}$ and
$EFMSE_{\boldsymbol{Y}_{i}^{H}},$  are displayed in Tables
\ref{A4:tab:20}--\ref{A4:tab:21}.

\bigskip

\begin{table}[H]
\caption[\hspace{0.7cm} Comparative study of the empirical functional mean square errors on the estimation of the response for case A.]{\small{$EFMSE_{\boldsymbol{Y}_{i}^{fMRI}}$ and
$EFMSE_{\boldsymbol{Y}_{i}^{H}}$ for case A.}}
\centering
\begin{small}
\begin{tabular}{|c||c|c|c|}
 \hline
  Slices $S_i$ & $EFMSE_{\boldsymbol{Y}_{i}^{fMRI}}$ & $EFMSE_{\boldsymbol{Y}_{i}^{H}}$  \\
  \hline \hline
   1 & $2.417(10)^-3$  &  $3.492(10)^-3$  \\
  \hline
    2 & $3.051(10)^-3$ & $3.119(10)^-3$ \\
  \hline
     3 & $4.293(10)^-3$ & $5.523(10)^-3$  \\
  \hline
     4 & $6.666(10)^-3$ & $7.690(10)^-3$ \\
  \hline
     5 & $8.986(10)^-3$ & $9.961(10)^-3$  \\
  \hline
      6 & $8.462(10)^-3$ &$9.434(10)^-3$ \\
  \hline
     7 & $1.108(10)^-2$ & $1.920(10)^-2$  \\
  \hline
    8 & $1.720(10)^-2$ & $2.720(10)^-2$  \\
  \hline
     9 & $1.499(10)^-2$ & $1.914(10)^-2$ \\
  \hline
     10 & $1.036(10)^-2$ & $1.851(10)^-2$  \\
  \hline
     11 & $1.308(10)^-2$ & $1.634(10)^-2$ \\
  \hline
      12 & $1.302(10)^-2$ & $1.300(10)^-2$ \\
  \hline
     13 & $7.850(10)^-3$ & $7.939(10)^-3$  \\
  \hline
    14 & $6.640(10)^-3$ & $6.730(10)^-3$ \\
  \hline
     15 & $3.511(10)^-3$ & $2.832(10)^-3$  \\
  \hline
     16 & $2.771(10)^-3$ & $3.540(10)^-3$ \\
  \hline
\end{tabular}
\end{small}
  \label{A4:tab:20}
\end{table}

\begin{table}[H]
\caption[\hspace{0.7cm} Comparative study of the empirical functional mean square errors on the estimation of the response for case B.]{\small{$EFMSE_{\boldsymbol{Y}_{i}^{fMRI}}$ and
$EFMSE_{\boldsymbol{Y}_{i}^{H}}$ for case B.}}
\centering
\begin{small}
\begin{tabular}{|c||c|c|}
 \hline
  Slices $S_i$ & $EFMSE_{\boldsymbol{Y}_{i}^{fMRI}}$ & $EFMSE_{\boldsymbol{Y}_{i}^{H}}$  \\
  \hline \hline
   1 & $2.417(10)^-3$  &  $2.592(10)^-3$  \\
  \hline
    2 & $3.051(10)^-3$ & $3.119(10)^-3$ \\
  \hline
     3 & $4.293(10)^-3$ & $4.733(10)^-3$  \\
  \hline
     4 & $6.666(10)^-3$ & $7.671(10)^-3$  \\
  \hline
     5 & $8.986(10)^-3$ & $9.065(10)^-3$  \\
  \hline
      6 & $8.462(10)^-3$ &$8.435(10)^-3$ \\
  \hline
     7 & $1.108(10)^-2$ & $1.120(10)^-2$  \\
  \hline
    8 & $1.720(10)^-2$ & $1.919(10)^-2$ \\
  \hline
     9 & $1.499(10)^-2$ & $1.524(10)^-2$  \\
  \hline
     10 & $1.036(10)^-2$ & $1.040(10)^-2$  \\
  \hline
     11 & $1.308(10)^-2$ & $1.481(10)^-2$  \\
  \hline
      12 & $1.302(10)^-2$ & $1.299(10)^-2$ \\
  \hline
     13 & $7.849(10)^-3$ & $7.929(10)^-3$  \\
  \hline
    14 & $6.640(10)^-3$ & $ 6.719(10)^-3$  \\
  \hline
     15 & $3.511(10)^-3$ & $ 2.829(10)^-3$ \\
  \hline
     16 & $2.771(10)^-3$ & $3.540(10)^-3$ \\
  \hline
\end{tabular}
\end{small}
  \label{A4:tab:21}
\end{table}

\bigskip

It can be observed, in Tables \ref{A4:tab:20}--\ref{A4:tab:21}, that the performance of the
two approaches is very similar. However, the advantage of the
presented approach relies on the important dimension reduction it
provides, since, as commented before, we have considered the
truncations orders $TR=2$ (Case A) and $TR=5$ (Case B). Note that,
for each slice, the parameter vector has dimension $2\times
\times (64 \times 64),$ in the model fitted by \textit{fmrilm.m} Matlab function.
While the presented approach fits the functional projected model, that, for the
the cases A and B studied, is defined in terms of a parameter vector $\boldsymbol{\beta }$
 with  dimension $2\times 2$ and
$2\times 5,$ respectively.  Furthermore, the iterative estimation method implemented in \textit{fmrilm.m} requires several steps,  repeated at each one of the $64\times 64$ voxels in  the $16$ slices (data pre--whitening, ordinary least-squares estimation of $\boldsymbol{\beta}$, and AR(1) correlation coefficient estimation iterations, jointly with the  spatial smoothing of the temporal correlation - reduced bias - parameter estimates).

\begin{figure}[H]
   \centering
    \includegraphics[width=0.8\textwidth]{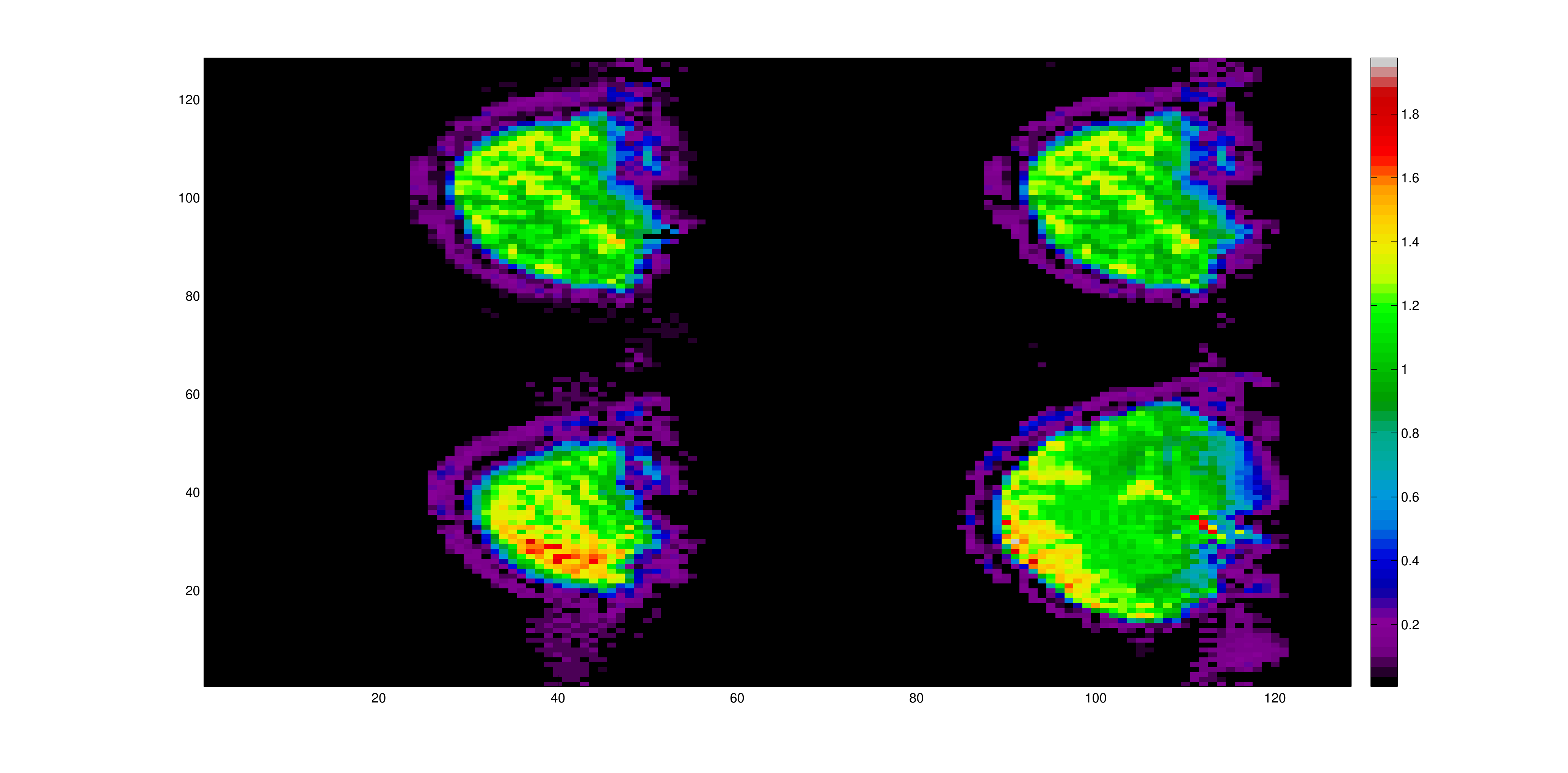}
    
    \vspace{-0.1cm}
    \caption[\hspace{0.7cm} Averaged in time (frames $5$--$68$) estimated response values for slices $1, 5, 10$ and $15,$ obtained by applying \textit{fmrilm.m} MatLab function.]{\small{Averaged in time (frames $5$--$68$) estimated response values for slices $1, 5, 10$ and $15,$ obtained by applying \textit{fmrilm.m} MatLab function.}}

\vspace{-0.2cm}
  \label{A4:fig:7}
\end{figure}

\begin{figure}[H]
   \centering
    \includegraphics[width=0.8\textwidth]{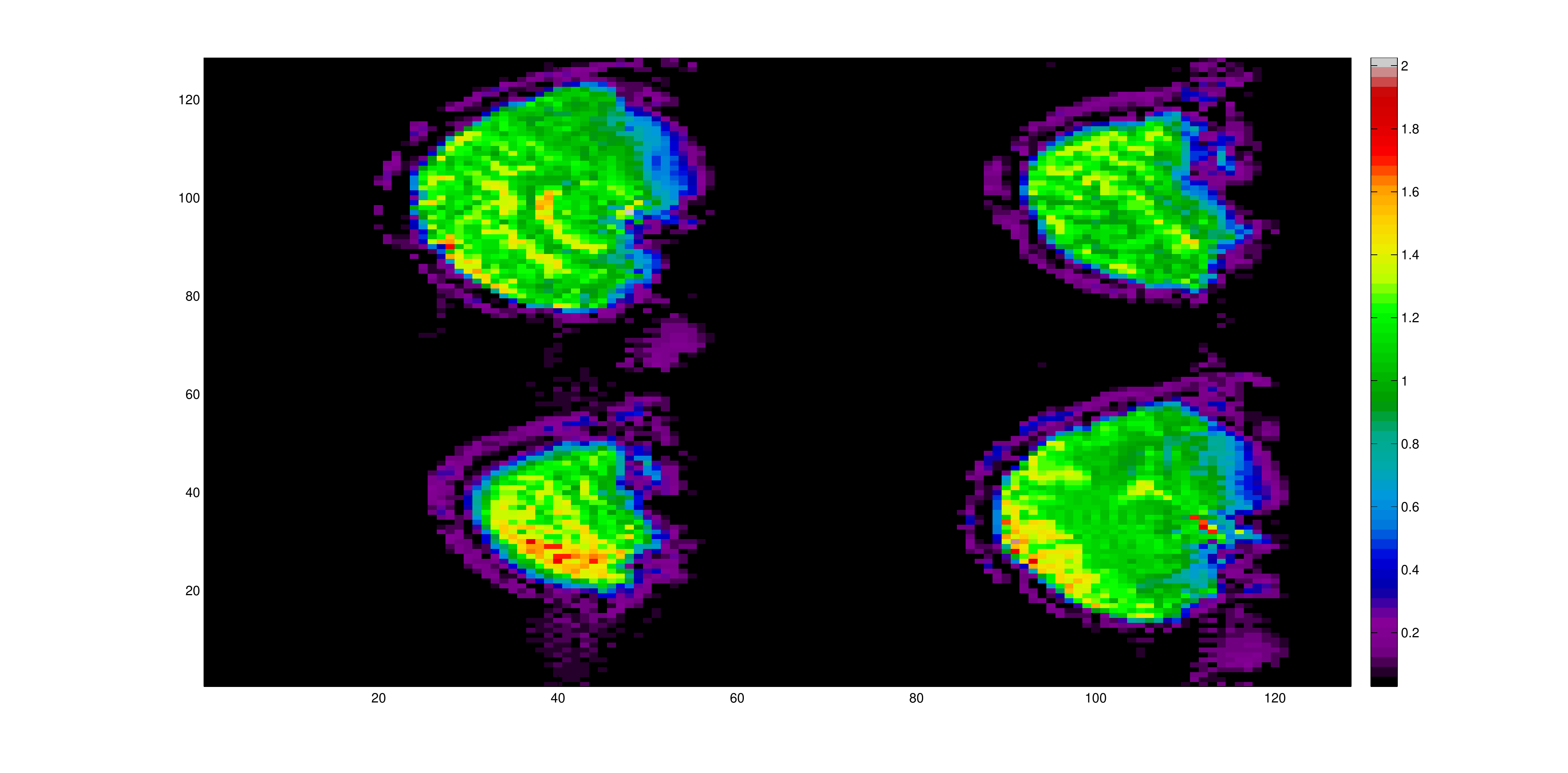}
    
    \vspace{-0.1cm}
    \caption[\hspace{0.7cm} Averaged in time (frames $5$--$68$) estimated response values for slices $1, 5, 10$ and $15,$ obtained by applying the  fixed effect approach with ARH(1) error  term, for case  A.]{\small{Averaged in time (frames $5$--$68$) estimated response values for slices $1, 5, 10$ and $15,$ obtained by applying the  fixed effect approach with ARH(1) error  term, for case  A.}}
  \label{A4:fig:9}
\end{figure}

\begin{figure}[H]
   \centering
    \includegraphics[width=0.8\textwidth]{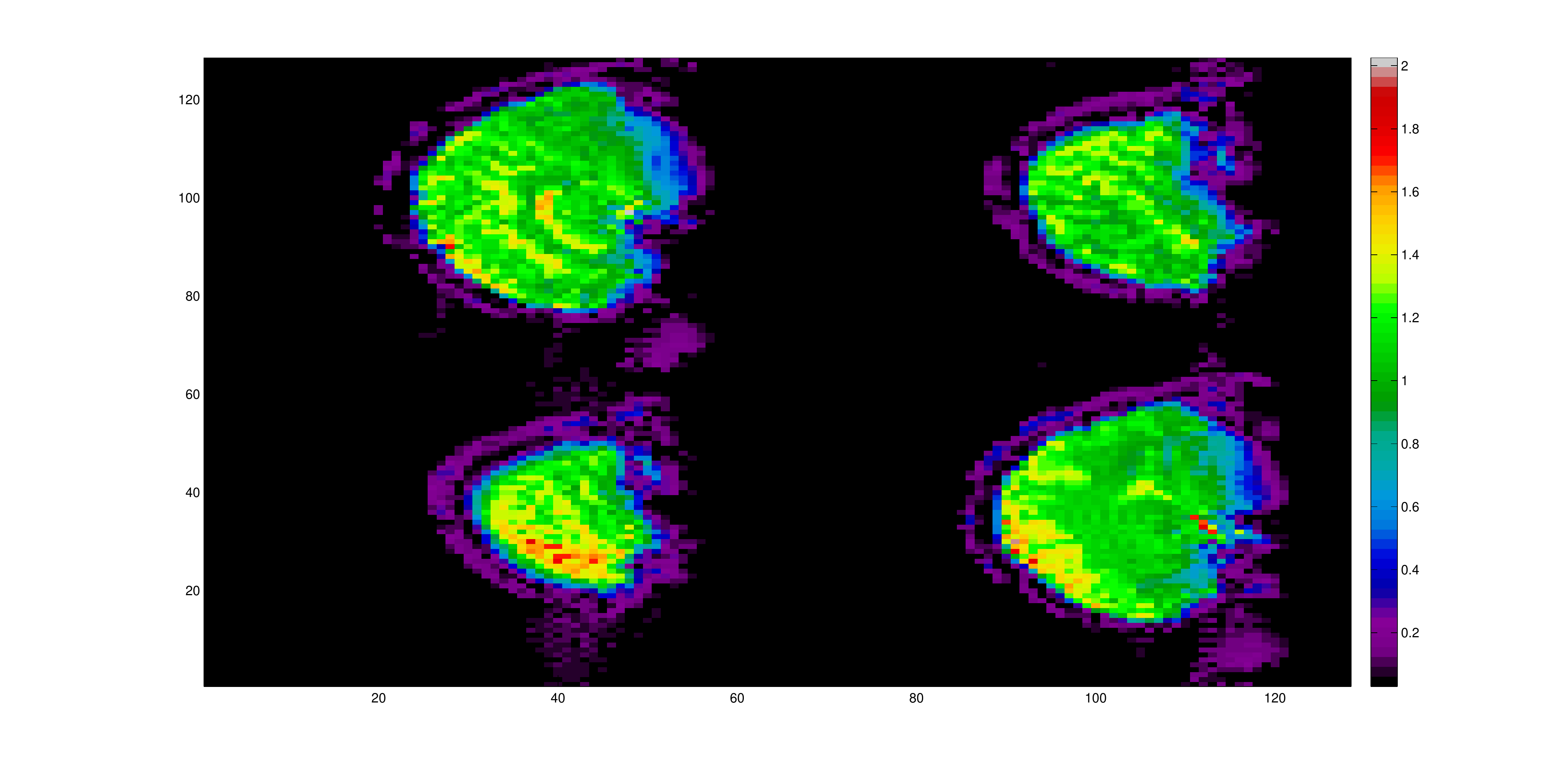}
    
    \vspace{-0.1cm}
    \caption[\hspace{0.7cm} Averaged in time (frames $5$--$68$) estimated response values for slices $1, 5, 10$ and $15,$ obtained by applying the  fixed effect approach with ARH(1) error term, for case  B.]{\small{Averaged in time (frames $5$--$68$) estimated response values for slices $1, 5, 10$ and $15,$ obtained by applying the  fixed effect approach with ARH(1) error term, for case  B.}}
  \label{A4:fig:12}
\end{figure}

\begin{figure}[H]
   \centering
    \includegraphics[width=0.8\textwidth]{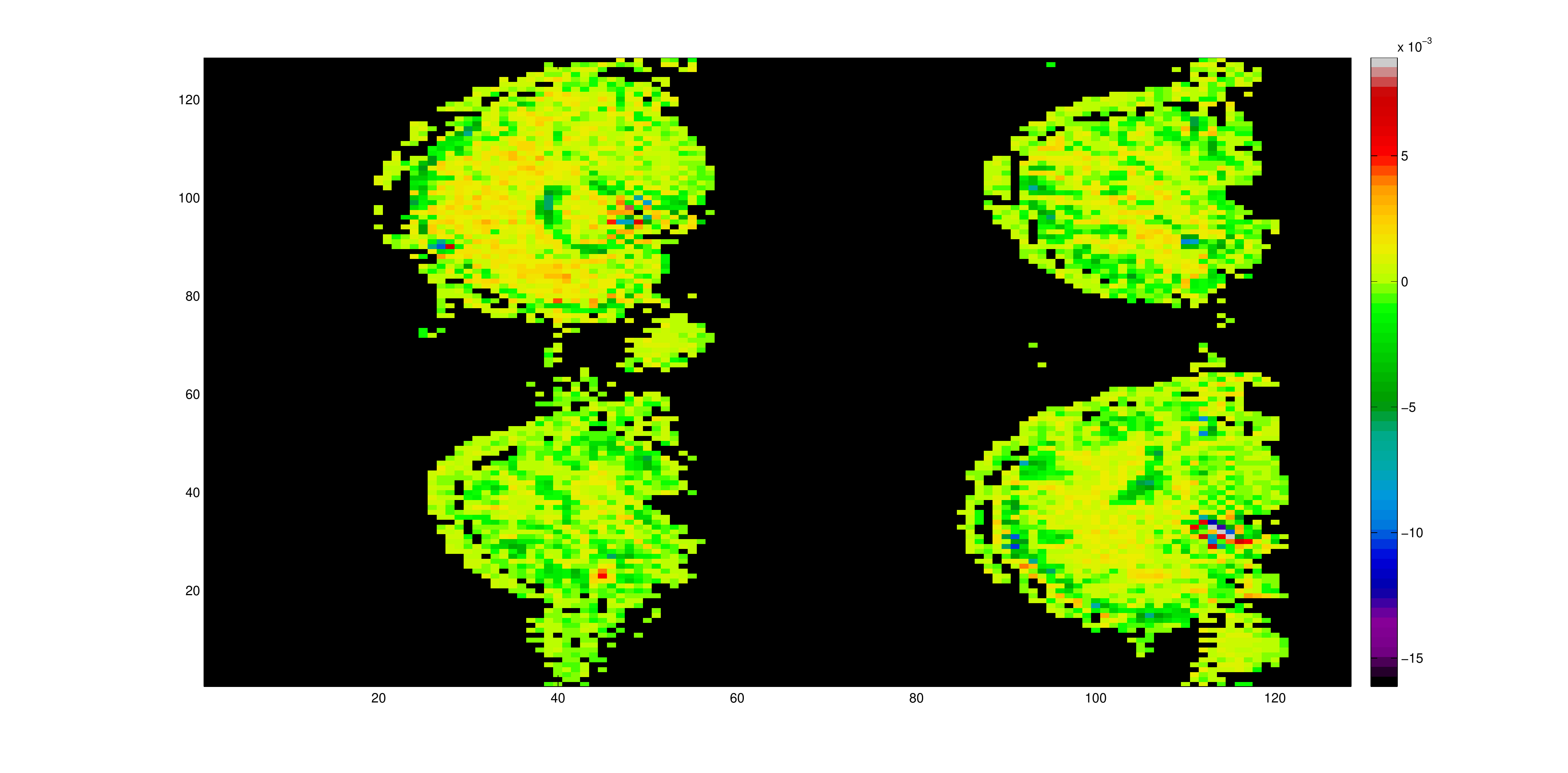}
    
    \vspace{-0.1cm}
    \caption[\hspace{0.7cm} Averaged in time (frames $5$--$68$) empirical errors  for slices $1, 5, 10$ and $15,$ obtained by applying \textit{fmrilm.m} MatLab function.]{\small{Averaged in time (frames $5$--$68$) empirical errors  for slices $1, 5, 10$ and $15,$ obtained by applying \textit{fmrilm.m} MatLab function.}}
  \label{A4:fig:8}
\end{figure}

\begin{figure}[H]
   \centering
    \includegraphics[width=0.8\textwidth]{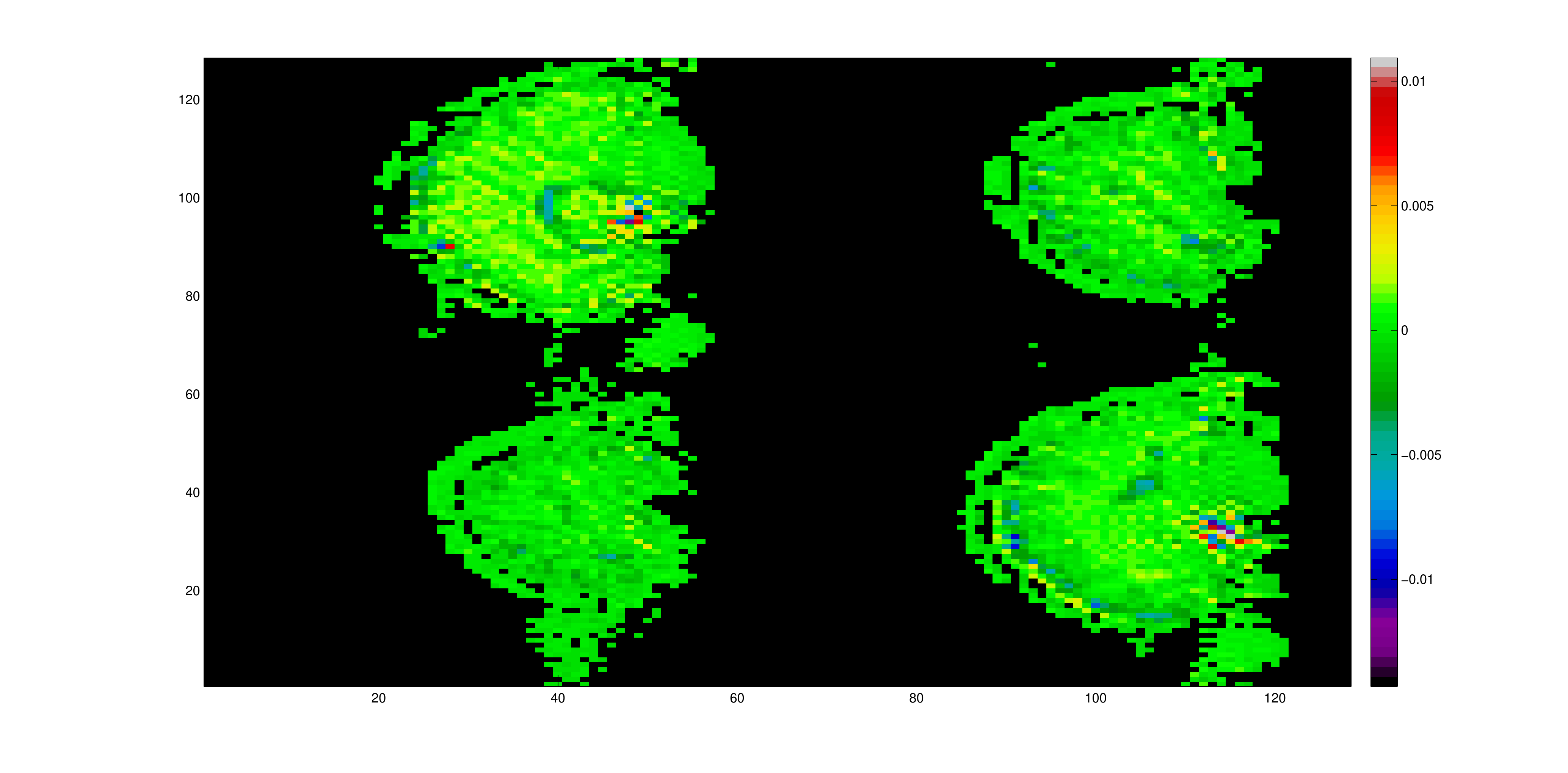}
    
    \vspace{-0.1cm}
    \caption[\hspace{0.7cm} Averaged in time (frames $5$--$68$) empirical errors  for slices $1, 5, 10$ and $15,$ obtained by applying the  fixed effect approach with ARH(1) error  term, for case  A.]{\small{Averaged in time (frames $5$--$68$) empirical errors  for slices $1, 5, 10$ and $15,$ obtained by applying the  fixed effect approach with ARH(1) error  term, for case  A.}}
  \label{A4:fig:10}
\end{figure}

\begin{figure}[H]
   \centering
    \includegraphics[width=0.8\textwidth]{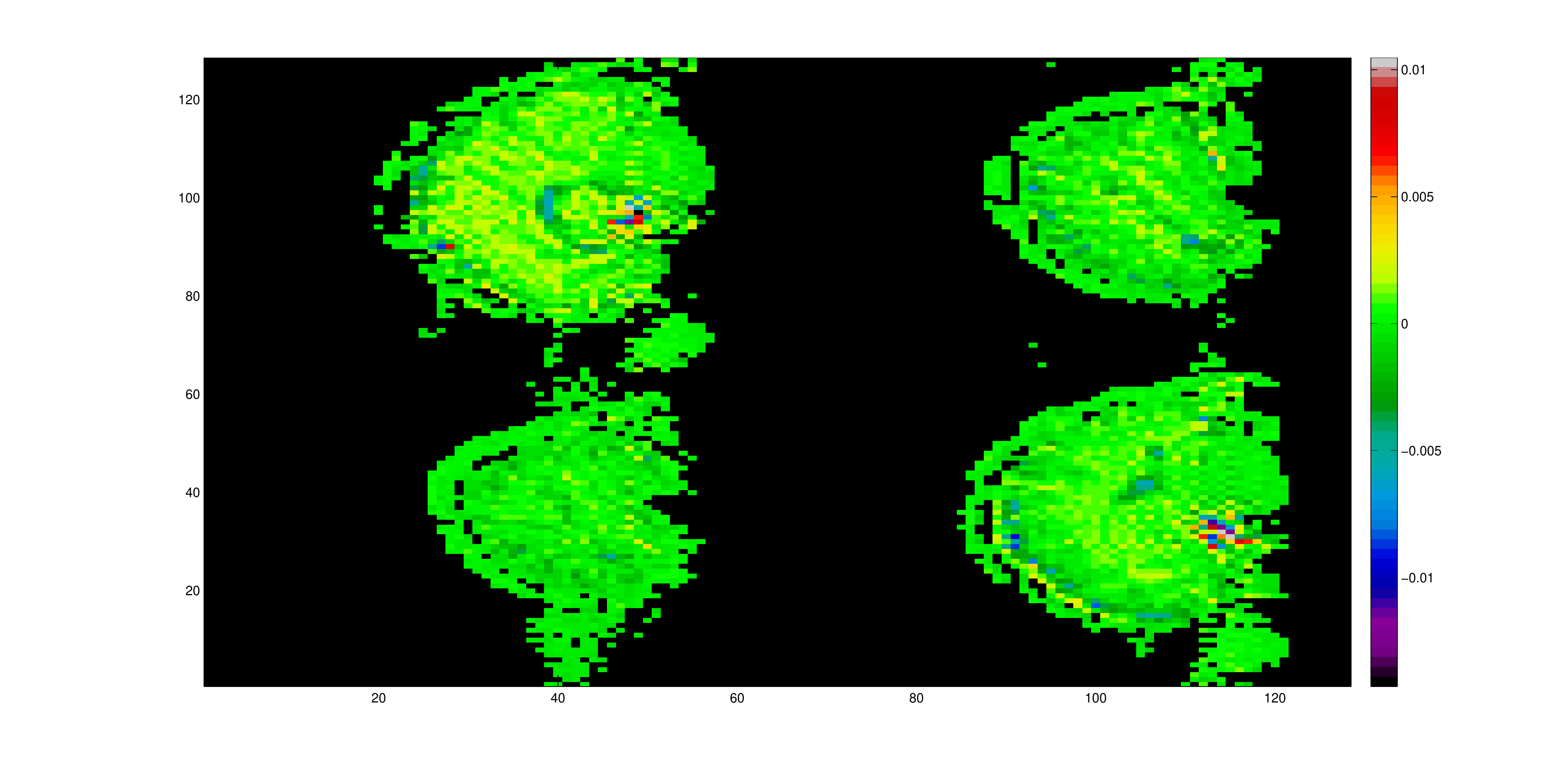}
    
    \vspace{-0.1cm}
    \caption[\hspace{0.7cm} Averaged in time (frames $5$--$68$) empirical errors  for slices $1, 5, 10$ and $15,$ obtained by applying the  fixed effect approach with ARH(1) error  term, for case  B.]{\small{Averaged in time (frames $5$--$68$) empirical errors  for slices $1, 5, 10$ and $15,$ obtained by applying the  fixed effect approach with ARH(1) error  term, for case  B.}}
  \label{A4:fig:11}
\end{figure}

For the slices $1,$ $5,$ $10$ and $15,$ the temporal averaged (frames $5$--$68$) estimated
values of the response, applying \textit{fmrilm.m}
MatLab function, and the fixed effect model with ARH(1) error term, in cases A and B, are respectively displayed in Figures  \ref{A4:fig:7}--\ref{A4:fig:12}.  The corresponding  empirical time-averaged quadratic  errors
are displayed in
Figures
\ref{A4:fig:8}--\ref{A4:fig:11}, respectively.

\bigskip

\textcolor{Crimson}{\subsection{Significance test}}

We are interested in contrast the significance of the spatial varying  parameter
vector $\boldsymbol{\beta}$ combining the effects of the two stimulus considered on the overall brain, in its real-valued, and $H^{2}$-valued version.
The  $F$ statistic in the MatLab function \textit{fmrilm.m} (fMRI
linear model), computed, as before,  from a single run of fMRI data, leads to the  results reflected in Table \ref{A4:tab:Tabla1fs},  on the percentage of brain voxels,   where
the real--valued fixed effect model with AR(1) term  is significative, for each one of the $16$ slices considered.

\bigskip

\begin{table}[H]
\caption[\hspace{0.7cm} Percentage of brain voxels per slice, where the real--valued fixed effect model with AR(1) error term, fitted by \textit{fmrilm.m} MatLab function, is significative.]{\small{Percentage of brain voxels per slice, where the real-valued fixed effect model with AR(1) error term, fitted by \textit{fmrilm.m} MatLab function, is significative.}}
\centering
\begin{small}
\begin{tabular}{|c||c|}
  \hline
 $S$ & \% voxels with rejection of $H_{0}$\\
  \hline \hline
   $1$ & $99.927 \%$ \\
  \hline
   $2$ & $99.927 \%$ \\
  \hline
   $3$ & $99.707 \%$ \\
  \hline
     $4 $& $99.902 \%$ \\
  \hline
     $5 $& $99.805 \%$ \\
  \hline
      $6$ & $99.951 \%$ \\
  \hline
     $7$ & $99.927 \%$ \\
  \hline
    $8$ & $99.976 \% $\\
  \hline
   $9 $&  $99.805 \%$ \\
  \hline
  $10$ & $99.951 \%$ \\
  \hline
    $ 11 $&$ 99.951 \% $\\
  \hline
    $12$ &  $99.902 \%$ \\
  \hline
    $13 $& $99.878 \%$ \\
  \hline
     $14$ & $99.951 \%$ \\
  \hline
    $15$ & $99.951 \%$ \\
  \hline
     $16$ & $100 \%$ \\
  \hline
\end{tabular}
\end{small}
  \label{A4:tab:Tabla1fs}
\end{table}

\bigskip

 As described in \textcolor{Crimson}{Appendix}  \ref{A4:scwth}, for each
slice, i.e., for $i=1,\dots,16,$ the value of the conditional chi--squared test
statistics $T_{\boldsymbol{h}},$ in equation (\ref{A4:78}), is computed, from four
realizations of a Gaussian random function $\boldsymbol{h},$ generated from a Gaussian random field
$\xi ,$ satisfying equation (\ref{A4:eqdlrect}) on the rectangle
containing each brain slice. As before,
the functional response sample size at
each slice is $64,$ since the first four frames are discarded.
It
can be observed, in the numerical results displayed in Table \ref{A4:tab:Tabla1fmri},  for $TR=16,$ and  in Table  \ref{A4:tab:Tabla2fmri}, for  $TR=4,$ that the null hypothesis is rejected, in  most of the random directions  in all the brain slices; i.e., the functional fixed effect model with ARH(1) error term is significative for $\alpha =0.05.$  Note that a very few $p$--values are slightly larger than $\alpha =0.05,$ with very small difference, that could be produced by the numerical errors accumulated, due to the presence of  small values to be inverted.
Thus, we can conclude the suitability of  our approach, to combine the effects of  the  scans hot stimulus, and the scans warm stimulus, in a functional spatially continuous framework.

\bigskip

\begin{table}[H]
\caption[\hspace{0.7cm} Functional testing at the $16$ slices, considering four random directions,  for $TR=16$.]{\small{$p$-values for $T_{\boldsymbol{h}}$ computed at the $16$ slices, considering four random directions,  for $TR=16$.}}

\centering
\begin{small}
\begin{tabular}{|c||c|c|c|c|}
  \hline
 $S$ & $D_1$ & $D_2$ & $D_3$ & $D_4$ \\
  \hline \hline
   $1$ &$ 0$ &$ 0 $& $0.082$ & $0.023$ \\
  \hline
  $ 2 $& $0.590(10)^{-2}$  & $0$ & $0$ & $ 0$ \\
  \hline
   $3 $& $0.018$ & $0.066$ & $0.049$ & $0.030$ \\
  \hline
    $ 4$ & $0$ & $0$  & $0$  & $0.170(10)^{-10}$ \\
  \hline
     $5$ & $0$ & $0.026$ &$ 0$ &$ 0$  \\
  \hline
      $6$ & $0$ & $0 $ & $0$ & $0 $\\
  \hline
     $7$ & $0.710(10)^{-7}$ & $0 $ & $0 $& $0$ \\
  \hline
    $8 $& $0$ & $0.006$ & $0$ & $0$ \\
  \hline
   $9$ & $0.049$ & $0$ & $0$  & $0.023$ \\
  \hline
  $10 $& $0.390(10)^{-7}$  & $0.031$   & $0$  & $0$\\
  \hline
   $11$ & $0.004$ & $0.006$ & $0.660(10)^{-6}$ & $0.052$  \\
  \hline
    $12$ & $0.046$ & $0$ &$ 0$ & $0.034$ \\
  \hline
    $13 $& $0.340(10)^{-7}$ & $0.028$ & $0$ & $0.440(10)^{-3}$ \\
  \hline
     $14$ & $0$ & $0.180(10)^{-6}$& $0.021$ & $0.050$ \\
  \hline
    $15 $& $0$ & $0.140(10)^{-7}$  & $0.044$   & $0.052$ \\
  \hline
    $ 16$ & $0.110(10)^{-4}$ & $0.230(10)^{-7}$ & $0$  & $0$ \\
  \hline
\end{tabular}
\end{small}
  \label{A4:tab:Tabla1fmri}
\end{table}

\bigskip

\begin{table}[H]
\caption[\hspace{0.7cm} Functional testing at the $16$ slices, considering four random directions, for  $TR=4$.]{\small{$p$-values for $T_{\boldsymbol{h}}$ computed at the $16$ slices, considering four random directions, for  $TR=4$.}}
\centering
\begin{small}
\begin{tabular}{|c||c|c|c|c|}
  \hline
 $S$ & $D_1$ & $D_2$ & $D_3$ & $D_4$ \\
  \hline \hline
  $ 1$ & $0 $  & $0.051$ & $0.071$  & $0.011$\\
  \hline
   $2 $&  $0.880(10)^{-4}$ &$ 0$ & $0$  &$ 0 $\\
  \hline
   $3$ &  $0.067$ & $0.034$ & $0$ & $0.037$\\
  \hline
     $4$ &  $0$  &  $0.250(10)^{-4}$ & $0.110(10)^{-4}$ & $0.016$\\
  \hline
     $5$ & $0.370(10)^{-6}$  &$0$ & $0.280(10)^{-6}$ & $ 0$\\
  \hline
     $ 6 $&  $0.001$ & $0$ & $0 $ & $0.220(10)^{-4}$ \\
  \hline
     $7$ & $0.064$  & $0.034$& $0.007$ & $0.044$\\
  \hline
    $8 $& $0.072$  & $0.079$ &  $0.035$ & $0$\\
  \hline
  $ 9$ &  $0.220(10)^{-5}$ & $0.470(10)^{-4}$ & $0.004$ & $0.220(10)^{-9}$\\
  \hline
  $10$ &  $0$ & $0.120(10)^{-3}$ & $0.370(10)^{-4}$ & $0.970(10)^{-7}$\\
  \hline
   $  11$ &  $0.081$  & $0.058$ & $0$&$ 0$ \\
  \hline
   $ 12 $&  $0.870(10)^{-4}$ & $0$ & $0 $ & $0.036$\\
  \hline
   $ 13$ &  $0.760(10)^{-3}$  & $0$ &$ 0 $ & $0.370(10)^{-3}$\\
  \hline
   $  14$ & $0.210(10)^{-6}$  & $0$ & $0$ & $0.037$\\
  \hline
   $ 15 $&  $0 $ & $0.650(10)^{-4}$ & $0.032$ & $0$\\
  \hline
    $ 16$ & $0.540(10)^{-6}$ & $0 $& $0 $& $0.520(10)^{-3}$\\
  \hline
\end{tabular}
\end{small}
  \label{A4:tab:Tabla2fmri}
\end{table}

Comparing results in Tables \ref{A4:tab:Tabla1fs}--\ref{A4:tab:Tabla2fmri}, we can conclude that both methodologies, the one presented in   \cite{Worsleyetal02}, and the  functional  approach introduced here,  lead to similar results regarding the significance of the models they propose, respectively based on  spatial varying real--valued multiplicative coefficients with AR(1) error term, and Hilbert--valued coefficients with ARH(1) error term.

%
%

\textcolor{Crimson}{\section{Conclusions} \label{A4:sec:64}}

As shown in the simulation study, the boundary conditions affect the decay velocity at the boundary of the  covariance kernels,
defining the functional entries of the matrix covariance operator of the  error term. Thus, the dependence
range of the error components is directly affected by the boundary conditions.
A better performance of the generalized   least--squares estimator of the parameter vector $\boldsymbol{\beta }$  is observed,  when a fast continuous decay is displayed by the error covariance kernels close to the boundary, as it is observed in the circular domains.
Furthermore,  in the simulation study undertaken, and in the real--data problem addressed,  a good performance of the computed generalized least--squares estimator, and of the test statistics is observed  for low truncation orders. Thus, an important dimension reduction is achieved with the presented approach. Summarizing, the proposed approach  allows the incorporation of temporal and spatial correlations in the analysis,  with an important  dimension reduction.

The derivation of similar results under alternative boundary conditions like Neumann and Robin boundary conditions constitutes an open research problem (see, for
example, \cite{GrebenkovNguyen13}). Another important research problem is to address the same analysis under a slow decay of the error covariance kernels at the boundary
(see, for example, \cite{Friasetal17,Jiang12,Jiang16,Tong11}, beyond the Gaussian context).

\textcolor{Crimson}{\section{Supplementary Material} \label{A4:Supp}}

The eigenvectors and eigenvalues of the Dirichlet negative Laplacian
operator on the regular domains defined by the rectangle, disk and
circular sector are  described here (see, for example,
\cite{GrebenkovNguyen13}). It is well--known that the negative Laplacian
operator $\left(-\boldsymbol{\Delta}_D\right)$ on a regular bounded open domain
$D\subset \mathbb{R}^{2},$ with Dirichlet boundary conditions, is
given by

\begin{equation}
-\boldsymbol{\Delta}_D (f)(x_{1},x_{2})=
-\frac{\partial^2}{\partial x_{1}^{2}}f(x_{1},x_{2})
-\frac{\partial^2}{\partial
x_{2}^{2}}f(x_{1},x_{2}), \quad  f(x_{1},x_{2})=0, \quad (x_{1},x_{2})\in \partial D, \quad D\subseteq \mathbb{R}^2, \nonumber
\end{equation}
\noindent where $\partial D$ is the boundary of $D.$ In the
subsequent development, we will denote by $\left\lbrace
\phi_k, \ k \geq 1 \right\rbrace $ and \linebreak $\left\lbrace \lambda_k
(-\boldsymbol{\Delta}_D), \ k \geq 1 \right\rbrace$ the respective
eigenvectors and eigenvalues of $\left(-\boldsymbol{\Delta}_D\right),$ that
satisfy

\begin{eqnarray}
-\boldsymbol{\Delta}_D \phi_k \left(\mathbf{x}\right) &=& \lambda_k (-\boldsymbol{\Delta}_D)
\phi_k
\left(\mathbf{x}\right)~\left(\mathbf{x} \in D \subseteq \mathbb{R}^2 \right), \nonumber \\ 
\phi_k \left(\mathbf{x}\right) &=& 0 ~\left(\mathbf{x}
\in
\partial D \right),~\forall k \geq 1,\nonumber
\end{eqnarray}
\noindent for $D$ being one of the following three domains:
$$D_{1}=\prod_{i=1}^{2} \left[a_i, b_i \right], \quad D_{2}=\left\lbrace
\mathbf{x} \in \mathbb{R}^2:~R_0 < \Vert \mathbf{x} \Vert < R
\right\rbrace,$$ and $$D_{3}=\left\lbrace \mathbf{x} \in
\mathbb{R}^2:~R_0 < \Vert \mathbf{x} \Vert < R ,\ \mbox{and} \ 0 <
\varphi < \pi \theta \right\rbrace.$$

\textcolor{Crimson}{\subsection{Eigenelements of Dirichlet negative Laplacian operator on rectangles}
\label{A4:sec:23}}

 Let us first consider domain $$D_{1}=\prod_{i=1}^{2}
\left[a_i, b_i \right].$$ The eigenvectors
$\left\lbrace \phi_{\mathbf{k}}, \ \mathbf{k}\in \mathbb{N}^{2}_{*} \right\rbrace$ and
eigenvalues
$\left\lbrace \lambda_\mathbf{k}(-\boldsymbol{\Delta}_{D_{1}}), \ \mathbf{k}\in
\mathbb{N}^{2}_{*} \right\rbrace$ of $-\boldsymbol{\Delta}_{D_{1}}$ are given by
(see \cite{GrebenkovNguyen13}):

\begin{eqnarray}
\phi_{\mathbf{k}}\left(\mathbf{x}\right)&=&
\phi_{k_1}^{(1)}\left(x_1\right)\phi_{k_2}^{(2)}\left(x_2\right), ~
\lambda_{\mathbf{k}}= \lambda_{k_1}^{(1)} + \lambda_{k_2}^{(2)},\nonumber  \\
\phi_{k_{i}}^{(i)}\left(x_i\right) &=&
\sin \left(\frac{\pi k_i
x_i}{ l_i}\right), \quad x_i \in \left[a_i, b_i \right],~i=1,2,\nonumber \\
\lambda_{k_{i}}^{(i)} &=& \frac{\pi^2 k_i^2}{l_{i}^{2}},\quad 
k_{i}\geq 1,~i=1,2, \nonumber \\ \label{A4:19}
\end{eqnarray}
\noindent where $l_i = b_i - a_i,$ for $i=1,2.$

\textcolor{Crimson}{\subsection{Eigenelements of Dirichlet negative Laplacian operator on disks}
\label{A4:secdisk}}

In general, for the circular annulus
$$\widetilde{D}_{2} = \left\lbrace \mathbf{x} \in \mathbb{R}^2:~R_0 < \Vert
\mathbf{x} \Vert < R \right\rbrace,$$  its rotation
symmetry allows us to define
$-\boldsymbol{\Delta}_{\widetilde{D}_{2}}$ in polar coordinates as

\begin{equation}
 -\boldsymbol{\Delta}_{\widetilde{D}_{2}} =
-\frac{\partial^2}{\partial r^2} -
\frac{1}{r}\frac{\partial}{\partial r} -
\frac{1}{r^2}\frac{\partial^2}{\partial\varphi^2}, \quad  x_1 = r \cos \varphi,~x_2 = r \sin \varphi. \nonumber 
\end{equation}

The application of variable separation method then  leads
to the following  explicit formula of its eigenfunctions (see, for
example, \cite{GrebenkovNguyen13})

\begin{equation}
\phi_{khl} \left(r,\varphi \right) = \left[J_k \left(\alpha_{kh} r /
R \right) + c_{kh} Y_k \left(\alpha_{kh} r / R \right) \right]
\times  C_k\left(l \right), \label{A4:21}
\end{equation}
\noindent with
\[C_k(l) = \left\{
  \begin{array}{l}
    cos\left(k\varphi\right) \text{ l=1},\\
    sin\left(k\varphi\right) \text{ l=2 } \left(k \neq 0 \right),
  \end{array} \right.\]
\noindent where $\{J_k\left(z\right)\}$ and $\{Y_k\left(z\right)\}$
are the Bessel functions of order $k$ of first and second kind,
respectively, $$\{\lambda_{kh}
\left(-\boldsymbol{\Delta}_{\widetilde{D}_{2}}\right)=
\alpha_{kh}^{2} / R^2\}$$ are the corresponding  eigenvalues, and the
sets   $\left\lbrace \alpha_{k,h}, \ k\geq
1,~h=1,\dots,M(k) \right\rbrace$ and \linebreak $\left \lbrace c_{k,h}, \ k\geq
1,~h=1,\dots,M(k) \right\rbrace$ are defined from the boundary conditions at
$r=R$ and $r=R_{0}.$

If we focus on domain $D_{2},$ the disk, i.e.,  $R_0 = 0$, the
coefficients $\left \lbrace c_{k,h}, \ k\geq
1,~h=1,\dots,M(k) \right\rbrace$ are set to $0.$ The eigenfunctions then adopt
the following expression:

\begin{eqnarray}
\phi_{khl}\left(r,\varphi \right)  &=& J_k \left(\alpha_{kh} r / R
\right)C_{k}(l), \quad l=1,2, \nonumber \\ 
\end{eqnarray}

\noindent with eigenvalues
\begin{equation}\lambda_{kh}\left(-\boldsymbol{\Delta}_{D_{2}}\right)
= \frac{\alpha_{kh}^{2}}{R^2},\quad k\geq
1,~h=1,\dots,M(k),\nonumber 
\end{equation} \noindent where  $\left\lbrace \alpha_{k,h}, \ h=1,\dots,M(k) \right\rbrace$ are the $M(k)$ positive roots of
 the Bessel function $J_k
\left( z \right)$ of order  $k.$   Note that we can also consider
truncation at parameter $M(k)$ for  $k\geq 1,$ since this parameter
increases with the increasing of the radius $R.$

\textcolor{Crimson}{\subsection{Eigenelements of Dirichlet negative Laplacian operator on circular sectors}
\label{A4:seccsec}}

Lastly, we consider domain $D_{3},$ the circular
sector of radius $R$ and angle $0 < \varphi < \pi \theta .$ The
eigenvectors and eigenvalues are given by the following expression
(see, for example, \cite{GrebenkovNguyen13}):

\begin{eqnarray}
 \phi_{kh} \left(r,\varphi \right) &=&  J_{k/\theta} \left(\alpha_{kh} r / R \right) \sin \left(k\varphi / \theta \right),\quad r \in \left[0,R\right],\nonumber \\ 
 \lambda_{kh}\left(-\boldsymbol{\Delta}_{D_{3}}\right) &=& \frac{\alpha_{kh}^{2}}{R^2}, \
 k\geq 1,~h=1,\dots,M(k),\nonumber\\
 \label{A4:eqcsa}
\end{eqnarray}
\noindent with  $M(k)$ and  $\left\lbrace \alpha_{k,h}, \ k\geq 1,~h=1,\dots,M(k) \right\rbrace$  being given as in the
previous section.

\textcolor{Crimson}{\subsection{Asymptotic behavior of eigenvalues} \label{A4:sec:22}}

\textcolor{Crimson}{\subsubsection{The rectangle} \label{A4:sec:221}}

The  functional data sets generated in \textcolor{Crimson}{Appendix}  \ref{A4:sec:6} must
have a  covariance matrix operator with functional entries
(operators)  in the trace class.  We then apply  the results in
\cite{Widom63} to study the asymptotic order of eigenvalues of the
integral equation
\begin{equation}
\displaystyle \int_{\mathbb{R}^{2}} V^{1/2}(\mathbf{t})l_{\varepsilon_{i}}(\mathbf{t}-\mathbf{s})
V^{1/2}(\mathbf{s})f(\mathbf{s})d\mathbf{s}=\lambda f(\mathbf{t}). \nonumber 
\end{equation}

In our case,  $V$ is the indicator function on the
rectangle, i.e., on domain
 $D_{1},$ and  $l_{\varepsilon_{i}}$ is  the covariance kernel defining the square root
\begin{equation}R^{1/2}_{\varepsilon_{i}\varepsilon_{i}}=f_{i}(-\boldsymbol{\Delta}_{D_{1}})=(-\boldsymbol{\Delta}_{D_{1}})^{-(d-\gamma_{i})},\quad \gamma_{i} \in (0,d/2),\nonumber 
\end{equation}
\noindent  of the autocovariance operator of the Hilbert-valued
error component $\left\lbrace \varepsilon_{i}, \ i=1,\dots,n \right\rbrace,$
 with $$R_{\varepsilon_{i}\varepsilon_{i}}= R^{1/2}_{\varepsilon_{i}\varepsilon_{i}}R^{1/2}_{\varepsilon_{i}\varepsilon_{i}}.$$
 Note that with the choice made of functions $V$ and $\left\lbrace l_{\varepsilon_{i}}, \ i=1,\dots,n \right\rbrace,$ the conditions  assumed in \cite{Widom63} are satisfied. In particular,  the following asymptotic
holds:

\begin{equation}\lambda_{k}(R_{\varepsilon_{i}\varepsilon_{i}}^{1/2})=\mathcal{O}(k^{-2(d-\gamma_{i} )/d}),\quad k\longrightarrow \infty,\quad i=1,\dots,n, \nonumber 
\end{equation}
\noindent (see \cite[p. 279, Eq. (2)]{Widom63}). Also, in
general,  the eigenvalues of the Dirichlet negative Laplacian
operator on a
 regular bounded open domain $D$ satisfy
\begin{equation}
\gamma_{k}(-\Delta_{D} )\sim 4\pi\frac{\left(\Gamma \left( 1+\frac{d}{2}%
\right)\right)^{2/d}}{|D|^{2/d}}k^{2/d},\quad k\longrightarrow
\infty .  \nonumber 
\end{equation}

\textcolor{Crimson}{\subsubsection{Asymptotic behavior of zeros of Bessel functions.}
\label{A4;sec:222}}

As before,  $J_k \left( z \right)$ denotes the Bessel function
of the first kind of order $k.$ Let $\left\lbrace j_{k,h}, \ h=1,\dots,M(k) \right\rbrace$ be its $M(k)$ roots.  In
\cite{Elbert01,Olver51,Olver52}, it is shown
that, for a  fixed $h$ and large $k$, the Olver's expansion holds

\begin{equation}
j_{kh} \simeq k + \delta_h k^{1/3} + \mathcal{O}(k^{-1/3}),\quad
k\rightarrow \infty. \nonumber 
\end{equation}

On the other hand, for fixed $k$ and large $h$, the McMahon's
expansion also is satisfied (see, for example, \cite{Watson66})
\begin{equation}
j_{kh} \simeq \pi \left(h + k/2 - 1/4 \right) +
\mathcal{O}(h^{-1}),\quad h\rightarrow \infty. \nonumber 
\end{equation}

These results will be applied in \textcolor{Crimson}{Appendix}   \ref{A4:sec:6}, in the
definition of the eigenvalues of the covariance operators
$\left\lbrace \boldsymbol{R}_{\varepsilon_{i}\varepsilon_{i}}, \ i=1,\dots,n \right\rbrace,$  on
the disk and circular sector, to ensure their rapid decay to zero, 
characterizing  the trace operator class.

\textcolor{Crimson}{\section*{\textbf{Acknowledgments}}}

\textcolor{Aquamarine}{\textbf{This work has been supported in part by project MTM2015--71839--P (co-funded by Feder funds), of the DGI, MINECO, Spain.}}

\vspace{0.5cm}
\renewcommand\bibname{\textcolor{Crimson}{\textit{\textbf{References}}}}

\bibliographystyle{dinat}
\bibliography{Biblio}

\end{document}